%% file: main.tex
\documentclass[sigplan,10pt,nonacm]{acmart}
\pdfoutput=1
\usepackage{amsmath}
\usepackage{proof}
\usepackage{graphicx}
\usepackage{array}
\usepackage{xspace}
\usepackage[all]{xy}
\usepackage{macros}
\usepackage{ebproof}
\usepackage{stmaryrd}
\usepackage{rotating,color,xcolor}
\usepackage{tikz}
\usepackage{caption}
\usepackage{multirow}
\usetikzlibrary{automata, positioning, arrows, matrix}
\usepackage{multicol}
\begin{document}
 \title{First-order tree-to-tree functions}
 \author{Miko\l{}aj Boja\'nczyk and Amina Doumane}
 \begin{abstract}
    We study tree-to-tree transformations that can be defined in   first-order logic or  monadic second-order logic. We prove a decomposition theorem, which shows that every transformation can be obtained from  prime transformations, such as tree-to-tree homomorphisms or pre-order traversal, by using combinators such as function composition. 
\end{abstract}

 \maketitle

\input{intro}

\input{trees-and-logic}

\input{types-and-functions}
\input{stt}

\input{forational}

\input{stt-derivable-amina}

\input{extensions}

\bibliographystyle{plain}
\bibliography{bib}

\pagebreak
\appendix
\input{appendix-unfold}
\input{appendix-examples}

\input{appendix-to-fo}
\input{appendix-forational}
\input{appendix-stt}
\input{appendix-one-register}
\input{appendix-factfor}
\input{appendix-chain}

\end{document}

%% file: intro.tex
\section*{Erratum}
In an early version of this paper, Theorem~\ref{thm:normalise} was stated without the restriction that $\lambda$-terms to be normalized need to use a unique variable as a bound variable. This old version is not correct, as pointed to us by Lê Thành Dũng (Tito) Nguy\~{\^e}n. His counter-example can be found in Example~\ref{ex:tito}.

\section{Introduction}

The purpose of this paper is to decompose tree transformations into simple building blocks. An important inspiration  is the Krohn-Rhodes theorem~\cite[p.~454]{Krohn1965}, which says that every string-to-string function recognised by a Mealy machine can be decomposed into certain prime functions. 

\paragraph*{Regular functions.} The transformations studied in this paper are the  regular functions.

In~\cite[Theorem 13]{engelfrietMSODefinableString2001}, Engelfriet and Hoogeboom proved that deterministic two-way transducers recognise the same string-to-string functions as \mso transductions. Because of this and other properties -- such as closure under composition~\cite[Theorem 1]{chytilSerialComposition2Way1977} and decidable equivalence~\cite[Th.~1]{gurariEquivalenceProblemDeterministic1982} --  this class of functions is now called the \emph{regular string-to-string functions}. Other  equivalent descriptions of the regular functions include: string transducers of Alur and {\v C}ern{\'y}~\cite{alurExpressivenessStreamingString2010}, and several models based on combinators~\cite{alur2014regular,daveGastinKrishna18, bojanczykRegularFirstOrderList2018}. 
 

There are also regular functions for trees, which can be defined using any of the following equivalent models: \mso tree-to-tree transductions~\cite[Section 3]{bloem_comparison_2000}, single use attributed tree grammars~\cite{bloem_comparison_2000}, macro tree transducers that are single use~\cite{ENGELFRIET199934} or of linear size increase~\cite[Theorem 7.1]{engelfriet_macro_2003}, and streaming tree transducers~\cite[Theorem 4.6]{alur2017streaming}. 

The goal of this paper is to prove a decomposition result for regular tree-to-tree functions. As in the Krohn-Rhodes theorem, we want to show that every such function can be obtained by combining certain prime functions.  

\paragraph*{First-order transductions. } Although \mso transductions are the more popular model, we work mainly with the less expressive model of first-order transductions. Why?

As we explain in Section~\ref{sec:mso-trans}, every \mso tree-to-tree transduction can be decomposed as: (a) first, a relabelling defined in \mso, which does not change the tree structure; followed by (b) a first-order tree-to-tree transduction. In this sense, as far as transformations of the tree structure are concerned,  first-order and \mso transductions have the same expressive power. Another argument for the importance of first-order tree-to-tree transductions is a connection with the $\lambda$-calculus. As we explain in Section~\ref{sec:stt-derivable}, first-order tree-to-tree transductions are expressive enough to capture evaluation of $\lambda$-terms (assuming the use of a single variable and linearity), and such evaluation turns out to be one of the core computational steps implicit in a tree-to-tree transduction. 

Another advantage of first-order logic on trees, compared to \mso, is  a better decomposition theory, in the sense of decomposing formulas into  simpler ones~\cite{haferthomas,bojanczykDecidablePropertiesTree2004,esik-weil1}. 
For our paper, the most useful decomposition is a remarkable theorem of Schlingloff, which says that first-order logic on trees is equivalent to a certain two-way variant  of {\sc ctl}~\cite[Th.~4.5]{schlingloff1992expressive}. In contrast, there are no such results for \mso. 


Summing up, we believe that first-order  tree transformations are  expressive, have a strong theory, and deserve to leave the shadow of their better known \mso cousin.

\paragraph*{Structured datatypes.} We present our main decomposition result in a formalism based on  functional programming (in a combinatory variant, i.e.~without variables), with structured datatypes such as pairs or co-pairs.  The motivation behind this approach  -- which is inspired by~\cite{bojanczykRegularFirstOrderList2018} -- is to avoid encoding datatypes in our constructions using syntactic annotation such as endmarkers and separators. Thanks to the  structured datatypes,  we can use established operations such as {\tt map}, and we can assign informative types to our functions, such as $\Sigma_1 \times \Sigma_2 \to \Sigma_i$ for projection, as opposed to saying that all functions input and output trees.  

The choice of datatypes for trees is harder than for the string case that was studied in~\cite{bojanczykRegularFirstOrderList2018}.  The difficulty is in splitting the input into smaller pieces. A piece of a string is also a string, but this is no longer true for trees, where the pieces have dangling edges (or variables). As a result, more complicated  datatypes are needed; and our design choices lead us to functions that operate on ranked sets, where each element has an associated arity.

This is a long paper. Given the limited space, we have decided to prioritise  explaining  design choices and intuitions, with  examples and many pictures. As a result, almost all of the proofs are in the appendix.

%% file: trees-and-logic.tex
\section{Trees and tree-to-tree functions}
\label{sec:trees-transductions}
 In this section, we describe the trees and tree-to-tree functions that are discussed in this paper. 
A \emph{ranked set} is a set where each element has an associated \emph{arity} in $\set{0,1,2,\ldots}$. If $a$ of a ranked set has arity $n$, then elements of $\set{1,\ldots,n}$ are called \emph{ports of $a$}. We adopt the convention that ranked sets are red, e.g.~$\rSigma$ or $\rGamma$, and other objects (elements of ranked sets, or unranked sets) are black.  We use ranked sets as building blocks for trees. The following picture describes the notion of trees that we use and some terminology:
\mypic{1}


We use standard tree terminology, such as ancestor, descendant, child, parent. We write $\trees \rSigma$ for the (unranked) set of trees over a ranked set $\rSigma$. This paper is about \emph{tree-to-tree functions}, which are functions of the type \begin{align*}
f : \trees \rSigma \to \trees \rGamma.
\end{align*}

\subsection{First-order logic and transductions}
To define tree-to-tree functions and tree languages, we use  logic, mainly first-order logic and monadic second-order logic \mso. The  idea is to view a tree as a model, and to use logic to describe properties and transformations of such models.

A \emph{vocabulary} is defined to be a set of relation names, each one with associated arity.  We do not use function symbols in this paper. A  vocabulary can be formalised as a ranked set, which is why we use red letters like $\ranked \sigma $ or $\ranked \tau$ for  vocabularies. 

\begin{definition}[Tree as a model]\label{def:tree-model}
   For a tree $t$  over a ranked alphabet $\rSigma$, its \emph{associated model} 
    is defined as follows. The  universe is the nodes of the tree, and it is equipped with the following relations:
   $$\begin{array}{lcll}
   x<y &  &   \text{$x$ is an ancestor of $y$} & \text{arity 2}\\
   \mathrm{child}_i(x) &  & \text{$x$ is an $i$-th child ($i\in \set{1,2,\ldots}$)} & \text{arity 1} \\
   a(x) &  &   \text{$x$ has label $a$ ($a \in \rSigma$)} & \text{arity 1}
   \end{array}$$
    \end{definition}

The $i$-th child predicates are only needed for $i$ up to the maximal arity of letters in the ranked alphabet, and hence the vocabulary in the above definition is finite. We refer to this vocabulary as \emph{the vocabulary of trees over $\rSigma$}.
 A sentence of first-order logic (or  \mso)  over this vocabulary   describes a tree language, namely the set of trees whose associated models satisfy the sentence.  For example, the sentence 
 \begin{align*}
 \forall x \ a(x) \Rightarrow \exists y \ x < y \land b(x)
 \end{align*} 
 is true in (the models associated to)  trees $t$ where every node with label $a$ has a descendant with label $b$. For more background about defining properties of trees using logic, see the survey of Thomas~\cite{thomas1997languages}.
 
 The regular tree languages are exactly those that can be defined in \mso, which was proved by Doner~\cite[Corollary 3.11]{Doner70}, and also Thatcher and Wright~\cite[p.~74]{thatcherGeneralizedFiniteAutomata1968}. The tree languages definable in first-order logic are a proper subset  of those definable in \mso, and it is an open problem whether or not one can decide if a regular tree language can be defined in first-order logic~\cite[Section 3]{bojanczyk2015automata}. This is in contrast to the case of words, where the decidable characterisation of  first-order logic by Sch\"utzenberger-McNaughton-Papert~\cite[Theorem 10.5]{McNaughtonPapert71} is  a cornerstone of algebraic language theory.
 
 \paragraph*{Tree-to-tree functions.}
 Apart from defining tree languages, logic can also be used  to define transformations on  models. In the context of this paper, we are interested mainly in first-order transductions, defined below.  Roughly speaking, a first-order transduction uses first-order logic to define a new tree structure on the input tree.


\begin{definition}[First-order tree-to-tree transduction] \label{def:fo-transduction} A tree-to-tree function is called a  \emph{first-order transduction} if it can be obtained  by composing any number of operations\footnote{There is a normal form of first-order transductions, where at two phases are used: first  item 1, then item 2. We do not need the normal form, so we do not prove it, but it can be shown similarly to~\cite[Section 7.1.5]{courcelle1991}. } of the following two kinds:
\begin{enumerate}
    \item \emph{Copying.} Let  $k \in \set{1,2,\ldots}$. Define  $k$-copying to be the operation which inputs a tree and outputs a tree where every node is preceded by a chain of $k-1$ unary nodes with a fresh label $\blueball$, as in the following picture:
    \mypic{94}
    After $k$-copying, the number of nodes grows $k$ times.
    \item \emph{Non-copying first-order transductions.} This is a tree-to-tree function which uses first-order logic to define a new tree structure over the nodes of the input tree. The syntax of such a transduction is given by:
     \begin{enumerate} 
        \item  \emph{Input and output alphabets} $\rSigma$ and $\rGamma$, which are finite ranked sets. We use the name \emph{input vocabulary} for the vocabulary of trees over the input alphabet $\rSigma$, likewise we define the \emph{output vocabulary}.
        \item \label{it:universe-formula} A first-order formula over the input vocabulary, with one free variable, called the \emph{universe formula}.
        \item \label{it:tree-structure} For each relation of the output vocabulary, of arity $n$, a corresponding first-order formula  over the input vocabulary with $n$ free variables.
    \end{enumerate}
    The transduction inputs a tree over the input alphabet, and outputs a tree over the output alphabet where:
    \begin{itemize}
        \item the nodes  are those nodes of the input tree that satisfy the universe formula in item~\ref{it:universe-formula};
        \item the labels, descendant, and child relations are defined by the formulas in item~\ref{it:tree-structure}.
    \end{itemize}
    In order for the transduction to be well defined, the formulas in item~\ref{it:tree-structure} must be such that they produce a tree model for every input tree.
 \end{enumerate}
\end{definition}

If we allowed  monadic second-order logic \mso  in items~\ref{it:universe-formula} and~\ref {it:tree-structure} (the free variables of the formulas would  still be first-order variables ranging over tree nodes), then we would get the \mso tree-to-tree transductions of Bloem and Ensgelfriet~\cite[Section 3]{bloem_comparison_2000}. We  discuss these in  Section~\ref{sec:mso-trans}.

We conclude this section with two examples of first-order tree-to-tree transductions. 

\begin{example}\label{ex:filter-first}
    Let the input and output alphabets be:\vspace{-15pt}
    \mypic{17}
    and consider the  function which removes the unary nodes:
\begin{center}
\includegraphics[scale=.35, page=19]{pics.pdf}
\end{center}
This is a non-copying first-order  transduction. The universe formula selects nodes which have non-unary labels. The descendant relation is inherited from the input tree. To define the child relation on the output tree, we  use the descendant relation in the input tree. A node $x$  satisfies the unary  $i$-th child predicate   in the output tree if it satisfies the following first-order formula in the input tree:
\begin{align*}
    \exists y \ \child i (y) \land \underbrace{y \le x \land   \forall z\ (y \le z < x \Rightarrow \blueball(z))}_{\substack{\text{$y$ is the farthest ancestor that can be} \\ \text{reached from $x$ using only unary nodes}}}.
\end{align*}
This example shows the usefulness of first-order logic with descendant, as opposed to child only as used in~\cite{benediktSegoufin2009}.
\end{example}

\begin{example}\label{ex:pre-order-main} Define  \emph{pre-order} on nodes in a tree as follows:  $x$ is before  $y$ if either $x \le y$, or there exist nodes $x'$ and $y'$ such that $x' \le x$, $y' \le y$, and $x'$ is a sibling of $y'$ with a  smaller child number.  Consider  the tree-to-tree function which transforms a tree into a list of its nodes in pre-order traversal, as explained in the following picture:
    \mypic{112}
    This function is a first-order tree-to-tree transduction, 
    because the pre-order is first-order  definable. Unlike Example~\ref{ex:filter-first}, we need copying, because a node of arity $n$ in the input tree corresponds to $n+2$ nodes in the output tree.
\end{example}

%% file: types-and-functions.tex
\section{Derivable functions}\label{sec:derivable-functions}
In this section, we state the main result of this paper, which says that the  first-order tree-to-tree transductions are exactly those that  can be  obtained by starting with certain prime functions (such as pre-order traversal from Example~\ref{ex:pre-order-main}) and applying certain  combinators (such as function composition). 

The guiding principle behind our approach is to describe tree-to-tree functions without using  any iteration mechanisms, such as states or {\tt fold} functions. This principle validates the choice of first-order logic. If we were to use \mso, at the very least we would need to have some mechanism for groups, which are a basic building block for Krohn-Rhodes decompositions, or for evaluating Boolean formulas.

\subsection{Datatypes}
\label{sec:datatype-constructors}
The prime functions and combinators  use  datatypes  such as pairs of trees, or pairs of trees of pairs, etc. Although these datatypes  could be encoded in trees, we avoid this encoding and use explicit datatype constructors. 

An important property of our datatypes is that they represent ranked sets, i.e.~each element of a datatype has an arity. The datatypes are obtained from the atomic datatypes by applying four datatype constructors, as described below.

\paragraph*{Atomic datatypes.} Every finite ranked set is an atomic datatype. Apart from finite ranked sets, we allow one more atomic datatype: the   \emph{terminal ranked set}  $\termset$ which contains exactly one element of every arity.
The set is called terminal because it admits a unique arity preserving function from every  ranked set. 
We use $\termset$ for partial functions: a partial function with output  type  $\rSigma$ can be seen as a total function of output type $\ranked{\rSigma + \termset}$, which uses  $\termset$  for  undefined values.

\paragraph*{Terms.} The central datatype constructor is the  \emph{term} constructor, which is a generalisation of trees to higher arities. A term is a tree with dangling edges, called ports. The dangling edges ares used to decompose trees (and other terms) into smaller pieces, as  illustrated by the figure below. 
\mypic{15}
Formally speaking, terms are defined by induction as follows. As term over a  ranked set $\rSigma$ is either the \emph{identity term} denoted by  $\portletter$, which consists of a port and nothing else, 
or otherwise it is an expression of the  form $a \tensorpair{t_1,\ldots,t_n}$ where $a \in \rSigma$ has arity $n$, and $t_1,\ldots,t_n$ are already defined terms. The arity of a term is the number of ports. Terms of arity zero are the same as trees. We write $\tmonad \rSigma$ for the ranked set of terms over a ranked set $\rSigma$.   Because the term constructor -- like other datatype constructors -- outputs a ranked set, it makes sense to talk about terms of terms, etc.

Terms are a monad, in the category of ranked sets and arity preserving functions\footnote{An almost identical monad is used in ~\cite[Section 9.2]{bojanczykRecognisableLanguagesMonads2015}, which differs from ours in  that it allows multiple uses of a single port.}. The unit of the monad,  an operation of type $\ranked{\rSigma \to \tmonad \rSigma}$, is illustrated in the following picture:
\mypic{98}
The product of the monad,  an operation of type $\ranked{\tmonad \tmonad \rSigma \to \tmonad \rSigma}$ that we call \emph{flattening}, is illustrated in the following picture:
\mypic{97}
This monad structure will be part of our prime functions.
\paragraph*{Products and coproducts.}
There are two binary datatype constructors
\begin{align*}
\underbrace{\ranked{\Sigma_1 \product \Sigma_2}}_{\text{product}} \qquad \underbrace{\ranked{\Sigma_1 + \Sigma_2}}_{\text{coproduct}}.
\end{align*}
An element of the product is a pair $\tensorpair{a_1,a_2}$ where $a_i \in \ranked{\Sigma_i}$. The arity of the pair is the sum of arities of its two coordinates $a_1$ and $a_2$. 
An element of the coproduct is a pair $(i,a)$ where $i \in \set{1,2}$ and $a \in \ranked{\Sigma_i}$. The arity is inherited from $a$. 

The set of terms can be defined in terms of  products and coproducts, as the least solution of the equation:
\begin{align*}
\tmonad \rSigma = \redset{\portletter}  \ranked{+\coprod_{\black {a \in} \rSigma}} 
\powersmall{(\tmonad \rSigma)} {\text{arity of $a$}}
\end{align*}  where $\ranked \coprod$ denotes possibly infinite coproduct  and $\powersmall X n$ denotes the $n$-fold product of a ranked set $\ranked X$ with itself.

\paragraph*{Folding.}
The final -- and maybe least natural -- datatype constructor called \emph{folding}. Folding has two main purposes: (1) reordering ports in a term; and (2) reducing arities by grouping ports into groups. 

Folding is not one constructor, but a family of unary constructors $\reduce k \rSigma$, one  for every $k \in \set{1,2,3,\ldots}$.  An $n$-ary element of $\reduce k \rSigma$, which is called a \emph{$k$-fold}, consists of an element      $a \in \rSigma$  together with an injective    \emph{grouping}  function
\begin{align*}
    f :  \underbrace{\set{1,\ldots,\text{arity of $a$}}}_{\substack{\text{an element of this set is}\\\text{called a  port of $a$}}} \to \underbrace{\set{1,\ldots,n} \times \set{1,\ldots,k} }_{\text{these pairs are called \emph{sub-ports}}}
\end{align*}
We denote such an element as $a/f$ and draw it like this: \mypic{53}

Already for $k=1$, the constructor $\reduce 1$ is non-trivial. For example,  $\reduce 1 \tmonad \rSigma$ is a generalisation of terms where ports are not necessarily ordered left-to-right (because the grouping function need not be monotone), and some ports need not appear (because the grouping function need not be total); in other words this is the same as terms in the usual sense of universal algebra, with the restriction that each variable is used at most once (sometimes called linearity).

When viewed as a family of datatype constructors,  folds have a monad-like structure: they are a graded  monad in the sense of~\cite[p. 518]{fujiShinyaMellies2016}. The unit is the operation 
\mypic{98}
of type $\ranked{\Sigma \to \reduce 1 \Sigma}$, while  the product (or flattening) in the graded monad is the family of operations of type 
\begin{align*}
    \ranked{\reduce {k_2} \reduce {k_1} \Sigma \to \reduce {k_1 \cdot k_2} \Sigma},
\end{align*}
indexed by $k_1,k_2 \in \set{1,2,\ldots}$, that is illustrated below:
\mypic{100}
More formally, the flattening of a double fold $(a/{f_1})/{f_2}$ has the grouping function defined by
\begin{align*}
i \mapsto (i_2, \pi(p_1,p_2)) \qquad \text{where} \begin{cases}
    (i_1,p_1) &= f_1(i)\\
    (i_2,p_2) &= f_2(i_1)
\end{cases}
\end{align*}
and $\pi$ is the natural bijection between $\set{1,\ldots,k_1} \times \set{1\ldots,k_2}$ and  $\set{1,\ldots,k_1k_2}$.

\smallskip
\newcommand{\funcitem}[3]{\ranked{#1  } &:& \ranked{#2} \rto  \ranked{#3}}

 This completes the list of  datatype constructors.  
 \begin{definition}[Datatypes]
    \label{def:types}  
        The \emph{datatypes} are the least class of ranked sets which contains all finite ranked sets, the terminal set, and which is closed under applying the constructors 
        \begin{align*}
        \ranked{\tmonad \rSigma \qquad \rSigma_1 \product \rSigma_2 \qquad \rSigma_1 + \rSigma_2 \qquad \reduce k \rSigma}.
        \end{align*}
            \end{definition}

 \newcommand{\combfunc}[4]
 {
     \ranked{#1} : & \ranked{#2  \to #3} & \text{ for }\ranked{#4}
 }

\begin{figure}
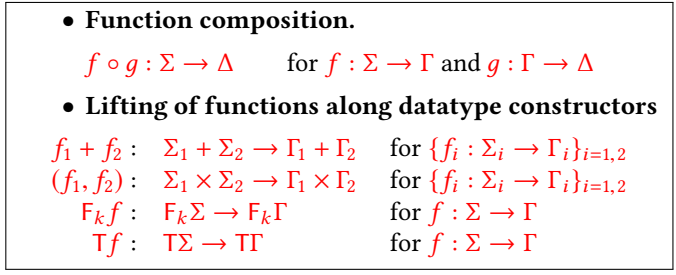

    \fbox{
    \begin{minipage}{1\linewidth}
    \begin{itemize}
    \item \textbf{Function composition.}
    $$
    \ranked{f \circ g : \Sigma \to \Delta} \qquad \text{for }\ranked{f : \Sigma \to \Gamma} \text{ and } \ranked{g : \Gamma \to \Delta}
    $$ 
    \item \textbf{Lifting of functions along datatype constructors}
    $$\begin{array}{rll}
        \combfunc{f_1 + f_2}{\Sigma_1 + \Sigma_2}{\Gamma_1 + \Gamma_2} {\set{f_i : \rSigma_i \to \rGamma_i }_{i =1,2}}\\
        \combfunc{\tensorpair{f_1,f_2}}{\Sigma_1 \product \Sigma_2}{\Gamma_1 \product \Gamma_2} {\set{f_i : \rSigma_i \to \rGamma_i }_{i =1,2}}\\
        \combfunc{\reduce k f}{\reduce k \Sigma}{\reduce k \Gamma} {f : \rSigma \to \rGamma }\\
        \label{eq:liftterm}
        \combfunc{\tmonad f}{\tmonad \Sigma}{\tmonad \Gamma} {f : \rSigma \to \rGamma}    
        \end{array}$$ 
    \end{itemize}
    \end{minipage}
    }
    \caption{Combinators}\label{fig:combinators}
    \end{figure}

\subsection{Derivable functions}
We now present the central definition of this paper.

\begin{definition}[Derivable function]\label{def:derivable-function}
    An  arity preserving function between two datatypes is called \emph{derivable}  if it can be generated, by using the combinators in Figure~\ref{fig:combinators}, from  the following prime functions:
    \begin{itemize}
    \item for every $\rSigma$, the unique arity preserving function $\ranked{\Sigma \to \termset}$;
    \item  all arity preserving functions with finite domain;
        \item  the prime functions in Figures~\ref{fig:monad},\ref{fig:product} and \ref{fig:not-explained};
         \end{itemize}

\end{definition}

\input{functions-reduced}

The combinators in Figure~\ref{fig:combinators} are  function composition, and the obvious liftings of functions along the datatype constructors. 
The prime functions in Figure~\ref{fig:monad} describe the monad structure of terms and folds, and were explained in Section~\ref{sec:datatype-constructors}. The prime functions in Figure~\ref{fig:product} are  simple syntactic transformations, which are intended to have no computational content.  Figure~\ref{fig:not-explained} contains   less obvious operations, whose definitions are  deferred to Section~\ref{sec:prime-and-combinators}. 

\input{example-derivable}

 We are now ready to state the main theorem of this paper. 
We say that a tree-to-tree function  
    \begin{align*}
        f : \trees \rSigma \to \trees \rGamma
    \end{align*}
    is \emph{derivable} if it agrees on arguments that are trees with some derivable partial function
    \begin{align*}
        \ranked {f : \tmonad \Sigma \to \tmonad \Gamma + \termset}.
    \end{align*}
The main result of this paper is the following theorem.
\begin{theorem}\label{thm:main} A tree-to-tree function is a first-order transduction if and only if it is derivable.
\end{theorem}

The right-to-left implication in the above theorem is proved by a relatively straightforward induction on the derivation. The general idea is that we associate to each datatype a relational structure; for example the relational structure associated to a pair $\tensorpair {a_1,a_2}$ is the disjoint union of the relational structures associated to $a_1$ and $a_2$.  In the appendix, we show that all prime functions are first-order transductions (adapted suitably to structures other than trees); and that this property is preserved under applying the combinators. There is one nontrivial step in the proof, which concerns monotone unfolding, and will be discussed below.  

The  left-to-right implication in the theorem, which says that every first-order transduction is derivable, is the main contribution of this paper, and is discussed in Sections~\ref{sec:stt}--\ref{sec:one-register}.

\subsection{The prime functions from Figure~\ref{fig:not-explained}}
\label{sec:prime-and-combinators}
In this section, we  describe the prime functions from Figure~\ref{fig:not-explained}. 
Each of these functions will play a key role in one of the main results of the paper.

\subsubsection{Factorisations}
\label{sec:factorisation-functions}
We begin with the two factorisation functions 
\begin{align*}
    \ancfact,\decfact  : \ranked{\tmonad(\Sigma_1+\Sigma_2) \to \tmonad(\tmonad \Sigma_1 + \tmonad \Sigma_2)},
\end{align*}
which are used to cut  terms into smaller parts. 
Define a \emph{factorisation} of a term   to be any term of terms that flattens to it.  An alternative view is that a factorisation is an equivalence relation on nodes in a term, where every equivalence class is connected via the parent-child relation.

Consider a  term $t \in \ranked{\tmonad(\rSigma_1 + \rSigma_2)}$.
        We say that two nodes have the \emph{same type} if both have labels in the same  $\ranked{\Sigma_i}$; otherwise we say that nodes have \emph{opposing type}. Define  two equivalence relations on  nodes in a term as follows: (a) nodes are called  \emph{$\uparrow$-equivalent}  if they have the same type and the same proper ancestors of opposing type; (b) nodes are  called  \emph{$\downarrow$-equivalent}  if they  are $\uparrow$-equivalent and have the same proper descendants of opposing type.
           Here is a picture of the equivalence classes, with $\ranked{\Sigma_1}$ being red and $\ranked{\Sigma_2}$ being blue: 
        \mypic{111}
        For both  equivalence relations, the equivalence classes are connected  under the parent-child relation, and therefore the equivalences can be seen as factorisations. These are the factorisations  produced by the functions $\ancfact$ and $\decfact$. 
    
        \subsubsection{Pre-order traversal.} The pre-order traversal function  
        \begin{align*}
            \ranked{\preorder : \tmonad \Sigma \to \reduce 1 \tmonad (\rSigma + \redset{\grayball,\grayballbin})}
        \end{align*}
        is the natural extension -- from trees to terms -- of the pre-order function in Example~\ref{ex:pre-order-main}. The fold in the output type is used to reorder the ports in a way which matches the input term, as illustrated in the following picture:
        \mypic{113}

\input{unfold}

%% file: functions-reduced.tex
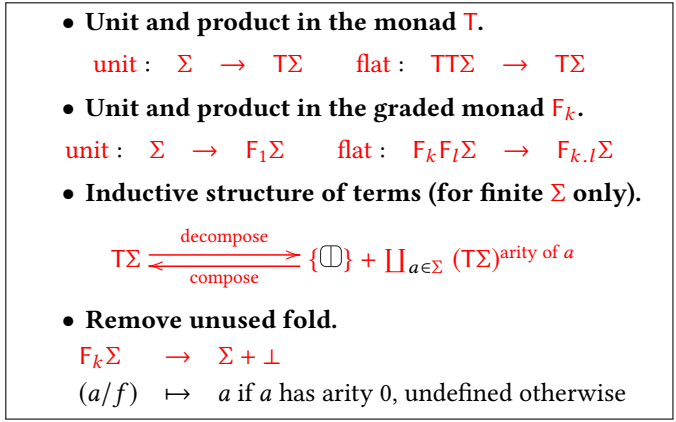
\begin{figure}
\fbox{
\begin{minipage}{1\linewidth}
\begin{itemize}
\item \textbf{Unit and product in  the monad $\tmonad$.} 
        $$\begin{array}{rlllcrlll}
\unit: &\rSigma &\ranked{\to} & \tmonad \rSigma & \qquad &
\flatt: &\tmonad\tmonad\rSigma &\ranked{\to} & \tmonad \rSigma
       \end{array}$$ 
\item \textbf{Unit and product in the graded monad $\reduce k$.} 
        $$\begin{array}{rlllcrlll}
\unit: &\rSigma &\ranked{\to} & \reduce 1 \rSigma & \qquad &
\flatt: &\reduce k \reduce l \rSigma &\ranked{\to} & \reduce {k.l} \rSigma 
       \end{array}$$ 
       \item \textbf{Inductive structure of terms (for finite $\rSigma$ only).}
       $$\begin{array}{cc}
        \ranked{
        \xymatrix@C=2cm{
\tmonad \Sigma \ar@<.5ex>[r]^-{\text{decompose}}
        & 
        \ar@<.5ex>[l]^-{\text{compose}}
        \redset{\portletter}  +\coprod_{\black {a \in} \rSigma}} 
        \powersmall{(\tmonad \rSigma)}{\text{arity of $a$}}
        } 
    \end{array}$$ 
    \item \textbf{Remove unused fold.}
    $$\begin{array}{rlll}
         &\reduce k \rSigma &\ranked{\to} &  \ranked{\rSigma + \termset}
         \\[2pt]
         &(a/f) &\mapsto& \text{$a$ if $a$ has arity 0, undefined otherwise}
               \end{array}$$ 
\end{itemize}
\end{minipage}
}
\caption{Prime functions for terms and fold.}\label{fig:monad}
\end{figure}

\begin{figure}
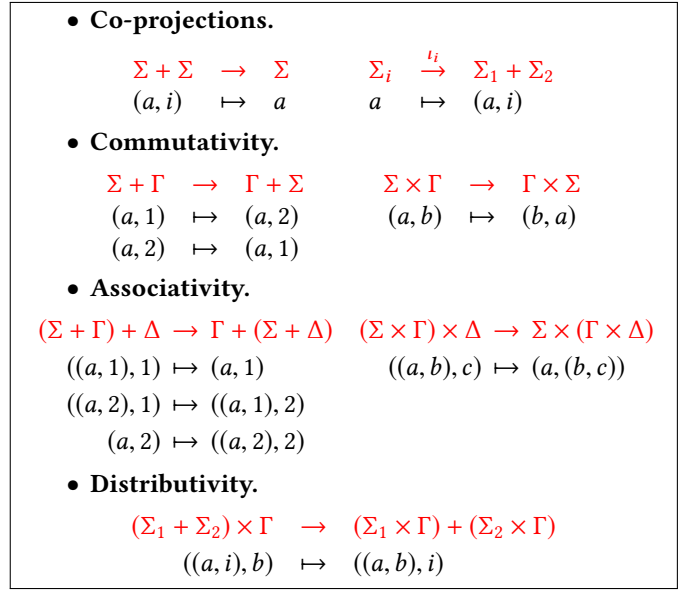

\fbox{
\begin{minipage}{1\linewidth}
\begin{itemize}
\item \textbf{Co-projections.} 
$$
\begin{array}{llllll}
\ranked{\rSigma + \rSigma} &\ranked{\to} & \rSigma \qquad & \qquad \ranked{\rSigma_i}  &\ranked{\overset{\iota_i}{\to}} & \ranked{\rSigma_1 + \rSigma_2} \\[0pt]
(a,i)&\mapsto& a   \qquad & \qquad  a &\mapsto& (a,i) 
       \end{array}
$$
\item \textbf{Commutativity.} 
$$
\begin{array}{rllrll}
\ranked{\rSigma + \rGamma} &\ranked{\to} & \ranked{\rGamma + \rSigma} \qquad & \qquad \ranked{\rSigma \product \rGamma} &\ranked{\to} & \ranked{\rGamma \product \rSigma} \\[0pt]
(a,1)&\mapsto& (a,2)  \qquad & \qquad
\tensorpair{a,b}&\mapsto& \tensorpair{b,a}\\
(a,2)& \mapsto& (a,1)
       \end{array}
$$
\item \textbf{Associativity.}
$$
\begin{array}{rllrll}
\ranked{(\rSigma + \rGamma) + \rDelta} &\hspace{-.2cm}\ranked{\to}&\hspace{-.2cm} \ranked{\rGamma + (\rSigma+\rDelta)}  & \ranked{(\rSigma \product \rGamma) \product \rDelta} &\hspace{-.2cm}\ranked{\to}&\hspace{-.2cm} \ranked{\rSigma \product (\rGamma \product \rDelta)} \\[2pt]
 ((a,1),1)&\hspace{-.2cm}\mapsto&\hspace{-.2cm} (a,1) & 
\tensorpair{\tensorpair{a,b},c}&\hspace{-.2cm}\mapsto&\hspace{-.2cm}\tensorpair{a,\tensorpair{b,c}}\\[2pt]
         ((a,2),1)&\hspace{-.2cm}\mapsto&\hspace{-.2cm}((a,1),2) & & &\\[2pt]
         (a,2) &\hspace{-.2cm}\mapsto&\hspace{-.2cm} ((a,2),2) & & &
       \end{array}
$$
\item \textbf{Distributivity.}
$$\begin{array}{rll}
\ranked{(\Sigma_1 + \Sigma_2)\product \Gamma }& \ranked{\to} &\ranked{(\Sigma_1 \product \Gamma) + (\Sigma_2 \product \Gamma)}\\[2pt]
\tensorpair{(a,i),b}&\mapsto& (\tensorpair{a,b},i)
\end{array}$$
\end{itemize}
\end{minipage}
}
\caption{Prime functions for product and coproduct.}\label{fig:product}
\end{figure}

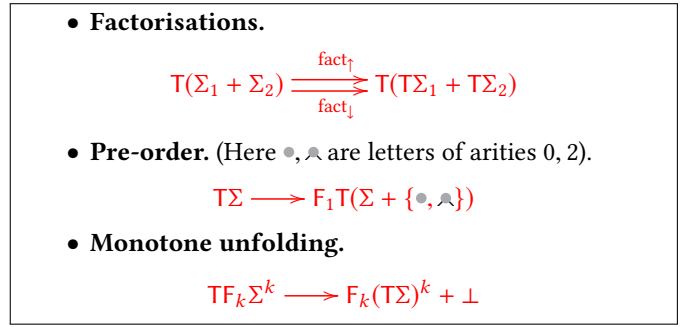
\begin{figure}
\fbox{
\begin{minipage}{1\linewidth}
\begin{itemize}
\item \textbf{Factorisations.}
$$\begin{array}{cc}
        \ranked{
        \xymatrix@C=1cm{
\tmonad(\Sigma_1+\Sigma_2) \ar@<.5ex>[r]^{ \ancfact}
        \ar@<-.5ex>[r]_{\decfact}& \tmonad(\tmonad \Sigma_1 + \tmonad \Sigma_2)
        }}
    \end{array}$$ 
\item \textbf{Pre-order.} (Here $\grayball, \grayballbin$ are letters of arities $0,2$).
$$\begin{array}{cc}
        \ranked{
        \xymatrix@C=.7cm{
\tmonad \Sigma \ar[r]& \reduce 1 \tmonad(\rSigma + \redset{\grayball,\grayballbin})
        }} 
    \end{array}$$ 
\item \textbf{Monotone unfolding.}
$$\begin{array}{cc}
        \ranked{
        \xymatrix@C=.7cm{
 \tmonad \reduce k \powersmall \Sigma k \ar[r]& \reduce k \powersmall{(\tmonad \rSigma)} k  +  \termset 
        }}
    \end{array}$$ 
\end{itemize}
\end{minipage}
}
\caption{Functions explained in Section~\ref{sec:prime-and-combinators}. }\label{fig:not-explained}
\end{figure}

%% file: example-derivable.tex
 
\noindent\begin{example}\label{ex:filter} 
   Define a \emph{term homomorphism}  to be any function of type $\tmonad \rSigma \to \tmonad \rGamma$ which is obtained by applying some function
   \begin{align*}
   \ranked{h : \rSigma \to \tmonad \rGamma}
   \end{align*}
   to every node of the input term. Examples of term homomorphisms include the function from Example~\ref{ex:filter-first} which removes all unary letters, or the $k$-copying function in item 1 of the definition of first-order tree-to-tree transductions.  We claim that every term homomorphism with a finite input alphabet is derivable. The function $\ranked{h}$ is a prime function, because it has a finite domain thanks to the assumption that the input alphabet is finite. We can lift $\ranked h$ to terms using the combinator of Figure~\ref{fig:combinators}, and then compose it with the product operation of terms monad, thus giving the homomorphism:
   \begin{align*}
   \ranked{\xymatrix{
\tmonad \rSigma \ar[r]^{\tmonad h} &
\tmonad \tmonad \rGamma \ar[r]^{\flatt} &
\tmonad \rGamma
   }}
   \end{align*} 
\end{example}

More examples of derivable functions are in Appendix.~\ref{sec:AppendixExamples}.

%% file: unfold.tex
\subsubsection{Unfolding of the matrix power}
\label{sec:unfolding}
The final  prime function is called monotone unfolding. The general idea is that unfolding unpacks a representation of several trees inside a single tree.  Before describing this function in more detail,  we introduce some notation,  inspired by the matrix power in  universal algebra~\citep[p.~268]{Taylor1975}. 
\begin{definition}
    [Matrix power] For $k \in \set{1,2,\ldots}$ define the $k$-th matrix power of a ranked set $\rSigma$, denoted by $\mati k \rSigma$, to be the ranked set $\reduce k \powersmall \rSigma k$.
\end{definition}
Here is a picture of elements in the third matrix power:
\mypic{102}

\begin{figure}[]
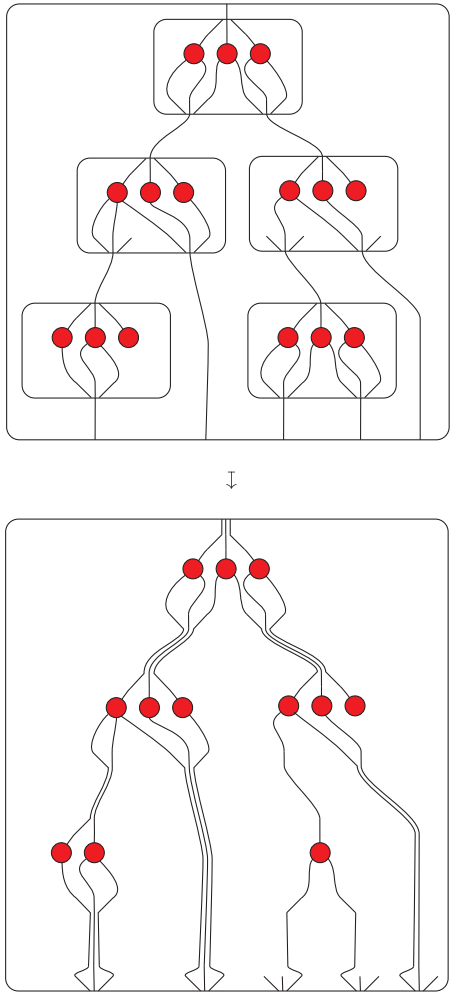

    \mypic{101}    
    \caption{Unfolding the matrix power}
    \label{fig:unfold}
\end{figure}

An element of the $k$-th matrix power  can be seen as having a group of $k$ incoming edges, and each of its ports  can be seen as a group of $k$ outgoing edges. The \emph{general unfolding} operation, which has type
\begin{align*}
    \ranked{\tmonad \mati k{\Sigma} \to \mati k{( \tmonad \Sigma)}},
\end{align*}
matches the $k$ incoming edges in a node with the $k$ outgoing edges in the parent port; it also removes the unreachable nodes. This operation is illustrated in Figure~\ref{fig:unfold}, and  a formal definition  is  in the appendix.
\paragraph*{Chain logic.}
The general unfolding operation is too powerful  to be included in the derivable functions, as we explain below. It does, however, admit a characterisation in terms of a fragment of \mso called \emph{chain logic},  see~\cite[Section 2]{thomas1992} or~\cite[Section 2.5.3]{bojanczykDecidablePropertiesTree2004}, whose expressive power is strictly between first-order logic and \mso. Chain logic is defined to be the fragment of  \mso where set quantification is restricted to sets  where all nodes are comparable by  the descendant relation. 

\begin{theorem}\label{thm:chain-transductions}
    The following conditions are equivalent for tree-to-tree functions:
\begin{itemize}
    \item is derivable, as in Definition~\ref{def:derivable-function}, except that general unfold is used instead of monotone unfold;
    \item is a transduction,  as in  Definition~\ref{def:fo-transduction},  except that chain logic is used instead of first-order logic. 
\end{itemize}
\end{theorem}


 To see why chain logic is needed to describe general unfolding, consider the following unfolding, where two coordinates are swapped in each node of the input tree:
 \mypic{108}
For inputs with an odd number of swaps, the output of unfolding has a white leaf in the first coordinate, and for inputs with an even number of swaps, the output has a white leaf in the first coordinate.  Checking if a path has even length can be done in chain logic, but not in  first-order logic.  

\paragraph*{Monotone unfolding}
To avoid the problems with cyclic swaps, the unfolding function in Figure~\ref{fig:not-explained} imposes a monotonicity requirement on the matrix power, described below.


 Let  $a \in \mati k \rSigma$ be an element of the matrix power,  let $p,q \in \set{1,\ldots,k}$, and let  $i$ be a port of $a$. Define the \emph{twist function of port $i$}, denoted by $\to_i$, as follows:
     $q \to_i p$ 
if coordinate $q$ in the $i$-th outgoing edge  is connected to  coordinate $p$ in root, as described in the following picture:
\mypic{125}
The twist function is partial. Call  an element of the matrix power  \emph{monotone} if for every port, its twist functions is monotone (when restricted to inputs where it is defined). In the picture above, $\to_1$ is monotone, while $\to_2$ is not. Also, the problems with an even number of swaps discussed earlier arise from  a non-monotone twist function:
\mypic{110}
The \emph{monotone unfolding} operation in Figure~\ref{fig:unfold} defined to be the restriction of general unfolding,  which  is undefined if the input contains at least one label which is non-monotone, and otherwise returns the output of the general unfolding.

\paragraph*{Is unfolding derivable?} The  prime functions in our main theorem  are meant to be simple syntactic rewritings. It is debatable whether the  unfolding operation -- even in its monotone variant -- is of this kind. For example, our proof  that monotone unfolding is a first-order transduction requires an invocation of the Sch\"utzenberger-McNaughton-Papert theorem about first-order logic on words being the same as counter-free automata. 

Is it possible to break down monotone unfolding into simpler primitives?
In the appendix, we devote considerable resources to answering this question. We propose one new  datatype
and seventeen additional prime functions, which can be called syntactic rewriting without straining the reader's patience. Then, we show that monotone unfolding can be derived using   the new datatype and functions. The proof of this result is one of the main technical contributions of this paper.

%% file: stt.tex
\section{Register tree transducers}
\label{sec:stt}
We now begin the proof of the harder implication in Theorem~\ref{thm:main}, which says that every first-order tree-to-tree transduction is derivable. Our proof passes  through an automaton model, which  is roughly based on existing transducer models for \mso transductions from~\cite{ENGELFRIET199934,alur2017streaming}.
The automaton  uses registers to store parts of the output tree. The semantics of the automaton involves two phases: (a) mapping the input tree to an expression that uses register updates; (b) evaluating the expression. These phases are described in more detail below.

\paragraph*{Register valuations and updates.} We begin by explaining how the registers work.  The registers store terms that are used to construct the output tree. Each register has an arity: registers of arity zero  store trees, registers of arity one store unary terms, etc.  

Fix two finite ranked sets: the \emph{register names} $\regnames$ and the \emph{output alphabet} $\rGamma$.
A \emph{register valuation} is defined to be any arity preserving function from the register names $\regnames$ to terms $\tmonad \rGamma$. 
To transform register valuations, we use \emph{register updates}.  A register update is an operation which inputs several  register valuations and outputs a single register valuation. For  $n \in \set{0,1,\ldots}$, an \emph{$n$-ary register update}  is defined to be any arity-preserving function
\begin{align*}
    \ranked {u : \regnames \rto \tmonad ( \rGamma + n\regnames)},
\end{align*}
where $\ranked{nR}$ stands for the disjoint union of $n$ copies of $\regnames$. The $i$-th copy of $\regnames$ represents the register contents in the $i$-th argument.
Here is a picture of a register update which has arity 3 and uses two registers $r$ and $s$:
\mypic{32}
An $n$-ary register update $\ranked u$ induces a operation, which inputs  $n$ register valuations and  outputs the  register valuation obtained by  taking $\ranked u$ and replacing the $i$-th copy of a register name with the  contents of that register in the $i$-th input register valuation. 
Register updates have arities, and therefore the ranked set of register updates is written in red, and can be used for labels in a tree. For such a  tree   
\begin{align*}
    t \in  \trees(\ranked{\text{register updates}}),
\end{align*}
define its \emph{evaluation} to be the register valuation defined by induction in the natural way. Note that register updates of arity zero are the same as register valuations, which gives the induction base.


\paragraph*{First-order relabellings.} Our automaton model has no states. Instead, it uses a first-order relabelling, as defined below, to  directly assign to each node of the input tree a register update that will be applied in that node. A similar model is used by    Bloem and Engelfriet~\cite[Theorem 17]{bloem_comparison_2000}, except that in their case, the first phase uses  \mso relabellings, and the second phase is   an attribute grammar.

\begin{definition}[First-order relabelling] \label{def:forat}  A \emph{first-order relabelling} is given by two finite ranked sets $\rSigma$ and $\rGamma$, called the \emph{input and output alphabets}, and a family 
    \begin{align*}
    \set{\varphi_a(x)}_{a \in \rGamma}
    \end{align*}
    of first-order formulas over the vocabulary of trees over  $\rSigma$. These formulas need to satisfy the following restriction:
    \begin{enumerate}
        \item[(*)] for every tree over the input alphabet and node in that tree, there is a unique output letter $a \in \rGamma$ such that $\varphi_a(x)$ selects the node; furthermore, the arity of $a$ is the same as the arity of (the label of) the  node. 
    \end{enumerate}
The semantics of a  first-order tree relabelling is a function 
\begin{align*}
\trees \rSigma \to \trees \rGamma,
\end{align*}
which changes the label of every node in the input tree to the unique letter described in  (*). 
      \end{definition}

A first-order tree relabelling is a very special case of a first-order tree-to-tree transduction, where only the labelling of the input tree is changed, while the universe as well as the child and descendant relations are not affected. 

\paragraph*{Register transducers.} Having defined registers, register updates, and first-order tree relabellings, we are now ready to define our automaton model.

\begin{definition}[First-order register transducer]\label{def:stt}
The syntax of a \emph{first-order register transducer} consists of: 
\begin{itemize}
    \item An \emph{input alphabet $\rSigma$}, which is a finite ranked set;
    \item An \emph{output alphabet $\rGamma$}, which is a finite ranked set;
    \item A set $\regnames$ of \emph{registers}, which is a finite ranked set;
    \item A total order on the registers.
    \item A designated \emph{output register} in $\regnames$, of arity zero.
    \item A \emph{transition function}, which is a  first-order  relabelling
    \begin{align*}
      \trees{\rSigma} \to \trees{\rDelta},
    \end{align*}for some finite set $\rDelta$ of  register updates over registers $\regnames$ and output alphabet $\rGamma$. We require  all register updates in  $\rDelta$ to be single-use and monotone, as defined below:
    \begin{enumerate}
        \item \emph{Single-use\footnote{The  single-use restriction  is a standard feature of transducer models with linear size increase~\cite{bloem_comparison_2000, alurStreamingStringTransducers2011,alur2017streaming}.   It prohibits iterated duplication of registers, which would lead to exponential size outputs.    
        }.}  An $n$-ary register update $\ranked{u}$ is  called  \emph{single-use} if   every $r \in \ranked{n \regnames}$
        appears in at most one term from $\set{\ranked u(s)}_{s \in \regnames}$, and it appears at most once in that term. 
        \item \emph{Monotone\footnote{This is notion of monotonicity corresponds to the one used in Section~\ref{sec:unfolding}, see the comments on page~\pageref{page:monotone-discussed}. A similar notion  appears in~\cite[p. 7]{bojanczykRegularFirstOrderList2018}.}.} This condition uses the total order on registers.  An $n$-ary  register update $\ranked u$ is called monotone  if for every $i \in \set{1,\ldots,n}$, the binary relation $\to_i$ on register names $r,s \in \regnames$  defined by
        \begin{align*} 
            r \to_i s \quad \text{if} \quad  \text{the $i$-th copy of $r$ appears in $\ranked u(s)$},
        \end{align*}
        which is a partial function from $r$ to $s$ when $\ranked u$ is single-use, is monotone:
        \begin{align*}
            r_1 \leq r_2 \land r_1 \to_i s_1  \land  r_2 \to_i s_2  \quad \Rightarrow \quad  s_1 \leq s_2
        \end{align*}
    \end{enumerate}
\end{itemize}
\end{definition}

The semantics of the transducer is  a tree-to-tree function, defined as follows. The input is a tree over the input alphabet. To this tree,  apply the transition function, yielding a tree of register updates. Next, evaluate the tree of register updates, yielding a register valuation. The output tree is defined to be the contents of the designated output register.

The main difference of our model with respect to prior work is that we want to capture  tree transformations defined in first-order logic, as opposed to \mso used in~\cite{bloem_comparison_2000,alurStreamingStringTransducers2011,alur2017streaming}. This is why we use first-order relabellings instead of  \mso relabellings.  For the same reason, we require the register updates to be monotone, see the discussion in Section~\ref{sec:unfolding}.  
\begin{proposition}\label{prop:unary-register-stt}
For every first-order register transducer, there is a first-order register transducer defining the same function, and whose registers are all unary.
\end{proposition}

The main result of this section is that first-order register transducers are expressively complete for first-order tree-to-tree transductions. 

\begin{theorem}\label{thm:stt}
    Every first-order tree-to-tree transduction is recognised by a first-order  register transducer. 
\end{theorem}

The proof, which is in Appendix~\ref{sec:stt-appendix}, uses the composition method for logic,  like similar proofs for~\cite[Theorem 4.6]{alur2017streaming} and~\cite[Theorem 14]{bloem_comparison_2000}. 
The converse  inclusion in the  theorem is also true. This is can be shown directly without much difficulty, following the same lines as in~\cite[Section 5]{bloem_comparison_2000}. The converse inclusion also follows from   other results in this paper: (a) we show in the following sections that every function computed by the transducer is derivable; and (b)  derivable functions are first-order tree-to-tree transductions by the easy implication in Theorem~\ref{thm:main}.

\paragraph*{Proof strategy for Sections~\ref{sec:fo-translation}--\ref{sec:stt-derivable}.} By Theorem~\ref{thm:stt}, to prove derivability of  every first-order tree-to-tree transduction, and thus finish the proof of our main theorem, it suffices to prove derivability for first-order register transducers. In a first-order register transducer,  the computation  has two steps: a first-order relabelling, followed by evaluation of the register updates. The first step is handled in Section~\ref{sec:fo-translation}, and the second step is handled in Section~\ref{sec:stt-derivable}.




%% file: forational.tex
\section{First-order  relabellings}\label{sec:fo-translation}
In this section we prove derivability of the first computation step used in first-order register transducers.
\begin{proposition} \label{prop:forat}    
    Every first-order relabelling is derivable.
\end{proposition}
To prove the proposition, we use a decomposition of first-order relabellings into simpler functions, in the style of the Krohn-Rhodes theorem. 
We use the name \emph{unary query} for a first-order formula with one free variable over the vocabulary of trees. This assumes some implicit alphabet $\rSigma$.
For a  unary query,  define its  \emph{characteristic function},  of type
\begin{align*}
 \trees \rSigma \to \trees (\ranked{\rSigma + \rSigma}),
\end{align*}
to be the function which replaces the label of each node by its first or second copy, depending on whether the node is selected by the query. This is a special case of a first-order relabelling. The key to Proposition~\ref{prop:forat} is the following  lemma, which decomposes first-order relabellings  into characteristic functions of certain basic unary queries.

\begin{lemma}\label{lem:schlingloff} Every first-order relabelling can be obtained by composing the following functions:
    \begin{enumerate}
        \item \label{it:relabelling} \emph{Letter-to-letter homomorphisms}. For  every finite $\rGamma,\rSigma$ and $\ranked {f : \rSigma \to \rGamma}$, its tree lifting $\trees \ranked f : \trees \rSigma \to \trees \rGamma$.
        \item \label{it:temporal-operators} For every finite  $\rSigma$ and its subsets $\rDelta, \rGamma \subseteq \rSigma$, the characteristic functions of the following unary queries over alphabet $\rSigma$:
        \begin{enumerate}
            \item \label{it:child} \emph{Child:} $x$ is an $i$-th child, for $i \in \set{1,2,\ldots}$
            \begin{align*}
            \child i (x); 
            \end{align*}
             \item \label{it:until} \emph{Until:}  $x$ has a descendant $y$ with label in $\rDelta$, such that all nodes strictly between $x$ and $y$ have label in $\rGamma$
             \begin{align*} 
                  \exists y\ y > x \land \rDelta(y) \land  \forall z \ (x < z < y \Rightarrow \rGamma(z));
             \end{align*} 
             \item \label{it:since}\emph{Since:} $x$ has an ancestor $y$ with label in $\rDelta$, such that all nodes strictly between $x$ and $y$ have label in $\rGamma$
             \begin{align*}
                  \exists y\ y < x \land \rDelta(y) \land  \forall z \ (y < z < x \Rightarrow \rGamma(z)).
             \end{align*} 
        \end{enumerate}
    \end{enumerate}
    
\end{lemma}

The  lemma uses a theorem of   Schlingloff~\cite[Theorem 2.6]{schlingloff1992expressive}, which says  that all first-order definable tree properties can be defined using a temporal logic with operators similar to the ones used in items~\ref{it:temporal-operators} of the lemma. Note that the temporal logic is a two-way logic, because  \emph{until} depends on the descendants of the node $x$, while \emph{since} depends on the ancestors. In fact, there is no temporal logic which characterises first-order logic, uses only descendants, and has finitely many operators~\cite[Theorem 5.5]{bojanczykWreathProductsForest2012}. 
The exact reduction to Schlingloff's theorem is  in Appendix~\ref{sec:AppendixForat}.

It remains to show that all of the functions from Lemma~\ref{lem:schlingloff} are derivable. 
The letter-to-letter homomorphisms from item~\ref{it:relabelling} are   a special case of homomorphisms discussed in Example~\ref{ex:filter}, and hence derivable. In Appendix~\ref{sec:AppendixForat}, we show that the functions from item~\ref{it:temporal-operators} are also derivable. In the proof, a key role is played by the factorisation functions discussed in Section~\ref{sec:factorisation-functions}. 

%% file: stt-derivable-amina.tex
\section{Evaluation of register updates}
\label{sec:stt-derivable}
In this section, we deal with the second computation phase in a first-order register transducer, namely evaluating register updates. As  discussed in the end of Section~\ref{sec:stt}, this completes the proof of our main theorem.

Our proof uses the language of $\lambda$-calculus.  In Section~\ref{sec:one-register},  we discuss derivability of normalisation of $\lambda$-terms. In Section~\ref{sec:updates-endgame}, we reduce evaluation of register updates to  unfolding the matrix power  and normalisation of $\lambda$-terms.  

\input{one-register}

\subsection{Evaluation of register updates}
\label{sec:updates-endgame}
Equipped with Theorem~\ref{thm:normalise}, we  prove derivability of  evaluation of  register updates. 
Fix a first-order register transducer. We suppose from now on that:
\begin{align}  \label{assumption}
\text{all its registers are unary} 
\end{align} 
which is possible by Proposition~\ref{prop:unary-register-stt}.
From now on, when speaking about register updates or register valuations, we mean those of the fixed transducer. 
Our goal is to prove the following lemma, which completes the proof of our main theorem.
\begin{lemma}\label{lem:derive-register-updates}
    Consider
    the   tree-to-tree function, which inputs a tree of  register updates, evaluates it, and outputs the  contents of the designated output register. This function is derivable. 
\end{lemma}

\paragraph*{Output letters in $\lambda$-terms.} We will   use $\lambda$-terms to represent register updates, which involve letters of the output alphabet $\rGamma$. Therefore, for the  rest of Section~\ref{sec:updates-endgame}, we use an extended notion of $\lambda$-terms, which allows  building $\lambda$-terms of the form 
\begin{align}\label{eq:non-pure}
a(M_1,\ldots,M_n) \qquad \text{for every $a \in \rGamma$ of arity $n$.}
\end{align}
The typing rules are extended as follows: if the arguments $M_1,\ldots,M_n$ all have type $\otype$ (no other type is allowed for arguments of $a$),  then~\eqref{eq:non-pure} has type $\otype$. These $\lambda$-terms can  be represented as trees, as in the following picture:
\mypic{118}
Theorem~\ref{thm:normalise} works without change for the extended notion of  $\lambda$-terms used in this section. Note that there is no $\beta$-reduction rule for $\lambda$-terms of the form~\eqref{eq:non-pure}.

\paragraph*{$\lambda$-representations of register updates.}
To prove Lemma~\ref{lem:derive-register-updates}, we represent  register updates using a matrix power of $\lambda$-terms. The idea is that the matrix power handles the parallel evaluation of registers.

Let $x$ be a  variable of type $\otype$. Define  $\ranked{\Gamma_\lambda}$ to be the output alphabet $\rGamma$ plus the following ranked alphabet:
\begin{align}
    \label{eq:alphabet-for-lambda-terms}
  \overbrace{\set{x }}^{\text{arity 0}} \cup \overbrace{\set{\lambda x }}^{\text{arity 1}} \cup  \overbrace{\set @}^{\text{arity 2}}
\end{align}

Recall that a register update -- of arity say $n$ -- consists of a family of terms  over alphabet $\ranked{\Gamma + n\regnames}$, one for each register $r \in \regnames$. We begin by explaining the $\lambda$-representation for terms in the family, which is a function 
 of type
\begin{align}\label{eq:placeholders}
\xymatrix@C=2cm{
    \ranked{\tmonad(\rGamma+n R)} \ar[r]^-{\text{$\lambda$-representation}} &
    \ranked{\tmonad {\Gamma_\lambda}}
}.
\end{align}
This function  is not arity preserving, which is why it is not written in red. Define a  \emph{placeholder} to be an element of $\ranked{n \regnames}$; we write placeholders as $r_i$ with $r \in \regnames$ and  $i \in \set{1,\ldots,n}$.
The function~\eqref{eq:placeholders}  is  explained in the following picture:
\begin{center}
\includegraphics[scale=0.9]{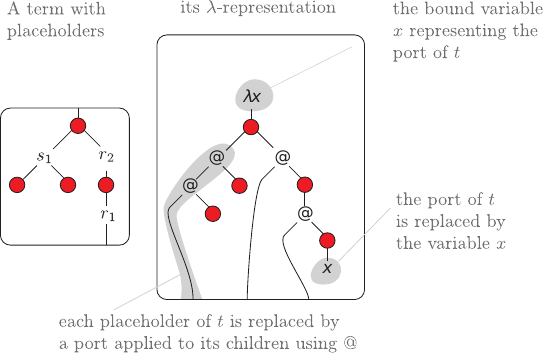}
\end{center}
Note how the  
 arities need not be preserved: the arity of the output is the number of placeholders in the input, but the input have always at most one port by assumption~(\ref{assumption}). The correspondence of ports in the output term with placeholders in the input term is defined with respect to some arbitrary order on the set $\ranked{n \regnames}$ of placeholders, say lexicographic with respect to the order on registers and  $\set{1,\ldots,n}$.


Having defined the $\lambda$-representation of terms with placeholders, we lift it to a $\lambda$-representation of  register updates
\begin{align}\label{eq:lambda-representation-regup}
\ranked{
    \xymatrix@C=2cm{
 \text{register updates}    \ar[r]^-{\text{$\lambda$-representation}} &
 \mati k {(\tmonad\Gamma_\lambda)}
}
},
\end{align}
where $k$ is the number of registers. This function is arity preserving. 

For a register update $(t_1,\ldots,t_k)$, where $t_i$ is the term with placeholders used in the $i$-th register,  its $\lambda$-representation is defined to be 
\begin{align*}
(\text{$\lambda$-representation of $t_1$},\ldots,\text{$\lambda$-representation of $t_k$})/f ,
\end{align*}
 where the grouping function $f$ connects a placeholder $r_i$ to the $r$-th sub-port of port $i$. 
Here is a picture
\begin{center}
\includegraphics[scale=0.9]{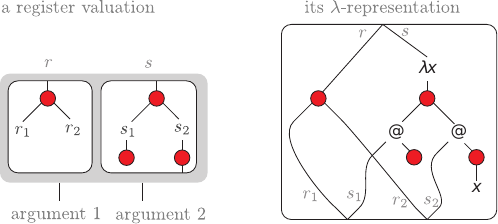}
\end{center}


The following  three properties of the $\lambda$-representation for register updates will be used later in the proof:
\begin{enumerate}
\item[(P1)]  If we restrict the domain to a finite set of register updates, e.g.~those used in the transducer, then it is a prime function, by virtue of having finite domain.
\item[(P2)] A register update is monotone  (as in Definition~\ref{def:stt})  if and only if its $\lambda$-representation is monotone (as defined in    Section~\ref{sec:unfolding} for the matrix power).
\item[(P3)] The $\lambda$-representation uses the unique variable $x$, every binder $\lambda x$ binds a unique occurrence of $x$, and the types that appear are of the form 
\begin{align*}
\overbrace{\otype \to \otype \to \cdots \to \otype \to \otype}^{\text{at most (maximal arity in $\rGamma$) times}} \to \otype,
\end{align*}
hence Theorem~\ref{thm:normalise} can be applied. 
\end{enumerate}

\label{page:monotone-discussed}

\paragraph*{Putting it all together.} To finish the proof of Lemma~\ref{lem:derive-register-updates}, we   observe  that the semantics of a register automaton are translated -- under the $\lambda$-representation -- to unfolding the matrix power and normalising a $\lambda$-term.  This observation is formalised by saying that the diagram in Figure~\ref{fig:lambda-representation-diagram} commutes, and it  follows directly from the definitions. Instead of giving a proof, we illustrate it on an example in Figure~\ref{fig:lambda-representation-proof}.

\begin{figure}[]
    \centering
    \begin{align*}
        \xymatrix@C=3cm{
            \trees \ranked{\text{(register updates)}} 
            \ar[dd]_{\substack{\text{evaluate}\\\text{register}\\\text{updates}}}^{\text{(a)}}
            \ar[r]^-{\trees(\text{\ranked{$\lambda$-representation}})}_{\text{(c)}}
            &
            \trees \ranked{(\mati k{(\tmonad \Gamma_\lambda)})}
            \ar[d]^{\substack{\text{unfold}\\\text{matrix}\\\text{power}}}_{\text{(d)}} \\
            & 
            \txt{{\tiny arity 0 elements of }\\
            $
            \mati k {\ranked{(\tmonad \Gamma_\lambda)}}
            $}
            \ar[d]^{\substack{\text{normalise}\\\text{$\lambda$-terms}}}_{\text{(e)}}\\
             \text{register valuations}
            \ar[r]_-{\ranked{\text{$\lambda$-representation}}}^{\text{(b)}}
            &
            \txt{{\tiny arity 0 elements of }\\
            $
            \mati k {\ranked{(\tmonad \Gamma_\lambda)}}
            $}
        }
    \end{align*} 
    \caption{}
    \label{fig:lambda-representation-diagram}
\end{figure}
\begin{figure}[]   
    \hspace{-0.5cm}
    \includegraphics[scale=.8]{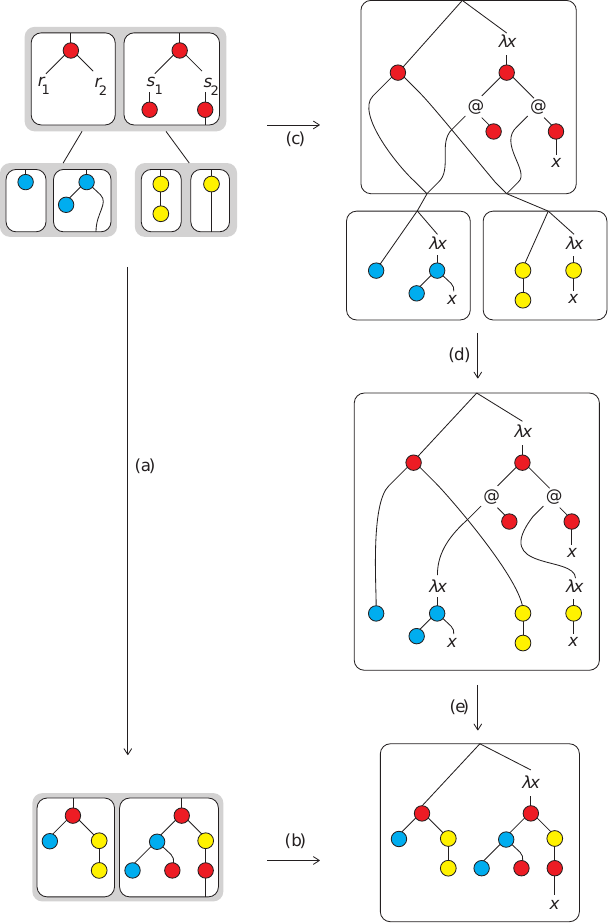}
    \caption{Example for Figure~\ref{fig:lambda-representation-diagram}.}
    \label{fig:lambda-representation-proof}
\end{figure}
 
\pagebreak 
We claim that all of the arrows (c), (d) and (e) on the  right-down path  in  Figure~\ref{fig:lambda-representation-diagram}  are derivable:
\begin{itemize}
    \item[(c)] Since we work with a fixed register transducer, there is a finite subset $\rDelta$ of register updates  used, and therefore  operation (a) in the figure is derivable by property (P1).
    \item[(d)] Arrow (d) represents the unfolding of the matrix power. By property (P2), the outputs of arrow (c) are monotone, and so we can use the monotone unfolding operation, which is a  prime function and therefore derivable. 
    \item[(e)] Finally, arrow (e) represents normalisation of $\lambda$-terms. This arrow is derivable by Theorem~\ref{thm:normalise}. The assumptions of this theorem are met by property (P3).
\end{itemize}
Since the arrows (c), (d), (e) are derivable, and the diagram commutes, it follows that  the composition of the arrows (a) and (b) is derivable. In other words, there is a derivable function which maps a tree of register updates to the $\lambda$-representation of the resulting register valuation (when viewing a register valuation as a special case of a register update of arity zero). Finally, to get the contents of the output register, we get rid of the fold in the matrix power by using the last function from Figure~\ref{fig:monad}, and  project onto the coordinate for the output register.


This completes the proof of Lemma~\ref{lem:derive-register-updates}, and therefore also of the main theorem.

%% file: one-register.tex
\subsection{Normalisation of simply typed linear $\lambda$-terms}
\label{sec:one-register}


We assume that the reader is familiar with the basic notions of the simply typed $\lambda$-calculus; more detailed definitions can be found in~\cite{sorensen_lectures_2006}. 
Define  \emph{simple types} to be expressions  generated from an atomic type $\otype$ using a binary arrow constructor, as in the following examples:
\begin{align*}
    \otype \qquad \otype \to \otype \qquad (\otype \to \otype) \to (\otype \to \otype) \qquad \cdots 
\end{align*}
In this paper,  the atomic type $\otype$ represents trees over the output alphabet.
Let $X$ be a set of variables, each one with an associated simple type.  A $\lambda$-term  is any expression that can be built from the variables, using $\lambda$-abstraction $\lambda x. M$ and term application $M N$. 
We say that a $\lambda$-term is \emph{well-typed} if one can associate  to it  a simple type according to the usual typing rules of simply typed $\lambda$-calculus,
see~\cite[Definition 3.2.1]{sorensen_lectures_2006}. Because the variables are typed, a  $\lambda$-term has  either a unique type, or is not be well-typed.  Here is an example of a well-typed $\lambda$-term, with the type annotation in blue: 
\begin{align*}
    \typecolor{\overbrace{
        \usualcolor{\lambda \typevar y {\otype \to \otype}. \ \lambda \typevar x \otype.}  \ \underbrace{\usualcolor{y (y x).}}_{\otype}}^{(\otype \to \otype) \to \otype \to \otype}}
    \end{align*}

    
We use the standard notion of $\beta$-reduction for $\lambda$-terms, see~\cite[Definition 1.2.1]{sorensen_lectures_2006}.  
Because of normalisation and confluence for the simply typed $\lambda$-calculus, every well-typed $\lambda$-term has a unique normal form, i.e~a $\lambda$-term to which it $\beta$-reduces (in zero or more steps), and which cannot be further $\beta$-reduced.

A $\lambda$-term  can be seen as a tree over the ranked alphabet
\begin{align}
    \label{eq:alphabet-for-lambda-terms}
  \overbrace{\set{x : x \in X}}^{\text{arity 0}} \cup \overbrace{\set{\lambda x : x \in X}}^{\text{arity 1}} \cup  \overbrace{\set @}^{\text{arity 2}}
\end{align}
where @ represents term application. Using this representation, and assuming that the set of variables is finite, it makes sense to view normalisation as a  tree-to-tree function
\begin{align*}
\text{$\lambda$-term} \qquad \mapsto \qquad \text{its normal form},
\end{align*}
and ask about its derivability.
 We show that this function is derivable, under three assumptions on the input $\lambda$-term. 
 
  The first assumption is that in the input $\lambda$-term, there is a unique fixed variable $x$ which can be bound. The second assumption is that the input $\lambda$-term is \emph{linear} in this variable: every binder $\lambda x$ bounds a unique occurrence of $x$\footnote{This restriction could easily be relaxed to ``at most once''.}.
     The third assumption is that  the input $\lambda$-term can be typed using a fixed finite set of types $\Tt$: it has type in $\Tt$, and the same is true for all of its  sub-terms.  In Appendix~\ref{sec:explaining-restrictions}, we explain why the  assumptions are needed.




\begin{theorem}\label{thm:normalise} Let $X$ be a set of simply typed variable, $x\in X$ and let $\typeset$ be a finite set of simple types.
    The following tree-to-tree function is derivable, assuming that $\lambda$-terms are represented as trees:
    \begin{itemize}
        \item{\bf Input.} A $\lambda$-term over $X$.
        \item {\bf Output.} Its normal form, if it is linear, the only bound variable is  $x$ and  can be typed using $\Tt$, and undefined otherwise.
    \end{itemize}
\end{theorem}

This is one of our main technical contributions, and its proof is in Appendix~\ref{sec:eval}. A key role in the proof is played by the pre-order function.

%% file: extensions.tex
\section{Monadic second-order transductions}
\label{sec:mso-trans}
We finish the paper by discussing a variant of our main theorem for   \mso tree-to-tree transductions. 
We simply add, as prime functions, all \mso relabellings, which are defined the  same way as the first-order relabellings from Definition~\ref{def:forat}, except that the unary queries can use \mso logic instead of first-order logic. 

\begin{theorem}\label{thm:mso-transductions}
    A tree-to-tree function is an \mso  transduction if and only if it can be derived  using  Definition~\ref{def:derivable-function}  extended by  adding  all \mso relabellings as prime functions. 
\end{theorem}
\begin{proof}
    In~\cite[Corollary 1]{colcombetCombinatorialTheoremTrees2007}, Colcombet shows that every \mso formula on trees can be replaced by a first-order formula that runs on an \mso relabelling of the input tree. Applying that result to transductions, we see that every \mso tree-to-tree transduction can be decomposed as: (a) an \mso relabelling; followed by (b) a first-order tree-to-tree transduction.  The theorem follows.
\end{proof}
The solution above is not particularly subtle, and contrasts our results for  first-order logic and chain logic, where  we took care to have a small number of primitives. This  was possible thanks in part to the decomposition of first-order queries into simpler ones that was is in Section~\ref{sec:fo-translation}, and the Krohn-Rhodes theorem that is used in the proof of Theorem~\ref{thm:chain-transductions} about chain logic.  In principle,  a decomposition of \mso relabellings could be possible, but proving it  would likely require developing a new decomposition  theory for regular tree languages, in the style of the Krohn-Rhodes theorem, which we feel is beyond the scope of this paper. One would expect a Krohn-Rhodes theorem  for trees to yield an effective characterisation of first-order logic -- as it does for words -- but finding such a characterisation remains a major open problem~\cite[Section 3]{bojanczyk2015automata}.

%% file: appendix-unfold.tex
\section{Unfolding the matrix power} 
\label{sec:appendix-unfold}
In this part of the appendix, we define formally the unfolding function
\begin{align*}
    \ranked{\tmonad \mati k{\Sigma} \to \mati k{( \tmonad \Sigma)}}
\end{align*}
that was described in Section~\ref{sec:unfolding}. We  present the definition in a slightly verbose manner,  by  decomposing unfolding into simpler operations. The  presentation highlights the inductive character of unfolding, and the reasons why we are uneasy about it being a prime operation. 

\subsection{Shallow terms}
\label{sec:shallow-terms}
We begin by defining unfolding  for terms of depth two, called \emph{shallow terms}. Later, we extend the definition to all other terms by induction. 
We describe shallow terms as a separate datatype, since this datatype will also be used later, in Section~\ref{ap:matrix-power}, to derive the (monotone) unfolding operation.
 For now, shallow terms are just an intermediate type used to define formally the unfolding function. 

 Let $\rSigma$ and $\rGamma$ be two ranked sets. The  shallow terms datatype, which is  denoted $\shallowterm \rSigma \rGamma$, consists of  expressions of the form $a\tensorpair{b_1,\dots,b_n}$ where $a$ is an $n$-ary element of $\rSigma$ and $b_1,\dots, b_n$ are elements of $\rGamma$. The arity of such an expression is the sum of arities of $b_1,\ldots,b_n$. We draw shallow terms as terms of depth two, where the root is from $\rSigma$ and  the children are from $\rGamma$:
\mypic{54}
An equivalent definition of shallow terms, in terms of products and co-products, is 
\begin{align}\label{eq:shallowterm-definition}
\shallowterm \rSigma \rGamma \quad \eqdef   \quad \ranked{\coprod_{\black{a \in} \rSigma} } \overbrace{\ranked{\Gamma \product \cdots \product \Gamma},}^{\text{arity of $a$ times}}
\end{align}

\subsection{Terms as an inductive datatype}
\label{sec:terms-induction-principle}
Using shallow terms, we can define the set of terms as the least solution of the equation
\begin{align*}
\ranked{\tmonad \rSigma = \set{\portletter} + \shallowterm \Sigma {(\tmonad \rSigma)}}.
\end{align*}
With this inductive definition, in order to define an operation of type $\ranked{ \tmonad \rSigma \to \Gamma}$
on terms, it is enough to explain the induction base for the  identity term and the induction step for shallow unfolding, as captured by two operations of types
\begin{align*}
        \underbrace{\ranked{\set \portletter \to \Gamma}}_{\text{induction base}}
        \qquad 
        \underbrace{\ranked{\shallowterm \Sigma \Gamma \to \Gamma}.}_{\text{induction step}}
\end{align*}
We use such an induction below to define general unfolding.  The crucial step is defining the induction step, which the unfolding for shallow terms defined in Section~\ref{sec:definition-of-shallow-unfolding} below. 

 As mentioned at the beginning of Section~\ref{sec:derivable-functions}, the guiding principle behind our approach is to avoid iteration mechanisms. The inductive definition of general unfolding could be seen as such an iteration mechanism; this is the reason for Section~\ref{ap:matrix-power}, where (monotone) unfolding is derived using simpler operations.  In contrast, we believe that iteration is indeed avoided by  the operations used in the induction step that are presented in Section~\ref{sec:definition-of-shallow-unfolding} below.

We do not formalise what we mean by  ``avoiding iteration''. One possible direction would be to say that an operation ``avoids iteration'' if it can be computed by a family of bounded depth circuits, as in the circuit class AC$^0$. A further requirement could be that the family of circuits not only exists, but it is also easy to see. 

\subsection{Unfolding for shallow terms} 
\label{sec:definition-of-shallow-unfolding}
The induction step in general unfolding is the operation
\begin{align*}
    \ranked{
        \xymatrix{
            \shallowterm{\mati k \rSigma} {\mati k \rGamma}  \ar[r] & \mati k {(\shallowterm \Sigma \Gamma)},
        }
    }
\end{align*}
which we call shallow unfolding, and   
which is explained in the following picture:
\mypic{121}
To define this operation formally, we further decompose it using  three functions manipulating shallow terms. These functions, which are used here as  intermediate functions in the definition of shallow unfolding, will become prime functions when we decompose the unfolding function in Appendix~\ref{ap:matrix-power}.

\subsubsection{Distribute shallow terms over fold}
 Let $\rGamma$ and $\rSigma$ be two datatypes.  Consider the function $\ranked{f_1}$
\begin{align*}
\ranked{\shallowterm  \Gamma {\reduce k \Sigma}} \ranked{\xrightarrow{\quad f_1 \quad}} \ranked{\reduce k(\shallowterm  \Gamma  \Sigma)}
\end{align*}
which distributes shallow terms over folding. This function is illustrated by the following picture
\mypic{55}
and  defined by 
$$\begin{array}{rcl} 
a(b_1/g_1,\dots,b_n/g_n)&\mapsto& a(b_1,\dots,b_n)/g
\end{array}$$
where $g$ is the function defined as follows. For every  $i \in\set{1,\dots,n}$,  if  $j\in\set{1,\dots,\arity{b_i}}$ then 
 \begin{align*}
 \xymatrix@C=5cm{
 \begin{array}{c}
 \overbrace{j+\underset{l<i}{\Sigma} \arity{b_l}}^{\text{Position of the $j$-th port of $b_i$ is shifted}}
 \end{array}
 \ar@{|->}[d]^{g}\\
 \begin{array}{c}
  \biggl(\ \ \ \underbrace{\pi_2(g_i(j))+\underset{l<i}{\Sigma} \arity{b_l/g_l}}_{\text{Position of the group is shifted}}\ \ , \underbrace{\pi_1(g_i(j))}_{\begin{array}{c}
{ \scriptsize\text{Position inside}}\\[-5pt]{\scriptsize \text{the group is unchaged}}
 \end{array}}\biggr)
 \end{array}}
 \end{align*}
 

\smallskip

\subsubsection{Matching function}
We now define a function 
\begin{align*}
\ranked{\shallowterm  {(\reduce k \Gamma)} {(\Sigma^k)} \xrightarrow{\quad f_2\quad} \reduce 1 (\shallowterm  \Gamma  \Sigma)} 
\end{align*}
which matches the $k$-th fold with the $k$-th power\footnote{In order to reduce the number of parentheses, in  the rest of the paper we assume a notational convention where the unary datatype constructors -- like folding, terms or powering -- have priority over the binary shallow term constructor. Under this convention, the operation $\ranked{f_2}$ is written as 
\begin{align*}
    \ranked{\shallowterm  {\reduce k \Gamma} {\Sigma^k} \xrightarrow{\quad f_2\quad} \reduce 1 (\shallowterm  \Gamma  \Sigma)} 
    \end{align*}}.
    The function $\ranked{f_2}$ is illustrated by the following  picture
\begin{center}
\includegraphics[scale=.35]{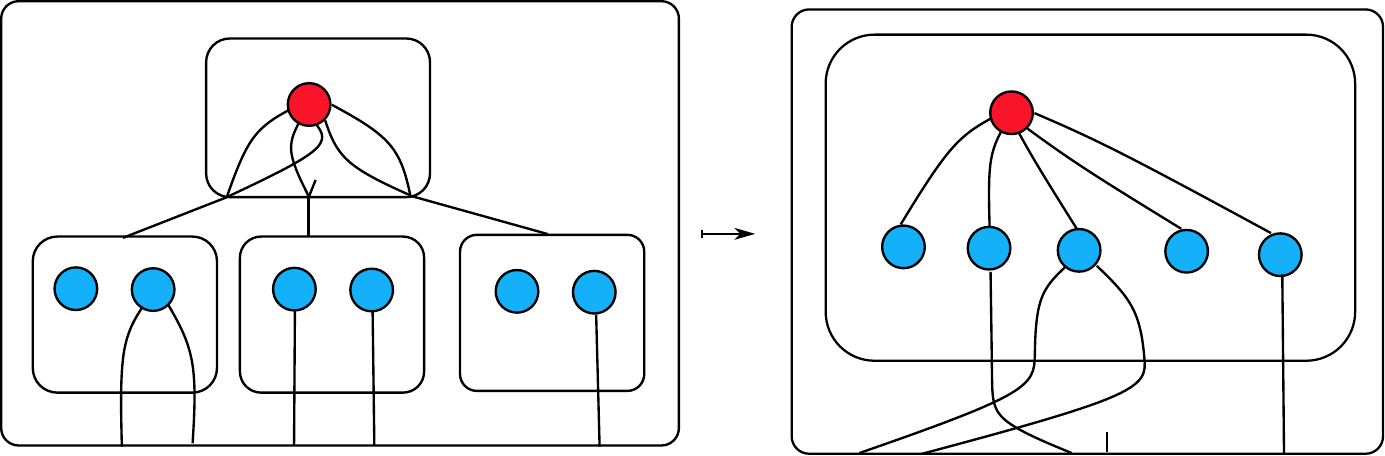}
\end{center}
and  defined by
\begin{eqnarray*}
    \xymatrix{
        (a/g)((b_{1,1},\dots,b_{1,k}),\dots, (b_{n,1},\dots,b_{n,k})) 
        \ar@{|->}[d]^{\ranked{f_2}} \\
         a(b_{g(1)},\dots,b_{g(m)})/g'
    }
\end{eqnarray*}
where $m$ is the arity of $a$ and the grouping function  $g'$ is  the natural embedding of ports 
\begin{align*}
\xymatrix{
    \text{ports of $a(b_{g(1)},\dots,b_{g(m)}))$}
    \ar[d]
    \\
 \displaystyle{\biggl(\coprod_{\substack{i \in \set{1,\ldots,n}\\ j \in \set{1,\ldots,k}}} \text{ports of $b_{i,j}$}, 1 \biggr)} 
}
\end{align*}
\smallskip

\subsubsection{Distribute shallow terms over product}
Finally, consider the function 
\begin{align*}
\ranked{\shallowterm  {\Gamma^k} {\Sigma} \xrightarrow{\quad f_3\quad} (\shallowterm  \Gamma  \Sigma)^k} 
\end{align*}
which distributes shallow terms over the $k$-th power.  This function is illustrated by the following  picture 
\mypic{67}
and defined by 
\begin{eqnarray*}
    \xymatrix{
        (a_1,\dots,a_k)(b_1,\dots,b_n)
        \ar@{|->}[d]^{\ranked{f_2}} \\
        (a_1(b_1,\dots,b_{\text{ar}_1}), a_2(b_{\text{ar}_1+1},\dots,b_{\text{ar}_2}),\dots ,a_k(b_{\text{ar}_{k-1}+1},\dots,b_{\text{ar}_k}))
    }
\end{eqnarray*}
where $\text{ar}_i$ is the arity of $a_i$ for $i\in\set{1,\dots,k}$.

\subsubsection{Unfolding shallow terms.} The following diagram defines  unfolding of shallow terms in terms of the   operations $\ranked{f_1}, \ranked{f_2}, \ranked{f_3}$   defined above:
  \begin{align*}
  \xymatrix@C=2.5cm{
          \ranked{\shallowterm{\mati k \rSigma} {\mati k \rGamma} = {\shallowterm{\reduce k {\Sigma^k}}{\reduce k {\Gamma^k}}} 
        \ar[d]_{\ranked{\substack{f_1}}}
        \ar[r]^-{\ranked{\text{Shallow unfold}}}}
        &
        \ranked{ \reduce k(\shallowterm{\Sigma}{ {\Gamma}})^k = \mati k {(\shallowterm \Sigma \Gamma)}}
        \\
       \ranked{  \reduce k(\shallowterm{\reduce k {\Sigma^k}}{ {\Gamma^k}})}
        \ar[r]_-{\ranked{\flatt\circ \reduce k f_2}}
        &
    \ranked{   \reduce k(\shallowterm{\Sigma^k}{ {\Gamma}}) } \ar[u]^{\ranked{\reduce k  f_3}}
    } 
\end{align*}     

\subsection{Definition of unfolding}
Having defined shallow unfolding, we apply the induction principle described in Section~\ref{sec:terms-induction-principle} to  define unfolding for general terms
\begin{align*}
    \ranked{\unfold : \tmonad \mati k \rSigma \to \mati k {(\tmonad \Sigma)} }.
    \end{align*}
If the input to general unfolding is the identity term $\portletter$, then  the output is:
\mypic{83}
Otherwise, if the input is a nonempty term $a(t_1,\ldots,t_n)$ then the output is obtained by first applying term unfolding to to the smaller terms $t_1,\ldots,t_n$, and then applying the shallow unfold. 

%% file: appendix-examples.tex
\section{Examples}\label{sec:AppendixExamples}

%
%
%
%
To illustrate derivable functions, we present a series of examples, some of them will be useful later. In the rest of this section, for every $k\in\set{1,2,\dots}$ the set $\ranked{k}$ designates the ranked set containing a single element of arity $k$ that we denote by simply by $k$.

\noindent \begin{example}[Parent and children]\label{ex:sibling}  Let $\rGamma$ be a finite type. We define $\ranked{\rGamma_0}$  to be the ranked set obtained from $\rGamma$ by setting the arity of every element to $0$.  
\medskip
Consider the function:
$$\ranked{ \mathsf{Parent}: \tmonad \rGamma \to \tmonad (\rGamma\product (\rGamma_0+0))}$$
which adds to every node of a term in $\tmonad \rSigma$ the label  of its parent if it has one, and $0$ if it is the root.

Let us explain how $\ranked{\mathsf{Parent}}$ can be derived. To illustrate this construction, we use the following alphabet $\rGamma$
\begin{center}
		 \includegraphics[scale=.4]{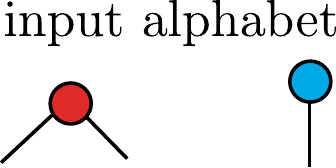}
		\end{center}
and the following term as a running example.
\begin{center}
		 \includegraphics[scale=.4]{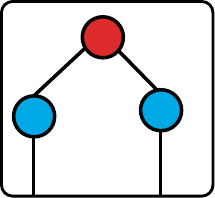}
		\end{center}
We denote by $\ranked{\Gamma_1}$ the ranked set obtained from $\rGamma$ by setting the arity of every element to $1$. If $a$ is a element of $\rGamma$, we denote by $a_1$ the corresponding element of $\ranked{\Gamma_1}$. In our example, the alphabet $\ranked{\Gamma_1}$ is
\begin{center}
		\includegraphics[scale=.4]{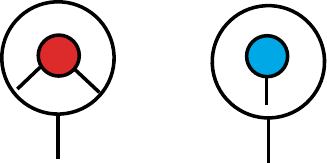}
		\end{center}
\begin{enumerate}
\item  First, we apply the homomorphism 
\begin{align*}
\ranked{\mathsf{Hom}_g:\tmonad\Gamma\to \tmonad(\Gamma+\Gamma_1+1)}
\end{align*}
where $\ranked{g}$ is defined on the elements of $\rGamma$ as follows
\begin{align*}
\ranked{g: \Gamma} & \ranked{\to  \tmonad(\Gamma+\Gamma_1+1 )}\\
      a & \mapsto a\tensorpair{1\tensorpair{a_1\tensorpair{\portletter}},\dots, 1\tensorpair{a_1\tensorpair{\portletter}}}
\end{align*}
In our example, the action of $\ranked{g}$ on the elements of $\rGamma$ looks like this
\begin{center}
		\includegraphics[scale=.4]{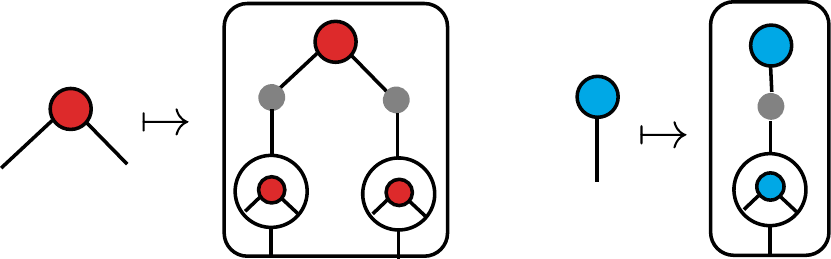}
				\end{center}
Hence, after the application of the homomorphism $\ranked{\mathsf{Hom}_g}$, our initial term becomes
\begin{center}
		\includegraphics[scale=.4]{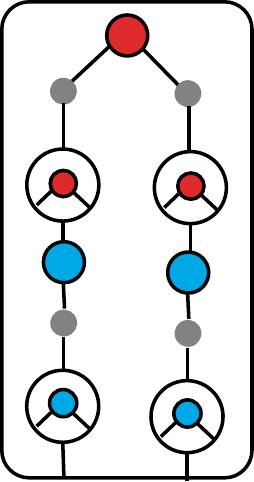}
		\end{center}
\item We apply the factorization 
\begin{align*}
\ranked{\ancfact: \tmonad(\Gamma+\Gamma_1+1) \to \tmonad(\tmonad(\Gamma+\Gamma_1)+\tmonad 1)}
\end{align*}
 to separate the symbol $1$ form the other symbols. After this operation, each node lies in the same factor as (the element of $\ranked{\Gamma_1}$ representing) its parent. In our example, the obtained term is the following
\begin{center}
		\includegraphics[scale=.4]{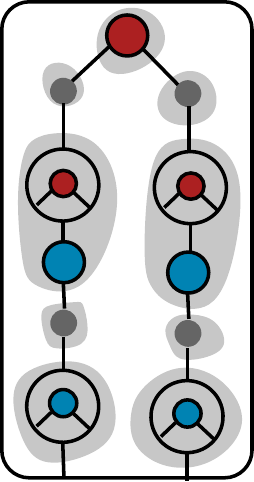}
		\end{center}
\item Consider the function 
\begin{align*}
\ranked{h: \tmonad 1 \to \tmonad((\Gamma+\Gamma_1)\product(\Gamma_0+0))}
\end{align*}
which is the empty term constant function. It is derivable by lifting the empty term constant function over $1$ to terms.
And let $k$ be the function
\begin{align*}
\ranked{k: \tmonad(\Gamma+\Gamma_1) \to \tmonad((\Gamma+\Gamma_1)\product(\Gamma_0+0))}
\end{align*}
which is the identity function, except for the following terms in which it is defined as follows
$$\begin{array}{rll}
a\tensorpair{\portletter,\dots, \portletter}& \mapsto & \tensorpair{a,0}\tensorpair{\portletter,\dots,\portletter}\\
b_1\tensorpair{a\tensorpair{\portletter,\dots,\portletter}}&\mapsto& \tensorpair{a,b_0}\tensorpair{\portletter,\dots,\portletter}\\
b_1\tensorpair{\portletter} &\mapsto& \portletter
\end{array}$$
\end{enumerate}
We apply the function $\ranked{h}$ to the factors $\ranked{\tmonad 1}$ and the function $\ranked{k}$ to the factors $\ranked{\tmonad(\rGamma+\rGamma_1)}$. Doing so, we obtain a term in $\ranked{\tmonad\tmonad((\Gamma+\Gamma_1)\product(\Gamma_0+0))}$, which we flatten, then we erase the symbols $\ranked{\Gamma_1}$ using the function $\mathsf{Filter}$ of Example~\ref{ex:filter} to obtain the desired term.

\smallskip
If $\rGamma$ is a finite ranked set, we define $\ranked{\Gamma^*}$ as
$$\coprod_{i \leq \text{ maximal arity in } \rGamma} \underbrace{\ranked{\rGamma\product \cdots \product \rGamma}}_{i\text{ times}}$$
Now consider the function $$\ranked{\mathsf{Children}:\tmonad\rGamma\to \tmonad (\rGamma\product (\rGamma_0+0)^*)}$$ which tags every node of a term in $\tmonad \rGamma$ by the list  of its children symbols. When a child is a port, it is marked by $0$ in the list.
The function $\ranked{\mathsf{Children}}$ can be derived using a similar construction as above.
\end{example}

\noindent  \begin{example}[Root and leaves] Let $\rSigma$ be a finite type and $\ranked{f:\rSigma \to \rGamma}$, $\ranked{g: \rSigma \to \rGamma}$ be derivable functions. The function $$\ranked{\mathsf{Root}_{f,g} : \tmonad\rSigma \to \tmonad\rGamma}$$
which applies $\ranked{f}$ to the root and $\ranked{g}$ to the rest of the tree is a derivable function.
To show this, we first start by applying the function $\ranked{\mathsf{Parent}}$. Doing so, the root can be distinguished from the other nodes since it will be tagged by $0$.  

The function $\ranked{h}$ defined below is derivable since its domain is finite. 
\begin{align*}
\ranked{h:\rSigma\product(\rSigma_0+0)}&\ranked{\to \rGamma}\\
  \tensorpair{a,0} &\mapsto f(a) \\
  \tensorpair{a,b} &\mapsto g(a) \text{ if } b\neq 0.
\end{align*}
We lift $\ranked{h}$ to terms to conclude.

\smallskip
Similarly, the function $$\ranked{\mathsf{Leaves}_{f,g} : \tmonad\rSigma \to \tmonad\rGamma}$$
 which applies $\ranked{f}$ to the leaves and $\ranked{g}$ to the rest of the tree is derivable. This is done using the same ideas as before, but invoking the function $\ranked{\mathsf{Children}}$ instead of the function $\ranked{\mathsf{Parent}}$: leaves can be distinguished from the other nodes since they are tagged either by a list of $0$ or the empty list.
\end{example}

\noindent\begin{example}[Descendants and ancestors]\label{ex:descendant} If $\rSigma$ is a finite type and $\ranked{\rGamma\subseteq \rSigma}$, then the functions 
\begin{itemize}
\item $\ranked{\mathsf{Descendant}_\rGamma: \tmonad \rSigma \to \tmonad (\rSigma+\rSigma)}$ which replaces the label of each node by its first or second copy, depending on whether it has a descendant in $\rGamma$,
\item $\ranked{\mathsf{Ancestor}_\rGamma: \tmonad \rSigma \to \tmonad (\rSigma+\rSigma)}$ which replaces the label of each node by its first or second copy, depending on whether it has a descendant in $\rGamma$,
\end{itemize}
are derivable.

To derive $\ranked{\mathsf{Descendant}_\rGamma}$, we start by applying the factorization $$\ranked{\decfact: \tmonad\rSigma\to \tmonad(\tmonad\rGamma+\tmonad(\rSigma\setminus\rGamma))}$$ which regroups the elements of $\rSigma$ and the elements of $\ranked{\rSigma\setminus\rGamma}$ into factors depending on whether they have the same ancestors of the same type.

Obviously, all the nodes of the $\rGamma$ factors have a descendant in $\rGamma$. 
In the $\ranked{\rSigma\setminus\rGamma}$ factors which are not leaves in the factorized term, all the nodes have a $\rGamma$ descendant in the original term. To show this, take $f$ to be one of these factors, and suppose by contradiction that one of its nodes does not have a descendant in $\rGamma$. By definition of $\ancfact$, all the elements of $f$ do not have a descendant in $\rGamma$ as well. Since $f$ is not a leaf, it has a child $g$. The factor $g$ cannot be a $\rGamma$
factor as the nodes of $f$ would have a descendant in $\rGamma$. The factor $g$ is then necessarily  a $\ranked{\rSigma\setminus \rGamma}$ factor. If a node of $g$ has a descendant in $\rGamma$, this would give a $\rGamma$ descendant to one of the node of $f$. Thus all the nodes of $g$ are in $\ranked{\rSigma\setminus \rGamma}$ and do not have a descendant in $\rGamma$, meaning that $f$ and $g$ are actually the same factor, which gives a contradiction. Finally, the $\ranked{\rSigma\setminus\rGamma}$ factors which are leaves do not have a descendant in $\rGamma$. With these observations, we can now implement $\mathsf{Descendant}_\rGamma$. 

Let us consider the functions 
$$\begin{array}{llll}
\ranked{\mathsf{Yes}_\rGamma :} & \rGamma &\ranked{\to} &\ranked{ \rSigma+\rSigma}\\
 \ranked{\mathsf{Yes}_{\ranked{\rSigma\setminus\rGamma}:}}& \ranked{\rSigma\setminus\rGamma}&\ranked{\to} &\ranked{ \rSigma+\rSigma}\\
\ranked{\mathsf{No}_{\ranked{\rSigma\setminus\rGamma}: }}&\ranked{\rSigma\setminus\rGamma} &\ranked{\to}& \ranked{ \rSigma+\rSigma}
\end{array}$$
which replaces the label of each node by its first copy for $\ranked{\mathsf{Yes}_\Gamma}$ and $\ranked{\mathsf{Yes}_{\rSigma\setminus\rGamma}}$, and by its second copy for $\ranked{\mathsf{No}_{\rSigma\setminus\rGamma}}$. The three functions are derivable as their domains are finite. 
Consider the functions 
\begin{align*}
\ranked{f:=} &\ranked{ \tmonad\mathsf{Yes}_\rGamma +\tmonad\mathsf{No}_{\rSigma\setminus \rGamma}}\ranked{: \tmonad\rGamma+\tmonad(\rSigma\setminus\rGamma) \to \tmonad (\rSigma+\rSigma)}
\\
\ranked{g:=} & \ranked{ \tmonad\mathsf{Yes}_\rGamma
+\tmonad\mathsf{Yes}_{\rSigma\setminus \rGamma}}\ranked{: \tmonad\rGamma+\tmonad(\rSigma\setminus\rGamma) \to \tmonad (\rSigma+\rSigma)} 
\end{align*}
The descendant function is obtained by applying $\ranked{\mathsf{leaves}_{f,g}}$ followed by a flattening.

\smallskip
To derive the function $\ranked{\mathsf{Ancestor}_\Gamma}$, we apply first a the factorization
\begin{align*}
\ranked{\ancfact: \tmonad\rSigma\to \tmonad(\tmonad\rGamma+\tmonad(\rSigma\setminus\rGamma))}
\end{align*} which regroups the elements of $\rSigma$ and the elements of $\ranked{\rSigma\setminus\rGamma}$ into factors depending on whether they have the same descendants of the same type. 
Using similar arguments as before, we can conclude that:
\begin{itemize}
\item The nodes inside $\rGamma$ factors have $\rGamma$ ancestors.  
\item If a $\ranked{\rSigma\setminus\rGamma}$ factor is the root of the factorized term, then its nodes do not have a $\rGamma$ ancestor.
\item   If a $\ranked{\rSigma\setminus\rGamma}$ factor is not the root of the factorized term, then its nodes do have a $\rGamma[Descendants and ancestors]\label{ex:descendant}$ ancestor.
\end{itemize}
The ancestor function is obtained by applying $\ranked{\mathsf{root}_{f,g}}$ followed by a flattening.
\end{example}

\begin{example}[Error raising.]\label{ex:error-raising}
We can think of the type $\ranked{\bot}$ as an error type. Indeed, the following raising error functions are derivable.
\begin{lemma}\label{lem:error-raising}
Let $\rSigma$ and $\rGamma$ be two datatypes. The functions
$$\begin{array}{rll}
\ranked{\tmonad(\Sigma+\bot)} &\ranked{\to }&\ranked{\tmonad \Sigma +\bot}\\
\ranked{(\Sigma+\bot) \product (\Gamma+\bot)} &\ranked{\to} &\ranked{\Sigma\product \Gamma+\bot }\\
\ranked{\shallowterm {(\Sigma+\bot)}{(\Gamma+\bot)}} &\ranked{\to} &\ranked{\shallowterm\Sigma\Gamma +\bot}\\
\ranked{\reduce k (\Sigma+\bot)} &\ranked{\to} &\ranked{\reduce k \Sigma +\bot} 
\end{array}$$
which are defined as follows
\begin{align*}
 t \text{ of arity } n \ \ \mapsto \  \begin{cases}
    t & 
    \text{if $t$ does not contain any element of $\ranked{\bot}$,}\\
    n & \text{otherwise.}
\end{cases}
\end{align*} are derivable.
\end{lemma}
These functions can be easily derived using Proposition~\ref{prop:forat} and distributivity prime functions. The details of the proof are left as an exercise to the reader.
\end{example}

\begin{example}[Partial functions.]
Thinking of $\ranked{\bot}$ as an error datatype, a function of type $\ranked{\Sigma \to \Gamma +\bot}$
can be seen as a partial function from $\rSigma$ to $\rGamma$. We write 
\begin{align*}
\ranked{\Sigma\rightharpoonup \Gamma}
\end{align*}
as a notation for the function type $\ranked{\Sigma \to \Gamma +\bot}$.  
Using the error raising mechanisms discussed earlier, we can manipulate transparently partial function. Indeed, all datatype constructors can be lifted to partial functions, by composing the liftings (1)--(4) with the error raising functions from Lemma~\ref{lem:error-raising}. For example, if $\ranked{f:\Sigma\rightharpoonup \Gamma}$ is a partial function, then $\ranked{\tmonad f:\tmonad\Sigma\rightharpoonup \tmonad\Gamma}$ is defined as the composition
\begin{align*}
\ranked{\tmonad\Sigma\xrightarrow{\tmonad f} \tmonad(\Gamma+\bot) \xrightarrow{\text{Error raising}} \tmonad \Gamma +\bot}.
\end{align*}
\end{example}

%% file: appendix-to-fo.tex
\section{Derivable functions can be described in first-order logic}
\label{sec:to-logic}
The goal of this section is to show the right-to-left implication of Theorem~\ref{thm:main}, which says that derivable functions can be implemented by first-order transductions. 

As discussed in the body of the paper, we proceed by induction on the derivation. During this induction, we will need to show that every prime function is a first-order transduction. Prime functions are not tree-to-tree functions, instead they transform dataypes into datatypes. This is the reason why we need 
\begin{itemize}
\item to generalize tree-to-tree transductions into  transductions that can transform models over arbitrary vocabularies (and not only the vocabulary of trees). 
\item show how datatypes (terms, pairs, copairs and folds) can be encoded as models over a well chosen vocabulary. More precisely, we will associate to every datatype $\rSigma$ a relational vocabulary that we call \emph{vocabulary of $\rSigma$}. Structures over this vocabulary will be called \emph{models over $\rSigma$}. Then we will define a function
\begin{align*}
    \xymatrix@C=2cm{
        \rSigma \ar[r]^-{x \mapsto \underline x} &
 \text{models over $\rSigma$}
    }
\end{align*} 
which assigns to each element $x \in \rSigma$ a corresponding model  over $\rSigma$, which is denoted by $\underline x$.
\end{itemize}
 
 Right-to-left implication of Theorem~\ref{thm:main} can be then generalized to the following statement, more suited to a proof by induction:

\begin{proposition}\label{prop:main-right-to-left}
Let $\rGamma$ and $\rSigma$ be two datatype.  For every derivable function $\ranked{f}$, there is a first-order transduction $g$ such that the following diagram commutes
  \begin{align*}
  \xymatrix@C=2.8cm{
          \rSigma 
        \ar[d]_{x \mapsto \underline x}
        \ar[r]^-{\ranked{f}}
        &
    \rGamma \ar[d]^{x \mapsto \underline x}
        \\
      \text{models over $\rSigma$}
        \ar[r]_-{g}
        &
    \text{models over $\rGamma$}   
    } 
\end{align*}  
\end{proposition}

The rest of this section is organized as follows. We define first-order transductions transforming arbitrary models in Section~\ref{sec:fo-transduction-def}. In Section~\ref{sec:data-as-models} we define the vocabularies for the  datatypes and the model representation $x \mapsto \underline x$. Finally, we prove Proposition~\ref{prop:main-right-to-left} which gives as a corollary the right-to-left implication of Theorem~\ref{thm:main}.

\subsection{First-order transductions}\label{sec:fo-transduction-def}
The following definition introduces first-order transductions, which generalizes tree-to-tree transductions given in Definition~\ref{def:fo-transduction} to arbitrary models.  

\begin{definition}[First-order transduction]\label{def:fo-transduction-gen}\ 
    A \emph{first-order transduction} is defined to be any  composition of the following two kinds of transformations on structures:
\begin{enumerate}
    \item \emph{Copying.} Fix some  relational vocabulary $\ranked \sigma$ and let $k \in \set{1,2,\ldots}$. Define $k$-copying to be the operation of type 
    \begin{align*}
    \xymatrix{
     \text{models over $\ranked \sigma$} \ar[d]\\
     \txt{models over $\ranked \sigma$  extended\\
with a $k$-ary relation $\mathrm{copy}$}
    }
    \end{align*}
which inputs a model $\mathbb A$, and outputs $k$ disjoint copies of $\mathbb A$, where the  $\mathrm{copy}$ relation is interpreted as the set of tuples $(a_1,\ldots,a_k)$ such that, for  some $a \in \mathbb A$, the first copy of $a$ is  $a_1$, the second copy of $a$ is $a_2$, etc. The $\mathrm{copy}$ relation  is not commutative, because we distinguish the copies.
\item    \emph{Non-copying first-order transduction.} The syntax of a \emph{non-copying first-order transduction}  is given by:
\begin{enumerate}
    \item Input relational vocabulary $\ranked\sigma$ and output relational vocalbulary $\ranked{\gamma}$.
    \item A first-order \emph{universe formula} $\varphi(x)$ over $\ranked{\sigma}$.
    \item For every relation $R$ in vacubulary $\ranked{\gamma}$, a first-order  formula $\varphi_R(x_1,\ldots,x_{\arity R})$ over $\ranked{\sigma}$.
\end{enumerate}
The semantics of a non-copying first-order transduction is  a function
\begin{align*}
    \xymatrix{
        \text{models over $\ranked \sigma$} \ar[d]\\
        \text{models over $\ranked \gamma$}
    }
\end{align*}
defined as follows. If the input model is $\mathbb A$, then the output model is defined as follows: the universe is elements of $\mathbb A$ which satisfy the universe formula, and each relation $R$ is interpreted as those tuples that satisfy $\varphi_R$. 
 \end{enumerate}
\end{definition}

The notion of copying used in the above definition is slightly different from the notion of copying used for tree-to-tree transductions in Definition~\ref{def:fo-transduction}, which was specifically tailored to stay within the realm of trees. Nevertheless, the two definitions are easily seen to define the same class of tree-to-tree functions.
 
\subsection{Datatypes  as models.}\label{sec:data-as-models}
Let us show how to encode datatypes as relational vocabularies and data as models over these vocabularies. 
\begin{definition}[Associated models for terms, pairs, co-pairs, folds.] \label{def:type-model} To each type  $\rSigma$ we associate a vocabulary, called the \emph{vocabulary of $\rSigma$}, and a map 
    \begin{align*}
        a \in \rSigma \qquad \mapsto \qquad \underbrace{\underline a \in \text{models over the  vocabulary of  $\rSigma$}}_{\text{associated model of $a$}}.
    \end{align*}
    Furthermore, for each $a \in \rSigma$ we  distinguish a  sequence (whose length is the arity of $a$) of elements in $\underline a$, which are called the ports of $\underline a$.   The definitions are by induction on the structure of $\rSigma$, as given below.
    \begin{itemize}
        \item \emph{Finite ranked sets.} Elements of a ranked set   \begin{align*}
        \rSigma =  \set{a_1,\ldots,a_k}
        \end{align*} are modelled  using a vocabulary which has unary relations $a_1,\ldots,a_k$ and $P_1,\ldots,P_m$ where $m$ is the maximal arity of elements in $\rSigma$. 
        For $a \in \rSigma$ of arity $n$, the  universe of $\underline a$ is $\set{0,1,\ldots,n}$, with the ports being $1,\ldots,n$. 
            The  relation $P_i$  is interpreted as $\set i$ when $i \in \set{1,\ldots,n}$ and as the empty set otherwise. The relation $a_i$ is interpreted as $\set 0$ when $a = a_i$ and as the empty set otherwise. 
        \item \emph{Coproduct.}  Elements of the coproduct $\ranked{\Sigma_1 + \Sigma_2}$ are modelled using the disjoint union of the vocabularies of $\ranked{\Sigma_1}$ and $\ranked{\Sigma_2}$. 
            If an element of the coproduct comes from $\ranked{\Sigma_1}$, then its associated model is defined as for the type $\ranked{\Sigma_1}$, with  the remaining relations from the vocabulary of   $\ranked{\Sigma_2}$ interpreted   as empty sets. The definition is analogous for  elements from $\ranked{\Sigma_2}$. 
        \item \emph{Product.}   Pairs in   $\ranked{\Sigma_1 \product \Sigma_2}$ are modelled
        using the disjoint union of the vocabularies of $\ranked{\Sigma_1}$ and $\ranked{\Sigma_2}$. 
            For  $\tensorpair{a_1,a_2}$, the associated model is    the disjoint union of models $\underline{a_1} + \underline {a_2}$, with the relations of $\underline {a_1}$ using the  vocabulary of ${\ranked{\Sigma_1}}$, and the relations of $\underline {a_2}$ using the vocabulary  ${\ranked{\Sigma_1}}$. 
            If $n_1$ is the arity of $a_1$, then the first $n_1$ ports are inherited from  $\underline {a_1}$ and the remaining ports are inherited from  $\underline {a_2}$.
        \item \emph{Folding.}   For $k \in \set{1,2,\ldots}$, elements of   $\reduce k \rSigma$ are modelled using the  vocabulary of $\rSigma$ plus two extra binary relations $\portord$ and $R$. If $a \in \rSigma$ has arity $nk$, then the model associated to $a/f$ -- which has arity $n$ --   is obtained from  $\underline{a}$ by adding a copy of the model below, where $\sqsubset$ is the natural ordering on integers
                \begin{align*}
                (\set{1,\ldots,n}, \portord),
                \end{align*}
whose elements are used as the ports, and interpreting the binary relation $R$ as
        \begin{align*}
        \set{(\text{$i$-th port of $\underline a$},f(i)) : i \in \set{1,\ldots,nk}}
        \end{align*}

        \item \emph{Terms.}   Terms in $\tmonad \rSigma$ are modelled using vocabulary of $\rSigma$ extended with two fresh binary relations $\anceord$ and $\portord$. 
          Let $t \in \tmonad \rSigma$. Consider the disjoint union of models
            \begin{align}\label{eq:non-port}
                 \coprod_{x \in \text{non-port nodes in $t$}} \underline{a(x)},
            \end{align}
         where  $\underline a(x)$ is the model over vocabulary of $\rSigma$ that  is defined by induction assumption.   In the above  disjoint union, the same vocabulary, namely the vocabulary of $\rSigma$,  is used  for all parts of the disjoint union. Next, consider  the model
            \begin{align}\label{eq:ports}
            (\set{1,\ldots,n}, \portord)
            \end{align}
            where $\portord$ is the natural ordering on $\set{1,\ldots,n}$. 
            The model of $t$ is defined by taking the disjoint union of the models in~\eqref{eq:non-port} and~\eqref{eq:ports}, and defining the descendent relation $\anceord$ as the set of pairs $(u,v)$ such that:
            \begin{itemize}
            \item either $u$ is the $i$-th port of $\underline{a(x)}$ for some node $x$ of $a$, $v$ is a port of $\underline{a(y)}$ for some node $y$ which is a descendent of the $i$-th child of $x$.
            \item or $u$ is the $i$-th port of $\underline{a(x)}$ for some node $x$ of $a$, $v=j\in\{1,\dots,n\}$ and the $j$-th port of $a$ is a descendent of the $i$-th child of $x$.
            \end{itemize} 

    \end{itemize}
\end{definition}

The above  definition creates a certain ambiguity for trees, because if $t$ is a tree over a finite ranked set $\rSigma$, then $\underline t$ can be understood in two ways: as per  Definition~\ref{def:tree-model} for trees, or as per Definition~\ref{def:type-model} when $t$ is viewed as a special case of a term $t \in \tmonad \rSigma$. Since we only use first-order transductions to transform relational structures,  this ambiguity is not a problem, because one can easily define first-order transductions which map one definition of $\underline t$ to the other.

\subsection{Proof of Proposition~\ref{prop:main-right-to-left}}
The proof proceeds by induction, following the definition of derivable functions. 
In the induction step, we have to deal with function composition and the lifting of function along the datatype constructors. First-order transductions are closed under composition by definition, while the liftings are immediate. 

In the induction base, we need to show that all of the prime functions are first-order transductions. 
All the cases are easy, and consist mainly on unfolding the definitions; this is the point of calling these functions prime. There is one exception, which requires some more explanation, namely monotone unfolding. We explain below just one of the easy functions, the unit function $\ranked{\Sigma \to \tmonad \Sigma}$, and the monotone unfolding. The other prime functions are left as an exercise. 
    
\subsubsection{A first-order transduction for the term unit}
\label{sec:transduction-unit}
    In the following, it will be convenient to use, as part of the vocabulary of $\rSigma$, a unary relation  $\mathsf{Port}_\rSigma$ which selects the ports of the structures over the vocabulary of $\rSigma$; and a binary relation $\sqsubset_\rSigma$ which orders these ports. By induction on $\rSigma$, we can show that both relations are definable by first-order formulas over  the vocabulary of $\rSigma$.

   Given an element $x$ of $\rSigma$, let us show how  $\unit(x)$ can be implemented using a first-order transduction.  The copying constant is 2,
    the first copy will contain the whole structure $\underline{x}$ and the second copy will select only the ports of $\underline{x}$ which will serve as the ports of the structure $\underline{\unit(x)}$, as illustrated by the following picture 
\begin{center}
    \includegraphics[scale=.18]{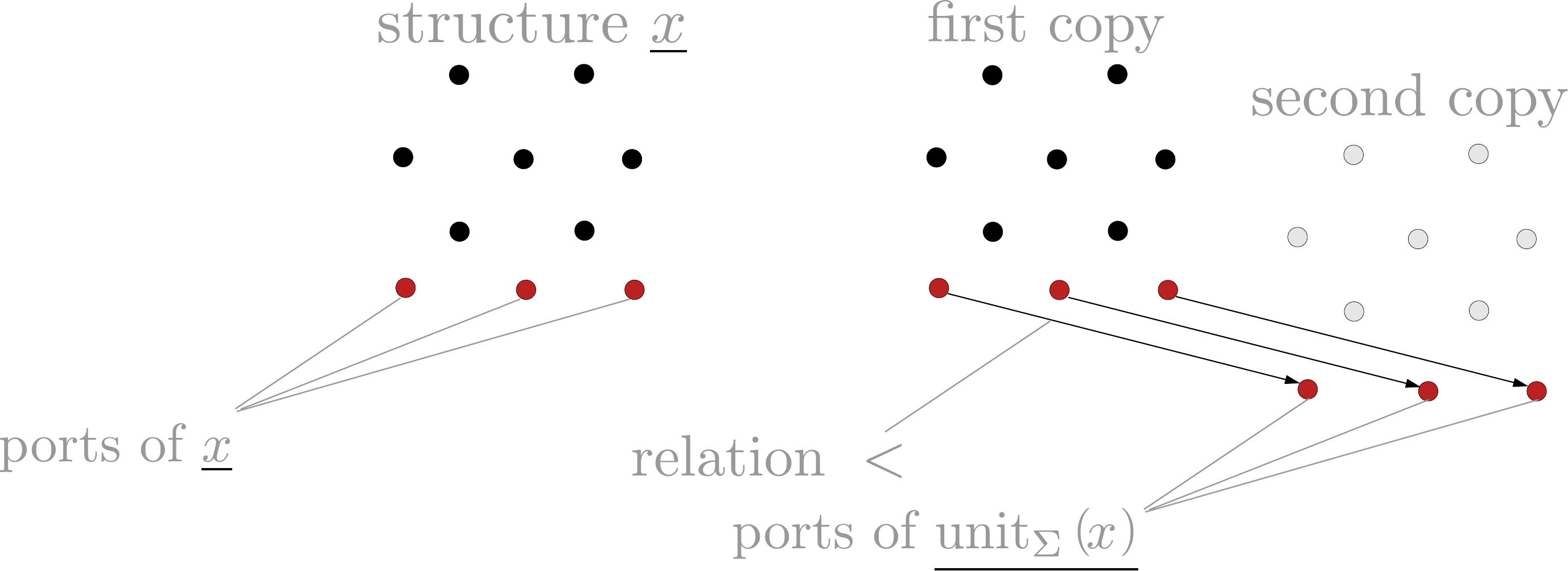}
    \end{center}    
      The universe formulas are then:
    \begin{align*}
    \varphi_1(x)=\mathsf{True} \qquad \varphi_2(x)=\mathsf{Port}_\rSigma(x)
    \end{align*}
    In the first copy, the vocabulary of $\rSigma$ will be interpreted as in the original structure, and as the empty set in the second copy. That is, for every unary relation $R$ and for every binary relation $S$ in the vocabulary of  $\rSigma$, we set:
    \begin{align*}
   \varphi_R^{1}(x)=R(x) \quad&\quad \varphi_S^{1,1}(x,y)=S(x,y)\\
   \varphi_R^{2}(x)=\mathsf{False} \quad&\quad \varphi_S^{2,2}(x,y)=\mathsf{False}
\end{align*}      
Let us interpret the relations $<$ and $\sqsubset$ of the vocabulary of $\tmonad\rSigma$. The  ports of $\underline{\unit(x)}$ inherit the order of the ports of $\underline{x}$, this is why we set:
\begin{align*}
\varphi_\sqsubset^{2,2}(x,y)=x\sqsubset_\rSigma y
\end{align*}
The descendant relation $<$ connects the $i^{th}$ port of $\underline{x}$ to the $i^{th}$ port of $\underline{\unit(x)}$. Since these nodes come from the same node in the original structure, we set:
\begin{align*}
\varphi_<^{1,2}(x,y)=x=y
\end{align*}

\subsubsection{A first-order transduction for monotone unfolding}
\label{sec:fo-transduction-for-unfolding}
Having illustrated the syntax of first-order transductions on the example of the unit function, we describe a first-order transduction for the monotone unfolding operation 
\begin{align*}
    \ranked{\tmonad \mati k{\Sigma} \to \mati k{( \tmonad \Sigma)} + \termset}.
\end{align*}
This is the only prime function whose corresponding first-order transduction is not obvious.
Unlike in Section~\ref{sec:transduction-unit},  we focus more  on the underlying conceptual difficulties than on the syntax of first-order transductions. 

Recall  that when  defining the  monotone  unfolding operation,  for each element $a \in \ranked{\tmonad \mati k{\Sigma}}$ of the matrix power, we used a family of (partial) twist functions
\begin{align*}
\to_i : \set{1,\ldots,k} \to \set{1,\ldots,k},
\end{align*}
one for each port $i$ of $a$. For the reader's convenience, we repeat a picture from Section~\ref{sec:unfolding}, which explains the twist functions:
\mypic{125}
In this example, the twist function $\to_1$ is monotone, but  $\to_2$ is not. The monotone unfolding operation works in the same way as general  unfolding, except that it uses the undefined value $\termset$ if the input term has at least one letter which uses at least one non-monotone twist. 

The following lemma, whose simple proof is left to the reader, shows that the twist functions can be defined using first-order logic.
\begin{lemma}
    Let $\rSigma$ be a datatype and let $k \in \set{1,2,\ldots}$. For every partial function
    \begin{align*}
    \tau : \set{1,\ldots,k} \to \set{1,\ldots,k}
    \end{align*}
    there is a first-order formula $\varphi_\tau(x)$ such that for every $a \in \mati k \rSigma$, 
    \begin{align*}
    \underline a \models \varphi_\tau(x)
    \end{align*}
     if and only if $x$ represents a port  with twist function $\tau$.  
\end{lemma}
By using the formulas from the above lemma, one can construct  a first-order formula which checks if a term in $\tmonad \mati k \rSigma$ uses only monotone twists, i.e.~whether or not the output of monotone unfolding should be $\termset$. 

We now proceed to the more interesting part of monotone unfolding, i.e.~actually doing the unfolding for  monotone inputs. Consider an input $t \in \tmonad \mati k \rSigma$ to monotone unfolding.
Define a \emph{sub-node} of $t$ to be a pair (node of $t$, number in $\set{1,\ldots,k$}), as explained in the following picture:
\mypic{124}

In the output of the monotone unfolding, which is of the form
\begin{align*}
(t_1,\ldots,t_k)/f \in \mati k {\tmonad \rSigma},
\end{align*}
the nodes of the  output terms $t_1,\ldots,t_k$ will correspond to the sub-nodes in the input $t$. The sub-nodes can be produced by  copying the input term $k$-times.

The most interesting part of the structure in the output is the descendant relation in the terms $t_1,\ldots,t_k$. This relation  can be viewed as a descendant relation on the sub-nodes. We only describe how the descendant relation on the sub-nodes can be defined in first-order logic, and the rest of the transduction is left to the reader.

When defining the descendant relation on sub-nodes, the crucial part is composing the twist functions.  Suppose that we want to check the descendant relationship between two sub-nodes
\begin{align}\label{eq:descendant-relationship}
(x,i) \stackrel?{\le} (y,j),
\end{align}
where $x,y$ are nodes on the input term and $i,j \in \set{1,\ldots,k}$. We will show that the descendant relationship~\eqref{eq:descendant-relationship} holds if and only if $x$ is an ancestor of $y$ in the input term,  and the twist functions on the path connecting $x$ and $y$ maps $j$ to $i$, as explained below. 

Consider a path in the input term, which connects  node $x$ with $y$, as  illustrated in the following picture
\mypic{123}    
Each edge in the input term corresponds to a chosen port in some node, which in turn corresponds to some twist function, and therefore it makes sense to talk about the twist function associated to an edge in the input term. Define 
\begin{align*}
\tau_y^x : \set{1,\ldots,k} \to \set{1,\ldots,k}
\end{align*}
to be the partial function, which is obtained by composing all of the twist functions corresponding to edges on the path connecting $y$ to $x$,  starting with $y$ and ending with $x$. In the example from the above picture, we compose two twist functions, which correspond to edges marked in yellow. 

Equipped with the above definitions, we can now characterise the descendant ordering on sub-nodes by
\begin{align*}
    (x,i) \le (y,j) \qquad \text{iff} \qquad x \le y \land \tau_y^x(j)=i.
\end{align*}
Therefore, to complete the proof, it remains to show the following lemma. This is where we use the monotonicity assumption.

\begin{lemma}\label{lem:counter-free}
    For every $i,j \in \set{1,\ldots,k}$ 
    there is a first-order formula $\psi_j^i(x,y)$ such that for every $t  \in \tmonad \mati k \rSigma$
    \begin{align*}
    \underline t \models \psi_j^i(x,y) \qquad \text{iff} \qquad \tau_y^x(j)=i.
    \end{align*}
\end{lemma}
\begin{proof}
    Let $F$ be the set of monotone partial functions  from $\set{1,\ldots,k}$ to itself. Define $L \subseteq F^*$ to be the set of those words $f_1 \cdots f_n$ such that the composition of functions $f_n\circ \cdots \circ f_1$ maps $i$ to $j$. We will show that -- thanks to the monotonicity assumption -- the language $L$ is definable in first-order logic. To get the conclusion of the lemma, we check if the sequence of twist functions on the path  from $x$ to $y$ satisfies the first-order formula defining the language $L$.

    The language $L$ is recognised by a finite automaton, which has states $\set{1,\ldots,k,\bot}$, and which simply applies the function in its input letter to the present state. We show below that this automaton is counter-free, in the sense of McNaughton and Papert~\cite[p.~6]{McNaughtonPapert71}, and therefore it can be defined in first-order logic. 

    Recall that a counter in an automaton is a sequence of at least two pairwise distinct states $q_1,\ldots,q_n$ such that
    \begin{align*}
    q_1  \stackrel w \to q_2 \stackrel w \to \cdots \stackrel w \to q_n \stackrel w \to q_1
    \end{align*}
    holds for some common input string $w$. In the automaton for the language $L$ that we have discussed above, there is no counter. Indeed, if we would have $q_1 \le q_2$, then by monotonicity of the function   $w \in F$ we would have 
    \begin{align*}
    q_1 \le q_2 \le \cdots \le q_n \le q_1
    \end{align*}
    and therefore all of $q_1,\ldots,q_n$ would be equal, contradicting the assumption that they are pairwise distinct. The same argument would work when $q_1 \ge q_2$. By~\cite[Theorem 10.5]{McNaughtonPapert71}, if an automaton has no counter, then its language is definable in first-order logic. 
    
\end{proof}

%% file: appendix-forational.tex
\section{Appendix on first-order relabelling}~\label{sec:AppendixForat}

The goal of this section is to show Proposition~\ref{prop:forat}, which says that first-order relabeling are derivable. As discussed in the body of the paper, the proof of this proposition is based on an equivalence result between first-order queries on trees  and a temporal logic, as stated in Lemma~\ref{lem:schlingloff}. While this result is deaply inspired from a similar result of Schlingloff \cite{schlingloff1992expressive}, our frameworks are not exactly the same (he uses for ainstance unranked trees).  
In the rest of this section, we provide more details about the reduction from Schilgloff's result to our lemma (Section~\ref{sec:reduction-schilingloff}). Then we show in Section~\ref{sec:relabeling}  how to use it in order to prove Proposition~\ref{prop:forat}.

\subsection{Reduction to Schilgloff's theorem}\label{sec:reduction-schilingloff}

Let us proceed to the proof of Lemma~\ref{lem:schlingloff}.
Clearly the functions in the lemma are first-order tree relabeling, and first-order tree relabeling are easily seen to be closed under composition, which gives the right-to-left inclusion in the lemma. The hard part is the left-to-right inclusion, which says that every first-order tree relabeling can be decomposed into functions as in items~\ref{it:relabelling},\ref{it:child}--\ref{it:since}. 
The first  step in the proof of  the right-to-left inclusion is the observation that  every first-order tree relabeling can be decomposed as 
\begin{align*}
    g \circ f_1 \circ \cdots \circ f_n
\end{align*}
where $g$ is a relabeling as in item~\ref{it:relabelling} of the lemma and each $f_i$ is a  characteristic function of some unary query (not necessarily of the simple form indicated in items~\ref{it:child} -- \ref{it:since} in the lemma). This is a simple observation: the functions $f_1,\ldots,f_n$ annotate the tree with the truth values of the unary queries used in the definition of the  first-order relabeling, and $g$ uses these truth values to select the appropriate output label. The hard part of the lemma is showing that each $f_i$  can be further decomposed into functions as indicated in the lemma. This is where we us the result of Schlingloff~\cite[Theorem 2.6]{schlingloff1992expressive}, which says that all first-order definable tree properties can be defined using a temporal logic that has operators similar to the ones used in items~\ref{it:child} -- \ref{it:since} of the lemma. 

The following table summarizes our framework (first column) and Schlingloff's one (second column). The first row describes the models under consideration, the second row the corresponding version of first-order logic, and the third row the corresponding temporal logic. 
\begin{center}
\begin{tabular}{|C{1.8cm}|C{2.9cm}|C{2.8cm}|}
\cline{1-3}
\textbf{Models} & \textbf{Trees over a finite ranked alphabet $\rGamma$}: finitely branching, ranked trees, labeled from $\rGamma$. & \textbf{Models over a set of propositions $P$}: finitely branching, unranked trees, labeled from $2^P$.\\
\hline
    \multirow{3}{1.8cm}{\begin{center}\textbf{First-order logic}\end{center}}&\multicolumn{2}{C{5.6cm}|}{Usual first-order connectives ($\exists, \vee, \neg$) with the descendant predicate $x\leq y$ and the following predicates:}\\
    \cline{2-3} 
                                 & $a(x)$:  $x$ is labeled $a$  $(a \in \rGamma)$.  & $p(x)$: label of $x$ contains $p$
                                  ($p\in P$). \\
                                  &\hspace{-.14cm}$\child i(x)$: $x$ is an $i$-th child. &\\
   &       \textbf{We call it $\rGamma$-\fo.}    &  \textbf{We call it $P$-\fo. }            \\
    \hline
    \multirow{4}{1.8cm}{\begin{center}
    \textbf{Temporal logic}
    \end{center}}& \multicolumn{2}{C{5.6cm}|}{Usual CTL connectives ($S$ (Since), $U$ (Until), $\vee$, $\neg$) together with:}\\
    \cline{2-3} 
    
     & $a \in \rGamma$, & $p \in P$, \\
    & $\odot_i \phi$: the $i$-th child satisfies $\phi$. & $X_i \phi$: at least $i$ children satisfy $\phi$.\\ 
    &       \textbf{ We call it 2-CTL. }   &   \textbf{We call it 4-CTL.            } \\
    \hline
\end{tabular}
\end{center}
What is named $\rGamma$-\fo in the table is what we simply called first-order logic along the paper. The operators of 2-CTL are those of Lemma~\ref{lem:schlingloff}.  Using the notation of the table, 
Schlingloff's theorem says that $P$-\fo formulas are equivalent to 4-CTL formulas, and  Lemma~\ref{lem:schlingloff} states that $\rGamma$-\fo formulas are equivalent to 2-CTL ones. To deduce the later from the former, we will show how to translate every ranked tree $t$ over $\rGamma$ into a model $[t]$ over a well chosen set of propositions $P$, then we will apply the following scheme
    $$\xymatrix@C=2.5cm{
        \varphi \in \rGamma\text{-\fo} 
        \ar@{<->}[r]^{\forall t, \ t\models \varphi \ \leftrightarrow \ [t] \models \psi}_{\text{Lemma}~\ref{lem:from-Gamma-to-P-FO}}
        &
        \psi \in P \text{-\fo}
        \ar@{<->}[d]^{\substack{\text{Schlingloff's}\\\text{theorem}}} \\
       \theta \in \text{2-CTL}
        \ar@{<->}[r]^{\forall t, \ t\models \theta \ \leftrightarrow \ [t] \models \delta}_{\text{Lemma}~\ref{lem:from-4-CTL-to-2CTL}}
        &
        \delta \in \text{4-CTL}
    }$$    
The translation $[\_]$ and Lemmas~\ref{lem:from-Gamma-to-P-FO} and \ref{lem:from-4-CTL-to-2CTL} are explained below.

\paragraph{From ranked trees to Schlingloff's models.} Let us fix a ranked alphabet $\rGamma$. Let $P$ be the following set of propositions
\begin{align*}
P\eqdef \rGamma\cup\set{i\text{-th-child}\ |\ i\in[1, \text{max arity of }\rGamma]}
\end{align*}
 Let $t$ be a ranked tree over $\rGamma$. The translation $[t]$ of $t$ is the model defined as follows. It has the same set of nodes and the same descendant relation as $t$. The label of a node contains $a$ if its label in $t$ is $a$. It contains the proposition $i\text{-th-child}$ if it is an $i$-th child in $t$.

\paragraph{First-order logic for ranked and unranked trees.} Let $\rGamma$ and $P$ be as above. Let us show that $\rGamma$-\fo and $P$-\fo are equivalent.
\begin{lemma}\label{lem:from-Gamma-to-P-FO}
For every $\rGamma$-\fo formula $\phi$, there is a $P$-\fo formula $\psi$ such that
$$ \forall t\in \trees\rGamma, \qquad t\models \phi \leftrightarrow [t]\models \psi$$
and conversely.
\end{lemma}
\begin{proof}
To show this lemma its is enough to show how to translate the specific predicates of each formalism into the other. The predicate $a(x)$ of $\rGamma$-\fo can be translated by the same predicate in $P$-\fo and conversely. The predicate $\child i(x)$ can be translated by $i\text{-th-child}(x)$ and conversely.  It is clear that these translations preserve the semantics.
\end{proof}

\paragraph{Temporal logic for ranked and unranked trees.}
Let $\rGamma$ and $P$ be as above. We show that 2-CTL and 4-CTL are equivalent.
\begin{lemma}\label{lem:from-4-CTL-to-2CTL}
For every $\rGamma$-\fo formula $\phi$, there is a $P$-\fo formula $\psi$ such that
$$ \forall t\in \trees\rGamma, \qquad t\models \phi \leftrightarrow [t]\models \psi$$
and conversely. \end{lemma}
\begin{proof}
Here again, it is enough to translate the specific connectives of each formalism into the other. The connective $X_i$ can be encoded in 2-CTL as follows, where $b$ is the maximal arity of $\rGamma$
\begin{align*}
\underset{\begin{array}{c}
{\scriptstyle I\subseteq [1,b]}\\ {\scriptstyle\#I=i}
\end{array}}{\bigvee} \underset{j\in I}{\bigwedge} \odot_j \phi
\end{align*}
Conversely, the connective $\odot_i$ can be encoded in 4-CTL as follows:
\begin{align*}
X_1(i\text{-th-child}\wedge \phi)
\end{align*}
\end{proof}
\subsection{First-order relabelling are derivable}\label{sec:relabeling}
To show Proposition~\ref{prop:forat}, saying that first-order relabelling are derivable, we will show that each function appearing in Lemma~\ref{lem:schlingloff} and corresponding to each operator of 2-CTL is derivable. This is the role of Lemmas\ref{lem:nextmod}--\ref{lem:sincemod} presented below.
\begin{lemma}\label{lem:nextmod}
  For every finite $\rSigma$, $\rGamma\subseteq \rSigma$ and $i \in \set{1,2,\ldots}$, the characteristic function $\ranked{f:\tmonad \rSigma\to\tmonad(\rSigma+\rSigma)}$ of the  unary query 
        \begin{align*}
\text{``The }i\text{-th child of }x\text{ is in }\rGamma\text{''}
        \end{align*}
        is derivable.
\end{lemma}
\begin{proof}
To show that $\ranked{f}$ is derivable, we start applying the children function \begin{align*}
\ranked{\mathsf{Children}:\tmonad\rSigma\to \tmonad (\rSigma\product (\rSigma_0+0)^*)}
\end{align*} from Example~\ref{ex:sibling} which tags every nodes by the list of its children. Consider the function $g$
$$\begin{array}{rlll}
\ranked{g:}  \ranked{\rSigma\product (\rSigma_0+0)^*} &\ranked{\to}& \ranked{\Sigma +\Sigma}&\\
            \tensorpair{a,l}                           &
     \mapsto& \tensorpair{a,1}              &\text{if } l[i]\in\ranked{\rGamma_0}, \\
                    &   \mapsto& \tensorpair{a,2} &\text{otherwise.}   
\end{array}$$
which maps an element of $\rSigma$ tagged by a list to the first copy of $\rSigma$ if the $i$-th element of the list is in $\rGamma$ and to the second copy otherwise. The function $g$ is derivable since its domain is finite.
We finally get $\ranked{f}$ by lifting $\ranked{g}$ to terms.
\end{proof}

\medskip
\begin{lemma}\label{lem:untilmod}
For every finite $\rGamma, \rDelta \subseteq \rSigma$,  the characteristic function $\ranked{f:\tmonad\rSigma\to\tmonad(\rSigma+\rSigma)}$ of the unary query
         \begin{align*}
              \underbrace{\exists y\ y \geq x \land \rDelta(y) \land  \forall z \ (x < z < y \Rightarrow \rGamma(z)).}_{\substack{\text{$x$ has a descendant $y$ with label in $\rDelta$, such that}\\ \text{all nodes between $x$ and $y$ have label in $\rGamma$}}} 
              \end{align*}
is derivable.
\end{lemma}
\begin{proof}
We start by applying the factorization
\begin{align*}
\ranked{\ancfact : \tmonad \Sigma \to \tmonad(\tmonad(\Sigma\setminus(\Gamma\cup\Delta)) +\tmonad(\Gamma\cup\Delta))}
\end{align*}
which decomposes our terms into factors, depending on whether their node labels are in $\ranked{\Gamma\cup\Delta}$ or not. 
Note that the value of a node w.r.t. the until query depends only on the node labels of its factor. 

The nodes of the $\ranked{\tmonad(\rSigma\setminus(\rGamma\cup\rDelta))}$ factors do not satisfy the query, thus we will apply to them the function $\ranked{\tmonad g}$ obtained by lifting the function 
$$\begin{array}{rrll}
 \ranked{g:}& \ranked{\rSigma\setminus(\rGamma\cup\rDelta)}& \ranked{\to} &\ranked{\rSigma+\rSigma}\\
&a&\mapsto& \tensorpair{a,2}.
\end{array}$$

Nodes of the $\ranked{\tmonad(\Gamma\cup\Delta)}$  factors satisfy the query if and only if they have a descendant in  $\rDelta$.
Consider the function $\ranked{h}$ obtained by composing the descendant function $\ranked{\mathsf{Descendant}_\Delta}$ from Example~\ref{ex:descendant} with an injection $\ranked{\tmonad(\iota+\iota)}$
\begin{align*}
\underbrace{\ranked{\tmonad(\Gamma\cup\Delta) \xrightarrow{\ranked{\mathsf{Descendant}_\Delta}} \tmonad(\Gamma\cup\Delta+\Gamma\cup\Delta) \xrightarrow{\ranked{\tmonad(\iota+\iota)}}\tmonad(\Sigma+\Sigma)}}_{\ranked{h}}
\end{align*}
Finally, to get the characteristic function $\ranked{f}$, we apply $\ranked{\tmonad{g}}$ to the $\ranked{\tmonad(\Sigma\setminus(\Gamma\cup\Delta))}$ factors and $\ranked{h}$ to the other factors using the co-pairing combinator, then we flat the obtained term. 
\end{proof}

\begin{lemma}\label{lem:sincemod}
For every finite $\rGamma, \rDelta \subseteq \rSigma$,    the characteristic function of the unary query
         \begin{align*}
              \underbrace{\exists y\ y \leq x \land \rDelta(y) \land  \forall z \ (y < z < x \Rightarrow \rGamma(z)).}_{\substack{\text{$x$ has a descendant $y$ with label in $\rDelta$, such that}\\ \text{all nodes strictly between $x$ and $y$ have label in $\rGamma$}}}  
         \end{align*} 
         is derivable.
\end{lemma}
\begin{proof}
The same proof as above, one only needs to replace the use of the function $\ranked{\mathsf{Descendant}_\Delta}$ by that of $\ranked{\mathsf{Ancestor}_\Delta}$, introduced in Example~\ref{ex:descendant}.
\end{proof}

%% file: appendix-stt.tex
\section{Proof of Theorem~\ref{thm:stt}}
\label{sec:stt-appendix}
In this part of the appendix, we prove Theorem~\ref{thm:stt}, which says that every first-order tree-to-tree transduction is recognised by a register transducer. 

According to Definition~\ref{def:fo-transduction}, a first-order transductions is a composition of any number of functions each of which is either  copying (item 1) or a non-copying first-order transduction (item 2). In other words:
$$\begin{array}{c}
\text{first-order transductions} \\ \eqdef (\text{copying} \cup \text{(non-copying first-order transductions)})^*
\end{array}$$
where the star denotes closure under composition. Although register transducers are closed under composition, this is not very   easy to show directly, and therefore we begin by simplifying the function composition in the definition of first-order transductions.  It is not hard to see that copying commutes with non-copying first-order transductions in the following sense:
\begin{align*}
    \text{copying} \circ  \text{(non-copying first-order transductions)}  \\ \subseteq   \text{(non-copying first-order transductions)} \circ \text{copying}.
    \end{align*}
Furthermore, since the class of copying functions is closed under composition, and the same is true for non-copying first-order transductions, we get the following normal form of first-order transductions:
$$\begin{array}{c}
    \text{first-order transductions} \\=  \text{(non-copying first-order transductions)} \circ \text{copying}.
    \end{array}$$
Therefore, in order to prove Theorem~\ref{thm:stt}, it suffices to show that a register transducer can compute any function which first copies the nodes of the input tree a fixed number of times, and then applies a non-copying first-order transduction. 

For the rest of this section, fix  a tree-to-tree function
\begin{align*}
f : \trees \rSigma \to \trees \rGamma
\end{align*}
which is a composition of first copying (some fixed number of times), followed by a non-copying first-order transduction. We will show that $f$ is computed by some register transducer.  

\newcommand{\origin}[1]{\mathrm{origin}_{#1}}
\newcommand{\orcol}[2]{\mathrm{orcol}_{#1}^{#2}}
In the proof, we use the origin information associated to $f$, i.e.~how nodes of the output tree can be traced back to nodes in the input tree. For an input tree $t \in \trees \rSigma$, define its origin map
to be the function of type
\begin{align*}
    \text{nodes in $f(t)$} \to \text{nodes in $t$}
   \end{align*}
which maps a node $x$ of the output tree to the node of the input tree that was used to define it. (The origin in a copying function is the node that is being copied, while the node in a non-copying transduction is the node of the input structure that represents the node of the output structure.) For a node $x$ in an input tree $t$, define the origin colouring of $x$ to be the function 
\begin{align*}
 y \in \text{nodes in $f(t)$} \ \ \mapsto \  \begin{cases}
    \text{below} & \begin{array}{l}
    \text{if the origin of $y$ is $x$}\\ \text{or a descendant of $x$}
\end{array}     \\
    \text{not below} & \text{otherwise.}
\end{cases}
\end{align*}
Define  the name \emph{origin factorisation of $x$ in $t$}, which is an element of $\trees \tmonad \rGamma$, to be the factorisation of  the output tree where the factors are connected parts of same type (``below '' or ``not below''). The origin factorisation is obtained by applying the ancestor factorisation $\ancfact$ to the output tree extended with its  origin colouring. 

The general idea behind the register transducer is that, after processing the subtree of a node $x$ in the input tree, its registers will store the ``below'' factors in the origin factorisation of $x$. We only store the ``below'' factors, and not the ``not below'' factors, because only the ``below'' factors can be computed using register updates based on the subtree of the node $x$ in the input tree. The key observation is the following lemma, which shows that the a constant number of registers will be enough.

\begin{lemma}\label{lem:composition-method}
    For every input tree $t$ and node $x$ in $t$, the origin factorisation of $x$ in $t$ has at most a constant (i.e.~depending only on the fixed transduction) number of  factors. 
\end{lemma}
\begin{proof}
    For an input tree $t$ and a node $x$ in it, we say that an edge in the output tree $f(t)$ is \emph{$x$-sensitive} if its  the two endpoints are in  different factors of  the origin colouring of $x$ in $t$.  The number of factors in the origin factorisation is one plus the  number of sensitive edges, and therefore to prove the lemma, it is enough to show that:
    \begin{itemize}
        \item[(*)] for every input tree $t$ and node $x$ in $t$,  there is at most a constant  number of $x$-sensitive edges.
    \end{itemize}
    
    Let us write  $\to$ for the image -- along the origin mapping -- of the  child relation in the output tree. In other words,  nodes $y,z$ in the input tree satisfy $y \to z$ if some node in the output tree with origin $z$ is a child of some node in the output tree with origin $y$.   
     It is not hard to  see that $\to$ can be defined in first-order logic, using the formulas from the transduction. Let $r$ be the quantifier rank of the first-order formula used to define $\to$. Using Ehrenfeucht-Fraisse argument, one can show that  if $x,y,z$ are nodes in the input tree such that $y$ and $z$ are on different sides of $x$ (i.e.~any path connecting $y$ and $z$ must necessarily pass through $x$),
    then the truth value of any rank $r$ first-order formula $\varphi(y,z)$   depends only on the following information:
    \begin{itemize}
        \item the $r$-type of  $(y,x)$ in the input tree, i.e.~the rank $r$ first-order formulas satisfied by $(y,x)$; and
         \item the $r$-type of  $(z,x)$ in the input tree, i.e.~the rank $r$ first-order formulas satisfied by $(y,x)$.
        \end{itemize}
    Since the relation $\to$ has constant outdegree and indegree, and it can be defined using quantifier rank $r$,  follows that if $y\to z$ are on different sides of $x$ then there can only be a constant number of nodes $y'$ such that $(y,x)$ and $(y',x)$ have the same $r$-type in the input tree. Since the number of $r$-types is constant, it follows that number of pairs $y \to z$ which are on different sides of $x$ is constant; these pairs are the sensitive edges. 
\end{proof}

Apply the above lemma, yielding an upper bound  $k \in \set{1,2,\ldots}$ on the number of factors in the origin factorisations.
Note that each of the factors in the origin factorisation has arity  $<k$, since the ports of the factors must lead to the other factors. It follows that, in order to store the ``below'' factors in registers, it is enough to have  $k$ groups of registers, with each group having one registers for every arity in $\set{0,\ldots,k-1}$:
\newcommand{\sttregval}[2]{\text{{(register valuation of $#1$ in $#2$)}}}
\newcommand{\rootnode}[1]{\mathrm{root}(#1)}
\begin{align*}
\regnames \quad \eqdef \quad \set{r_i^j \text{ of arity $j$} : i \in \set{1,\ldots,k}, j \in \set{0,\ldots,k-1}} 
\end{align*}

We now define the invariant that will be satisfied by the register transducer. (We use a slightly extended model of register transducers, where some register contents can be undefined; this model is easily seen to reduce to the original one, by filling the undefined registers with  some fixed nonces. 
\begin{itemize}
    \item {\bf Invariant.} Let $t$ be an input tree,  let $x$ be a node in $t$, and let $s_1,\ldots,s_n$ be the ``below'' factors of $x$, viewed as subsets of nodes in the output tree, ordered so that 
    \begin{align*}
       \rootnode{s_1} \preceq  \cdots \preceq \rootnode{s_n} 
    \end{align*}
    where $\preceq$ is the pre-order on nodes in the output tree and $\rootnode{s_i}$ denotes the  unique node in $s_i$ which is an ancestor of all other nodes in $s_i$. After processing the subtree of $x$ in the input tree, the register valuation of the register transducer is 
    \begin{align*}
        r_i^j  \mapsto \begin{cases}
            \begin{array}{l}
            \text{$s_i$ viewed as a term} \\[-2pt] \text{over the output alphabet}
            \end{array} & \text{if $s_i$ has arity $j$}\\[10pt]
            \text{undefined} & \text{otherwise}.
        \end{cases}
    \end{align*}
\end{itemize}

The output register of the transducer is $r_1^0$. When $x$ is the root of the input tree, then there is only one ``below'' factor, namely the entire output tree (which has arity $0$) and therefore -- thanks to the invariant -- the output tree will be found in the output register. 

The following lemma gives the  register updates of the transducer. 



\begin{lemma}\label{lem:register-updates-in-stt}
    There is a finite set $\rDelta$ of register updates with the following property. For every  input tree $t$ and every node $x$ in $t$,   there is some $u \in \rDelta$ such that the register valuation of $x$ (as defined in the invariant) is obtained by applying $u$ to the register valuations of the children $x_1,\ldots,x_n$ of $x$, in listed in left-to-right order. 
Furthermore, there is a family
$\set{\varphi_u(x)}_{u \in \rDelta}$  of unary queries over the input alphabet such that the update associated to a node $x$ is $u$ if and only if the node satisfies $\varphi_u(x)$ in the input tree. 
\end{lemma}
\begin{proof}
    The crucial observation is that each of the ``below'' factors in the origin factorisation for $x$  -- seen as subsets of nodes in the output tree -- is a (disjoint) set union of the   the ``below'' factors in the origin factorisations for the children of $x$, plus the nodes in the output tree which have origin in $x$. Since there is at most a constant number of children and nodes with origin $x$, there is a finite number -- depending only on the transduction -- of ways in which these factors can be combined; this finite set of possible combinations is the set $\rDelta$. The ``furthermore'' part of the lemma, about computing the update using first-order queries, follows from a simple inspection of the first-order formulas used in defining the  transduction. 
\end{proof}

The above lemma completes the definition of the register transducer. Its register updates are $\rDelta$ as in the lemma, and its transition function assigns label $u \in \rDelta$ to each node that satisfies $\varphi_u(x)$. The final part of the proof is showing that the register updates are monotone. We use the  following order on the registers:
\begin{align*}
\underbrace{r_1^0 < r_1^1 < \cdots < r_1^{k-1} < r_2^0 < r_2^1 < \cdots < r_k^{k-2} < r_k^{k-1}}_{\text{lexicographic, with the lower index having priority}}.
\end{align*}

Let $x$ be a node in an input tree $t$, and let $s_1,s_2$ be a ``below'' factor in the origin factorisation of $x$, which are register contents in register valuation of $x$. The registers storing $s_1$ and $s_2$ will be ordered -- according to the invariant -- with respect to the pre-order on the root nodes of $s_1$ and $s_2$. Let $x'$ be the parent of $x$. By the reasoning in the proof of Lemma~\ref{lem:register-updates-in-stt}, there are ``below'' factors in the origin factorisation of $x'$ which contain the factors $s_1$ and $s_2$; call these factors  $s'_1$ and $s'_2$ (possibly $s'_1=s'_2$). Since $s'_1$ contains $s_1$ (as a set of nodes in the output tree), and the same is true for $s'_2$ and $s_2$, we have 
\begin{align*}
\rootnode{s_1} \preceq \rootnode{s_2} \quad \text{implies} \quad \rootnode{s'_1} \preceq \rootnode{s'2}
\end{align*}
which establishes monotonicity of the register updates. 

This completes the proof of Theorem~\ref{thm:stt}.

%% file: appendix-one-register.tex
\section{Normalisation of $\lambda$-terms is a first-order transduction}
\label{sec:eval}

\newcommand{\lambdaterm}{$\lambda$-term }
\newcommand{\lambdaterms}{$\lambda$-terms }

\newcommand{\NonLinTerms}[2]{\Lambda_{#1} #2}
\newcommand{\LinTerms}[2]{\mathsf{Lin}_{#1} #2}

 \newcommand{\rlambda}{\ranked{\Lambda}}
 \newcommand{\rlambdalin}{\ranked{\Lambda^{\sf{lin}}}}
 \newcommand{\rlambdathin}{\ranked{\Lambda^{\sf{thin}}}}

\newcommand{\thinterm}[1]{\ranked{\mathsf{Thin}_{#1}}}

In this part of the appendix, we show Theorem~\ref{thm:normalise}, which says that under some restrictions, normalisation of $\lambda$-terms is a first-order transduction. Before proving this result in~\ref{sec:evaluation-lambda}, we will first explain in~\ref{sec:explaining-restrictions} why these restrictions are unavoidable. Then we show in \ref{sec:restrictions-are-fo} that the set of $\lambda$-terms satisfying these restrictions form a first-order tree language. This result will be useful for the proof of Theorem~\ref{thm:normalise}.

\subsection{Explaining the restrictions}\label{sec:explaining-restrictions}

Recall that Theorem~\ref{thm:normalise} says that normalisation of $\lambda$-terms is derivable under three assumptions:
the input term should be linear, uses a unique variable $x$ and could be typed using a fixed finite set of types. 

If the linearity condition is removed, and because of iterated duplication, the normal form of a well-typed $\lambda$-term can be exponential (or worse, see~\cite[Section 3.6]{sorensen_lectures_2006}), as shown by the following example.
 
\begin{example}\label{ex:exponential}
    Assume that we have two variables $\typevar x  \otype$ and $\typevar y {\otype \to \otype \to \otype}$ and consider the $\lambda$-terms defined by:
    \begin{align*}
        M_0 \eqdef \typevar x \otype \qquad M_{n+1} = (\lambda \typevar x  \otype . \typevar y {\otype \to \otype \to \otype}  \typevar x  \otype \typevar x  \otype)M_n.
    \end{align*}
    The $\lambda$-term $M_n$ is well-typed and of type $\otype$. It has size linear in $n$, but its normal form has size at least $2^n$. 
\end{example}
If there was a first-order transduction normalising these terms, it would be exponential-size increase, which is not possible since all first-order transductions are linear-size increase. 

Being linear alone is not enough to normalise terms with first-order transductions. Another obstacle is terms that use types of unbounded complexity, as illustrated in the following example. 

\begin{example}\label{ex:affine-not-enough}
Consider the following $\lambda$-terms, which have types of 
unbounded size: 
\begin{align*}
M_n = \overbrace{\lambda \typevar x \otype. \lambda \typevar  x \otype. \cdots \lambda \typevar  x \otype.}^{\text{$n$ times}} \typevar  x \otype
\end{align*}
This is a well-typed affine term, whose type is 
\begin{align*}
\otype^n \to \otype \qquad \eqdef \qquad  \overbrace{\otype \to \otype \to \cdots \to \otype}^{\text{$n+1$ arrows}}
\end{align*}
    To $M_n$, apply  $m$ arguments of type $\otype$:
    \begin{align}\label{eq:complicated-term}
    M_n \overbrace{\typevar y \otype \ \typevar y {\otype} \cdots \typevar y{\otype} }^{\text{$m$ times}}.
    \end{align}
    We claim that the above $\lambda$-term cannot be normalised using a first-order transduction, or even a monadic second-order transduction. In order to normalise, a transduction would need to be able to compare the numbers $n$ and $m$ as follows:  if $m < n$  the normal form contains $\lambda$, if $m=n$  the normal form does not contain $\lambda$, and if $m > n$ then the normal form is undefined because the $\lambda$-term is not well-typed.  Whether or not a $\lambda$-term (seen as a tree over a finite alphabet) contains $\lambda$ is a first-order definable property, and first-order definable properties are preserved under inverse images of first-order transductions. Therefore, if normalisation would be a first-order transduction,
then there would be a first-order formula which would be true for terms of the form~\eqref{eq:complicated-term} with $m>n$ and which would be false for terms of the form~\eqref{eq:complicated-term} with $m=n$. Such a formula cannot exist, which can be shown using a pumping argument or Ehrenfeucht-Fra\"iss\'e games. 
\end{example}

\begin{example}\label{ex:tito}
Let $\mathsf{not}$ and $\mathsf{id}$ be the following terms:
\begin{align*}
\mathsf{not} \eqdef \lambda \typevar b {\otype \to \otype \to \otype} . \lambda \typevar x {\otype} . \lambda \typevar y {\otype} . \typevar b {\otype \to \otype \to \otype} \typevar y {\otype} \typevar x {\otype}  \\
 \mathsf{id} \eqdef \lambda \typevar b {\otype \to \otype \to \otype}. \lambda \typevar x {\otype}. \lambda \typevar y {\otype}. \typevar b {\otype \to \otype \to \otype} \typevar x {\otype} \typevar y {\otype}
\end{align*}
For every $n\geq 1$, we let $\mathsf{not}_n$ be the following term:
\begin{align*}
\mathsf{not}_n = \lambda \typevar b {\otype \to \otype \to \otype}. \overbrace{\mathsf{not} \cdots \mathsf{not}}^{\text{$n$ times}} \typevar b {\otype \to \otype \to \otype}
\end{align*}
The normal form of $\mathsf{not}_n$ is $\mathsf{not}$ when $n$ is odd, and  $\mathsf{id}$ when $n$ is even. 

The $\lambda$-terms $\mathsf{not}_n$ cannot be normalised using
a first-order transduction. Otherwise we would have a first-order formula which is true for those terms $\mathsf{not}_n$ where $n$ is even and false when $n$ is odd. This formula cannot exist for the same reason as the example above.  

Note that $\mathsf{not}_n$ is linear and its subterms can be typed using only the types $\otype, \otype\to\otype$ and ${\otype \to \otype \to \otype}$. By restricting ourselves to terms which uses a unique bound variable, we avoid this situation. 
\end{example}
\subsection{Restrictions of Theorem~\ref{thm:normalise} are first-order definable} \label{sec:restrictions-are-fo}

In this section, we show that the restrictions of Theorem~\ref{thm:normalise} discussed above, are first-order definable, as stated in the following theorem.

\begin{proposition}\label{prop:WellTypedFo}
   Let $X$ be a finite set of simply typed variables and let $\Tt$ be a finite set of simple types.      The tree language of linear $\lambda$-terms which can be typed
using $\Tt$ is first-order definable. 
\end{proposition}

In the rest of this appendix, we denote by $\NonLinTerms \Tt X$ this tree language.
To prove Proposition~\ref{prop:WellTypedFo}, we first show that for $\lambda$-terms in $\NonLinTerms \Tt X$, checking if their type is $\tau$, where $\tau$ is a type in $\Tt$, is a \fo property:
\begin{lemma}\label{lem:IsTypeTauFo}
For every type $\tau$ in $\Tt$, there is a first-order query $\varphi_\tau$ such that:
$$ \forall M\in \NonLinTerms \Tt X \qquad\quad M,u \models \varphi_\tau \longleftrightarrow M|_u:\tau$$
where $M|_u$ is the sub-tree of $M$ rooted in $u$. 
\end{lemma}
Before establishing this lemma, let us see how Proposition~\ref{prop:WellTypedFo} can be derived from it. Linearity can be easily seen as a first-order property. The hard part is to show that the set of $\lambda$-terms which can be typed using $\Tt$ is first-order. Suppose for convenience that $\Tt$ is downward closed. For every type $\tau$ in $\Tt$, let $\varphi_\tau$ be the formula given by Lemma~\ref{lem:IsTypeTauFo}. In the following, we use the binary formula
 $\mathsf{Succ}_{i}(u,v)$ which is valid when $v$ is the $i$-th child of $u$, and which is easily expressible in first-order logic.
 \smallskip
 
Consider the unary formula $\mathsf{Lambda}(u)$, which expresses that $u$ is a binder node, that its type and the type of its child match well and both belong to $\Tt$: 
$$\begin{array}{ll}
\mathsf{Lambda}(u) &:= \underbrace{\lambda x(u)}_{\substack{\text{$u$ has label $\lambda x$}}} \wedge\bigvee_{\substack{\sigma\rightarrow\tau \in \Tt\\x:\sigma}}  \underbrace{\varphi_{\sigma\rightarrow\tau}(u)}_{\substack{\text{$u$ has type $\sigma\rightarrow\tau$}}}  \\ & \wedge \underbrace{\exists v\ \ \mathrm{Succ}_1(u, v) \wedge \varphi_\tau (v)}_{\substack{\text{the child of $u$ has type $\tau$}}}
\end{array}$$
Similarly, consider the unary formula $\mathsf{Application}(u)$ which checks that a node is an application node, that the type of its children match well and that both belong to $\Tt$: 
$$\begin{array}{l}
\mathsf{Application}(u) := \underbrace{@(u)}_{\substack{\text{$u$ has label @}}} \wedge \\ \exists v, w \underset{\sigma\rightarrow\tau \in \Tt}{\bigvee}  \underbrace{\mathrm{Succ}_1(u, v) \wedge \varphi_{\sigma\rightarrow\tau} (v)}_{\substack{\text{the left child of $u$ has type $\sigma\rightarrow\tau$}}} \wedge \underbrace{\mathrm{Succ}_2(u, w) \wedge \varphi_{\sigma} (w)}_{\substack{\text{the right child of $u$ has type $\sigma$}}}
\end{array}$$
Finally, consider the formula $\mathsf{Variable}(u)$, which expresses that $u$ is a variable node, whose type is in $\Tt$:
\begin{align*}
\mathsf{Variable}(u) :=  \bigvee_{x:\sigma\in \Tt}\underbrace{x(u)}_{\substack{\text{$u$ has label $x$}}} 
\end{align*}  
We claim that the following (nullary) formula $\phi$ recognizes the tree language $\NonLinTerms \Tt X$
\begin{align*}
\phi = \forall u.\ \mathsf{Variable}(u)\ \vee\ \mathsf{Lambda}(u)\ \vee\  \mathsf{Application}(u)
\end{align*}
If a $\lambda$-term is in $\NonLinTerms \Tt X$, then it clearly satisfies $\phi$. Suppose by contradiction that there is a $\lambda$-term $M$ which is not in $\NonLinTerms \Tt X$ and yet satisfies $\phi$. Let $u$  be the deepest node of $M$ which is not in $\NonLinTerms \Tt X$ (we identify in this proof a node $u$ and the sub-term $M|_u$). In particular, the descendants of $u$ are all in $\NonLinTerms \Tt X$. The node $u$ cannot be a variable, since variable nodes are well-typed and their type is in $\Tt$ by the first disjunct of $\phi$.  If $u$ was labeled by $\lambda x$, where $x$ is of type $\sigma$, then by the second disjunct of $\phi$ there is a type $\tau$ such that $\sigma\rightarrow\tau\in \Tt$ and the child $v$ of $u$  satisfies $\varphi_\tau$. Since $v$ is in $\NonLinTerms \Tt X$, its type is $\tau$ by Lemma~\ref{lem:IsTypeTauFo}. Hence $u$ is well-typed and its type is $\sigma\rightarrow\tau\in \Tt$. As a consequence $u$ is in $\NonLinTerms \Tt X$ which is a contradiction.  Finally, if $u$ was labeled by $@$, then by the third disjunct of $\phi$, its two children $u_1$ and $u_2$
would satisfy respectively $\varphi_{\sigma\rightarrow\tau}$ and  $\varphi_{\sigma}$ and by Lemma~\ref{lem:IsTypeTauFo} they are of type $\sigma\rightarrow\tau$ and $\sigma$ respectively. The node $u$ is then well-typed and its type is $\tau$ (which is a type of $\Tt$ thanks to downward closeness). As a consequence, $u$ is in $\NonLinTerms \Tt X$, which gives a contradiction and concludes the proof.
 \smallskip
 
We can go back now to the proof of Lemma~\ref{lem:IsTypeTauFo}.
\begin{proof}[Proof of Lemma~\ref{lem:IsTypeTauFo}]
Let us show that the following unary query is expressible in first-order logic
\begin{center}
''if $t$ is a $\lambda$-term of $\NonLinTerms s X$, then its type is $\tau$ ``:
\end{center}
 For that, notice that the type of a well-typed term depends only on its left-most branch. In fact, the type of a term is exactly the type of its left-most branch in the following sens.

Consider the (unranked) alphabet $ X^\lambda:= X\cup \{@, \lambda x | x\in X\}$. We can equip the words over $X^\lambda$ with the following typing rules:
$$\frac{}{x: \sigma} \qquad \frac{u:\tau}{u\lambda x: \sigma\rightarrow \tau} \qquad \frac{u:\sigma\rightarrow\tau}{u@:\tau}$$
where $x$ is of type $\sigma$ and $\sigma,\tau \in \Tt$.

We say that $w$ is of type $\tau$ and write $w:\tau$ if there is a typing derivation for $w:\tau$.

We can associate to every branch of a $\lambda$-term a word over $X^\lambda$ corresponding to the sequence of its labels read bottom-up. By induction on $\lambda$-terms, we can easily show that the type of a $\lambda$-term is the type of the word corresponding to its leftmost branch. 

By this last observation, we can reduce the query asking if the type of a term is $\tau$, to the same query but on $X^\lambda$ words. To show that the former is a first-order query, it is then sufficient to show that the following word language 
\begin{align*}
W_\tau = \{w\in X.\{@, \lambda x | x\in X\}^*\ |\ w:\tau \} 
\end{align*}
is first-order definable, or equivalently that  $W_\tau$ is recognized by a counter-free  finite automaton. For that we proceed as follows: first, we show that $W_\tau$ is recognized by a pushdown automaton $P_\tau$. Then we will show that the stack height of $P_\tau$ is bounded, thus it can be turned into a deterministic finite automaton $D_\tau$. Finally, we show that the obtained automaton $D_\tau$ is actually counter-free.  


Consider the pushdown automaton $P_\tau$ whose
\begin{itemize}
\item  set of states is $\{i, p, f\}$, where $i$ is the initial state and $f$ the accepting state;
\item input alphabet is the alphabet $X^\lambda$;
\item stack alphabet is the set of types $\Tt$;
\item and whose transition function is described as follows:
\begin{itemize}
\item If the automaton is in the initial state $i$ with an empty stack, and if the symbol it reads is a variable $x$ of type $\sigma_1\rightarrow\dots\rightarrow\sigma_n$, then we go to the state $p$ and push the symbols $\sigma_n,\dots,\sigma_1$  in the stack in this order. The top-level symbol of the stack is then $\sigma_1$.
\item If the automaton is in the state $p$ and it reads the symbol $\lambda y$, where $y$ is of type $\sigma$, then push the symbol $\sigma$ in the stack, and stay in the state $p$.
\item  If the automaton is in the state $p$,  if it reads the symbol $@$ and if the stack is non empty, then pop the top-level symbol and stay in the state $p$.
\item If the automaton reaches the end of the word being in state $p$, and if the stack contains the symbols $\tau_1,\dots\tau_m$ in this order, $\tau_1$ being the top-level symbol, where $\tau_1\rightarrow\dots\rightarrow\tau_m$ is the type $\tau$, then pop them all and go to the final state $f$.  
\end{itemize}
\end{itemize}
A word $w$ is accepted by $P_\tau$ if there is a run that reaches the end of $w$ in the accepting state $f$ with an empty stack. We write $(r, s)\xrightarrow{w} (r', s')$ if there is a run over the word $w$ which starts in the state $r\in\set{i, p,f}$ and with a stack $s$ and ends up in the state $r'\in\set{i, p,f}$ and with a stack $s'$.

By induction on the length of the word $w$, we can easily show that:
\begin{lemma}
For every word $w\in X.\{@,\lambda x | x\in X \}^*$, we have that:
 $$(i, \epsilon)\xrightarrow{w} (p, \sigma_n\dots\sigma_1) \qquad\text{iff} \qquad 
w:\sigma_1\rightarrow\dots\rightarrow\sigma_n$$ 
\end{lemma}

A direct consequence of this lemma is that $P_\tau$ recognizes $W_\tau$. Another direct consequence is that the stack height of $P_\tau$ is bounded by $m$, the size of the longest type in $\Tt$. Thus $P_\tau$ can be turned into a DFA $D_\tau$, by encoding the stack information in the states. More precisely, the states of $D_\tau$ are pairs $(r,s)$ where $r\in \set{i,p,f}$ and $s$ is a stack of height at most $m$, the initial state is $(i,\epsilon)$ and there is a transition $(r,s)\xrightarrow{a}(r',s')$ where $a\in X^\lambda\cup\{\epsilon\}$ if there is a corresponding run in $P_\tau$. We show in the following that $D_\tau$ is counter-free. 

Let us start with some observations. In the pushdown automaton $P_\tau$, the effect of a word $w$ on a stack $s$, starting from the state $p$ is the following: it erases the first $n$ top level elements of $s$, and replaces them by a word $u$. The number $n$ and the word $u$ do not depend on the stack $s$ but only on the word $w$. This is exactly what the following lemma claims.

\begin{lemma}
For every word $w$ over ${X^\lambda}^*$, there is a natural number $n$ and a word $u\in \Tt^*$ such that if $(p,s)\xrightarrow{w}(p,s')$ then $s$ and $s'$ can be decomposed as follows:
$$s=t.v,\qquad s'=t.u\qquad \text{ and }\qquad |v|=n.$$
\end{lemma}
The proof is an easy induction on the length of $w$. As a consequence we have that:
\begin{itemize}
\item If $(p,s_1)\xrightarrow{w}(p,s_2)\xrightarrow{w}(p,s_3)$ and $|s_2|>|s_1|$ then $|s_3|>|s_2|$.
\item If $(p,s_1)\xrightarrow{w}(p,s_2)\xrightarrow{w}(p,s_3)$ and $|s_2|<|s_1|$ then $|s_3|<|s_2|$.
\item If $(p,s_1)\xrightarrow{w}(p,s_2)\xrightarrow{w}(p,s_3)$ and $|s_2|=|s_1|$ then $s_3 =s_2$.
\end{itemize}

Let us show that $D_\tau$ is counter-free. Suppose by contradiction that there is a word $w$ and pairwise distinct stacks $s_1,\dots, s_n$ such that 
\begin{align*}
(p,s_1)\xrightarrow{w}(p,s_2)\xrightarrow{w}\dots(p,s_n)\xrightarrow{w}(p,s_1).
\end{align*} 
By the first two properties above, we have necessarily that 
\begin{align*}
|s_1|=\dots=|s_n|
\end{align*}
 Thus by the third property, we have that 
\begin{align*}
 s_1=\dots=s_n
\end{align*} 
 which concludes the proof.
\end{proof}

\subsection{Normalisation of $\lambda$-terms}\label{sec:evaluation-lambda}


This section is dedicated to the proof of Theorem~\ref{thm:normalise}. Let us first introduce some terminology. In a $\lambda$-term, we call \emph{redex} a pattern of the following form 
\begin{center}
\includegraphics[scale=.5]{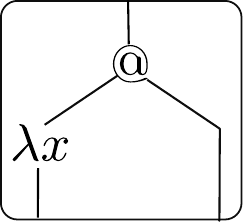}
\end{center}
that is, an application node whose left child is an abstraction node. In a linear $\lambda$-term, we call \emph{full redex} a set of nodes  containing a redex, the node of its variable, together with the set of nodes between them, as illustrated below
\begin{center}
\includegraphics[scale=1.5]{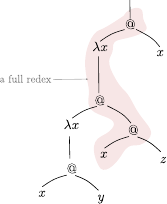}
\end{center}
\medskip

Let us go back to the proof of Theorem~\ref{thm:normalise}. Before normalising $\lambda$-terms, the first thing to do is to discriminate those $\lambda$-terms satisfying the restrictions of Theorem~\ref{thm:normalise} from the others. This amounts to pre-processing the normalisation process by the function
\begin{align*}
\trees \ranked{X^\lambda} \to \trees \ranked{X^\lambda} +\bot
\end{align*}
which is the identity for inputs satisfying the restrictions and is undefined otherwise. Let us see how this function can be derived. Thanks to Proposition~\ref{prop:WellTypedFo}, the restrictions of Theorem~\ref{thm:normalise} are first-order definable, say by a first-order query $\phi$. By virtue of Proposition~\ref{prop:forat}, the characteristic function of $\phi$ is derivable. Now following the label's root of the input, we either output the input tree if the label says that it satisfies the query $\phi$, or outputs the undefined symbol otherwise. This last function can be easily derived. From now on, we suppose that our $\lambda$-terms satify the restrictions of Theorem~\ref{thm:normalise}.  

To normalise $\lambda$-terms satisfying the conditions of Thm.~\ref{thm:normalise}, the main observation is that the evaluation of a redex does not create new redexes. Hence, it suffices to reduce all the available redexes in such term to reach the normal form. 



To show Thm.~\ref{thm:normalise}, we will factorise (via a derivable function) our \lambdaterms into factors satisfying the following properties:
\begin{itemize}
\item[(P1)] Every full redex falls into one of the factors. 
\item[(P2)] Each factor have a a very specific shape called \emph{thin}. These factors are those $\lambda$-terms with ports whose normal form have the shape of a word (by opposition to trees, which is the general case).
\end{itemize}
By properties (P1) and (P2), it is enough to show that normalisation of thin \lambdaterms (with ports) is derivable. For this purpose, our strategy will be to prove that the word obtained by normalising a thin \lambdaterm results from a pre-order traversal. Since pre-order traversal is a prime function,  this implies that normalisation of thin $\lambda$-terms is derivable.
 
 The last ingredient to conclude the proof is to notice that $\beta$-reducing the factors of (a factorisation of) a $\lambda$-term, then applying a flattening, is the same thing as $\beta$-reducing the original $\lambda$-term, which follows directly from the fact that $\beta$-reduction is a congruence on terms. This concludes the proof. 
 
 In the rest of this section, we develop on each of the two main steps of the proof, namely proving Properties (P1) and (P2).  In Section~\ref{subsub:thin} we present thin \lambdaterms with ports and show how to normalise them. Then we show in Section~\ref{subsub:facto} how to factorise a \lambdaterm into thin factors.

\subsubsection{Normalisation of thin $\lambda$-terms}\label{subsub:thin}

As discussed earlier, we will need to normalise \lambdaterms with ports (the factors of our factorisation). In the following, we will denote by $\ranked{\ranked{\lamrank X}}$ the ranked set 
\begin{align*}
     \overbrace{\set{y : y \in X}}^{\text{arity 0}} \cup \overbrace{\set{\lambda x }}^{\text{arity 1}} \cup  \overbrace{\set @}^{\text{arity 2}}
\end{align*}
With this notation, \lambdaterms with ports are the inhabitants of $\ranked{\tmonad \ranked{\lamrank X}}$. Normalisation of these terms generalizes that of usual \lambdaterms in a straightforward way: the $i$-th port is replaced by a fresh variable $x_i$, the obtained \lambdaterm (without ports) is evaluated as usual, then the variable $x_i$ is replaced back by the port $i$, as one can see in the following example.  
  \begin{center}
\includegraphics[scale=.4]{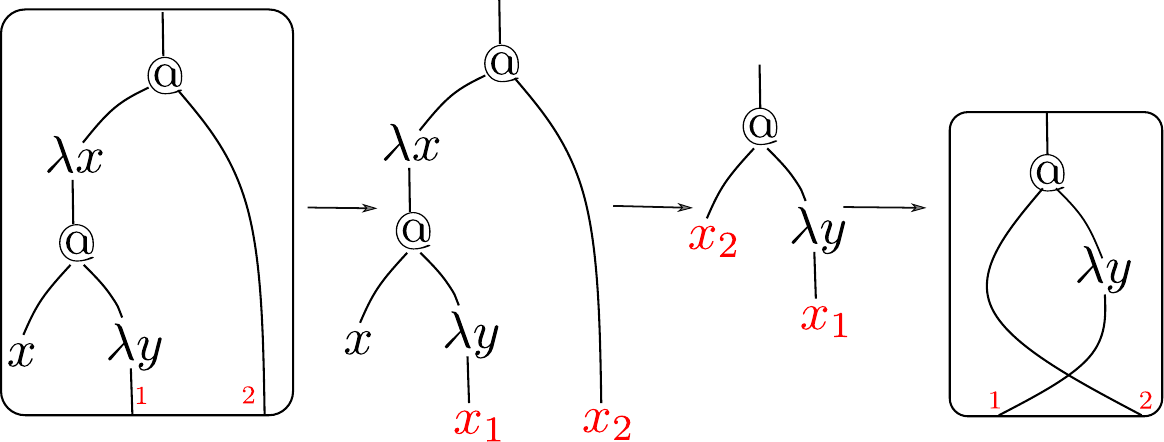}
\end{center}
  Note that when a \lambdaterm is linear, its normal form has the same number of ports. Note also that respecting the original order of ports in the normal form (which is important for compositionality) may twist ports, as in the example above. As a consequence, normalisation of linear \lambdaterms with ports is an arity preserving function of  type:
  \begin{align*}
  \ranked{\tmonad \ranked{\lamrank X} \to \reduce 1 \tmonad \ranked{\lamrank X} +\bot} 
  \end{align*}
   
Let us present now the class of \emph{thin \lambdaterms with ports}.
 
\begin{definition}
We say that the node of a $\lambda$-term is \emph{branching} if its has at two distinct children which are not ports.
 
A \emph{thin $\lambda$-term with ports} is a term from $\tmonad \ranked{\lamrank X}$ in which every branching node is the application node of a redex. 
\end{definition}
In the remaining of this section we will omit the mention ``with ports'' if clear from the context.

Since thin $\lambda$-terms branch only on redexes, the result of their normalisation is a ``word'', in the sens that every node has at most one non-port child.  We will show that this word can actually be obtained by a pre-order traversal of the original $\lambda$-term. We will then use the prime $\preorder$ function to show that normalisation of thin $\lambda$-terms is derivable. 

The left $\lambda$-term below is linear and thin. The rednodes are the ones which are not redexes nor the variables of these redexes. The right $\lambda$-term is its normal form: we can see that  nodes appear top-down in the pre-order  of the original $\lambda$-term.  
\begin{center}
		\includegraphics[scale=.45]{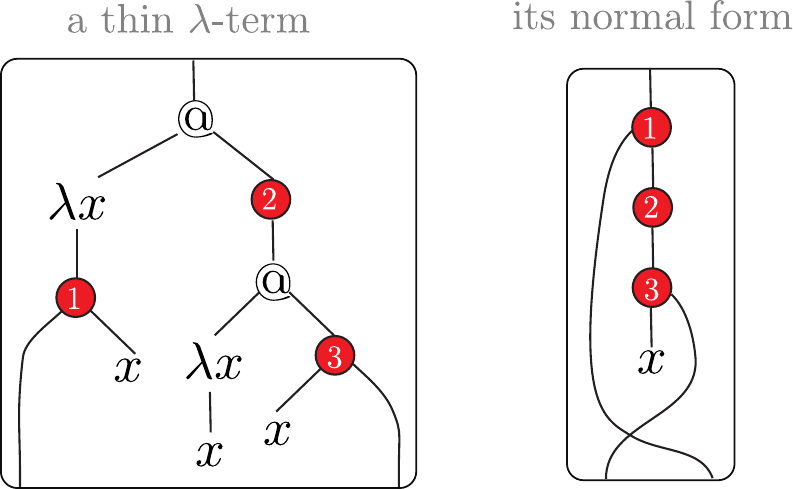}
		\end{center}

\begin{proposition}\label{prop:EvaluateThin}
    Let $X$ be a finite set of simply typed variables, $\typeset$ be a finite set of simple types.
    The following tree-to-tree function is derivable:
    \begin{itemize}
        \item{\bf Input.} A $\lambda$-term $t$ over variables $X$.
        \item {\bf Output.} The normal form of $t$, if it is  thin and satisfies the conditions of Thm.~\ref{thm:normalise}, and undefined otherwise.
    \end{itemize}
\end{proposition}

Let $t$ be a thin $\lambda$-term and let $u$ be its normal form. As noticed before, $u$ has the shape of a word. Moreover, since $t$ is linear, the nodes of $u$ are exactly the nodes of $t$ which are not redexes, nor their variables.

\begin{proposition}\label{prop:normal-form-depth-first} Let $t$ be a linear thin \lambdaterm and let $u$ be its normal form. 
The order in which the inner nodes (ie. non ports) of $u$ appear top-down is the pre-order of $t$.  
\end{proposition}

\begin{proof}
To establish this proposition, we need the following lemma.
\begin{lemma}\label{lem:internalLemma}
Let $t$ be a linear thin $\lambda$-term which binds only the variable $x$ and let $r$ be one of its redexes. Consider $m$ to be the binder node of $r$ and $n$ to be the node of the variable it binds.

The node $n$ is the greatest (that is the right-most) node in the sub-term $t|_m$ w.r.t. the pre-order. 
\end{lemma}

\begin{proof} Suppose by contradiction that  there is a node $o$ which is strictly greater than $n$ in the subterm $t|_m$.  Since $t$ is thin, the least common ancestor $l$ between $n$ and $o$ is an application node of a redex. By hypothesis on the term $t$, the binder of this redex is $\lambda x$.  Since $n$ is smaller than $o$, $n$ is the left descendant  of $l$, in other words it is the descendant of the left child $p$
of $l$, which is a binder $\lambda x$. The node $m, n, o$ and $p$ are illustrated below:
\begin{center}
		\includegraphics[scale=.3]{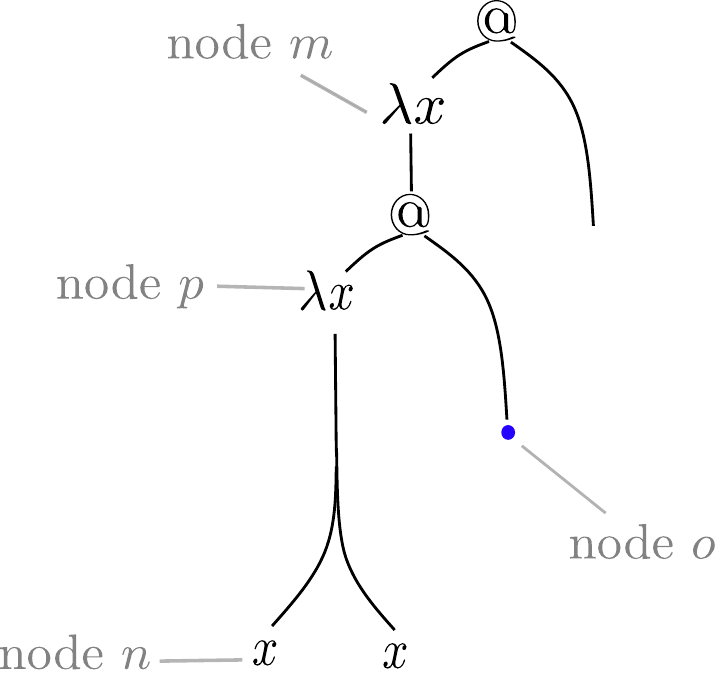}
\end{center} 
 The variable of the node $n$ is under the scope of the binder of the node $p$, which contradicts the fact that it is bound by the binder of the node $m$, which concludes the proof. 
\end{proof}

Let us go back to the proof of our proposition. Consider two inner nodes $n, m$ of $t$ which are also nodes of $u$, and such that $m$ is smaller than $n$ in the pre-order of $t$ (we well call it simply pre-order in the rest of the proof). We show that $n$ is a descendant of $m$ in $u$. There is two cases to consider:
\begin{itemize}
\item Either $n$ is a descendant of $m$ in $t$, in this case we can conclude easily since $\beta$-reduction preserves the descendant relation. Indeed, by a small analysis of $\beta$-reduction, one can notice that a reduction step may extend the descendant relation, but can never change (or break) the order of two comparable nodes in the original $\lambda$-term.  
\item  Otherwise, let us consider the lowest common ancestor $p$ of $m$ and $n$. We proceed by induction on the length of the path between $m$ and $p$. By definition of thin $\lambda$-terms, since $p$ is branching it is necessarily an application node, whose left child $q$ is a binder node  $\lambda x$. 
By Lemma~\ref{lem:internalLemma}, $m$ is smaller w.r.t. the pre-order than the node $r$ of the variable bound by $q$ . 
We are then left with the following two situations. The first case, illustrated by the left figure below, is when $r$ is a descendant of $m$ in $t$. In this case, after one reduction step, $n$ will be a descendant of $m$. The other case is when $m$ is in the left of $r$ in $t$, as illustrated by the right figure below. In this case, after one reduction step, the lowest common ancestor between $m$ and $n$ will be a descendant of $p$, and we can conclude by induction hypothesis. 
\begin{center}
\includegraphics[scale=.3]{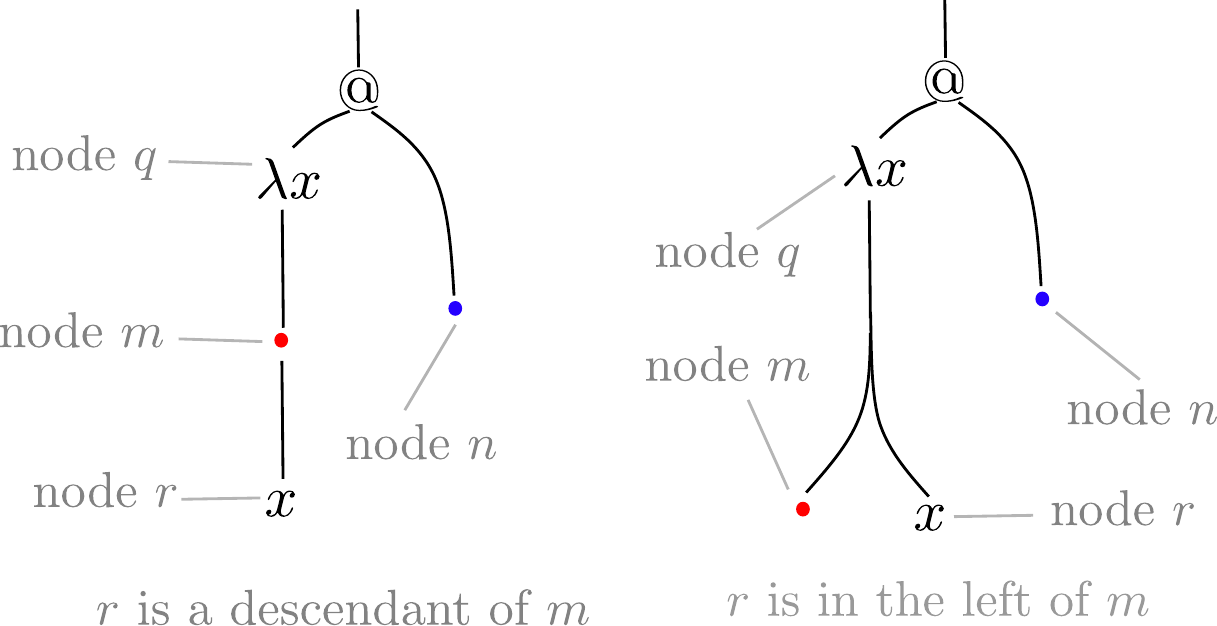}
\end{center}
\end{itemize}
This concludes the proof of the first claim.
\end{proof}

Let us construct now a derivable function which computes the normal form of linear thin $\lambda$-terms binding a single variable $x$. We illustrate this construction on the term $t$ below which will be our running example in this proof. 
\begin{center}
\includegraphics[scale=.4]{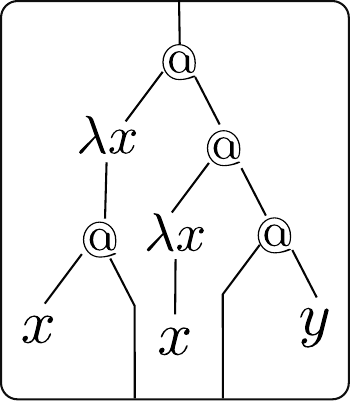}
\end{center}

\begin{proof}[Proof of Proposition~\ref{prop:EvaluateThin}] Let $t$ be a linear thin $\lambda$-term in $\tmonad \ranked{\lamrank X}$.


\begin{enumerate}
\item
We start by distinguishing the redexes of $t$ and their variables from the other nodes. For that, we apply the characteristic function of the following first-order query $\varphi$:
    \begin{center}
    ``The node $u$ is a redex or a variable of a redex''
    \end{center}
    This query is first-order expressible. Indeed it is the disjunction of the following queries
$$\begin{array}{rl}
@\mathsf{Redex}(u) = & @(u) \wedge \exists v \ \mathrm{Child}_1(u,v) \wedge \lambda x(v)\\[8pt]
\lambda\mathsf{Redex}(u)=& \lambda x(u) \wedge \exists v \ \mathrm{Child}_1(v,u) \wedge @(v) \\[8pt]
X\mathsf{Redex}(u) = & x(u) \wedge \exists v\ \lambda\mathsf{Redex}(v) \wedge v\ \mathsf{binds}\ u
\end{array}$$
where $@\mathsf{Redex}(u)$ says that $u$ is the application node of a redex, $\lambda\mathsf{Redex}(u)$ says that it is the abstraction node of a redex and $X\mathsf{Redex}(u)$ says that it is the variable of a redex. 
The formula $u\ \mathsf{binds}\ v$, defined below,  is a binary first-order query expressing that the node $u$ is an abstraction node that binds $v$.
 \begin{align*}
 &\lambda x(u) \wedge x(v) \wedge(u<v)\ \\
 \wedge \ & \forall u,v,w.\ u<w<v\Rightarrow \neg \lambda x(w)
 \end{align*}

The formula $\varphi$ being a first-order query, its characteristic function is derivable thanks to Proposition~\ref{prop:forat}. 

When we apply this function to $t$, we get a term in $\ranked{\tmonad(\ranked{\lamrank X}+\ranked{\lamrank X})}$. 
Below is the effect of this first step on our running example. We colored in red the nodes belonging to the first copy of $\ranked{\lamrank X}$, that is the nodes satisfying the query $\varphi$. These nodes are the ones that will disappear in the normal form of $t$. 
\begin{center}
\includegraphics[scale=.4]{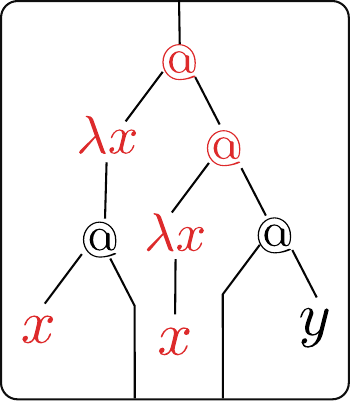}
\end{center}
\item After that, we apply the $\preorder$ function 
\begin{align*}
            \ranked{\preorder : \tmonad (\ranked{\lamrank X}+\ranked{\lamrank X}) \to \reduce 1\tmonad( \ranked{\lamrank X}+\ranked{\lamrank X} + 0 +2)}
\end{align*}
After this step, our initial term becomes
\begin{center}
\includegraphics[scale=.3]{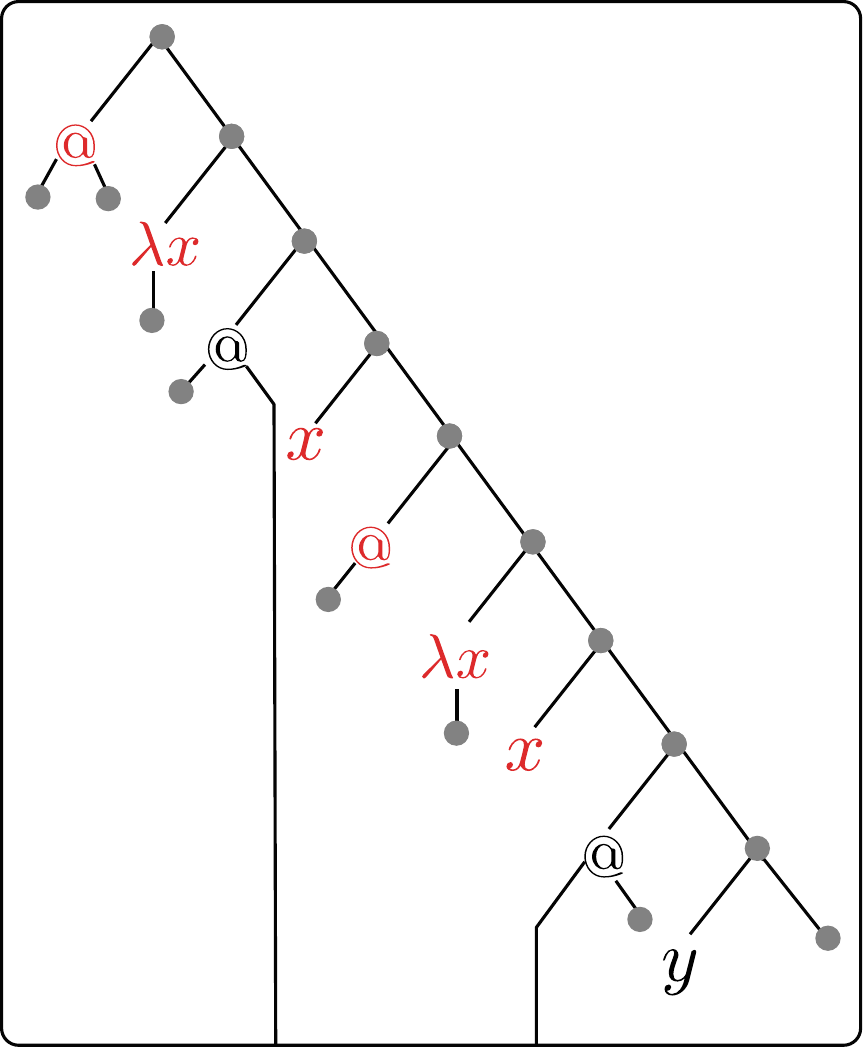}
\end{center}  
In this term, the nodes of the normal form appear in the right order thanks to Prop.~\ref{prop:normal-form-depth-first}. Now, we only need to get rid of the redexes and the variable nodes that participated in the computation of the normal form (that is the ones colored in red) together with the nodes $\grayball$ and $\grayballbin$ introduced by the $\preorder$ function. 
\item For this purpose, we apply the function 
\begin{align*}
\ranked{\ranked{\tmonad (X^\lambda+X^\lambda + 0+
2)} \to \ranked{\tmonad (X^\lambda+X^\lambda + 0+2+1) }}
\end{align*}
 which adds the unary symbol $1$ as the parent of every node $2$. This function can be easily  derivable. Then we apply the factorisation $\ancfact$
 to separate the symbol $1$ from the others:
 \begin{align*}
 \ranked{\ancfact : \tmonad (X^\lambda+X^\lambda + 0+2+1) \to \tmonad (\tmonad(X^\lambda+X^\lambda + 0+2)+\tmonad 1))}
 \end{align*}
 
 After this step, our example term becomes like this 
\begin{center}
\includegraphics[scale=.3]{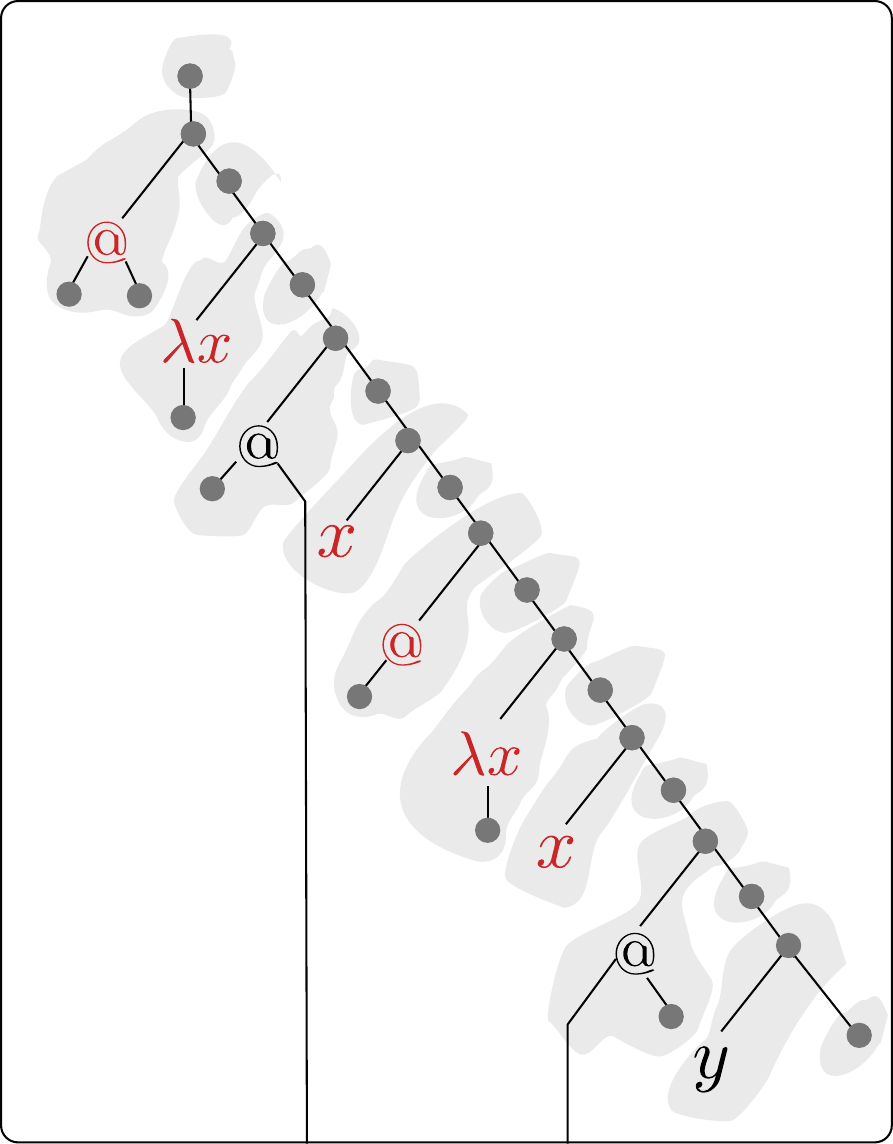}
\end{center}

\item Now consider the function 
 \begin{align*}
 \ranked{g: \tmonad(X^\lambda+X^\lambda+0+2) \to \reduce 1\tmonad(X^\lambda+X^\lambda+0+2)}
 \end{align*}
 which is the identity function, except for the following finite set of terms for which it is defined in figure~\ref{fig:definition-g}.
\begin{figure*}
\begin{center}
\includegraphics[scale=.3]{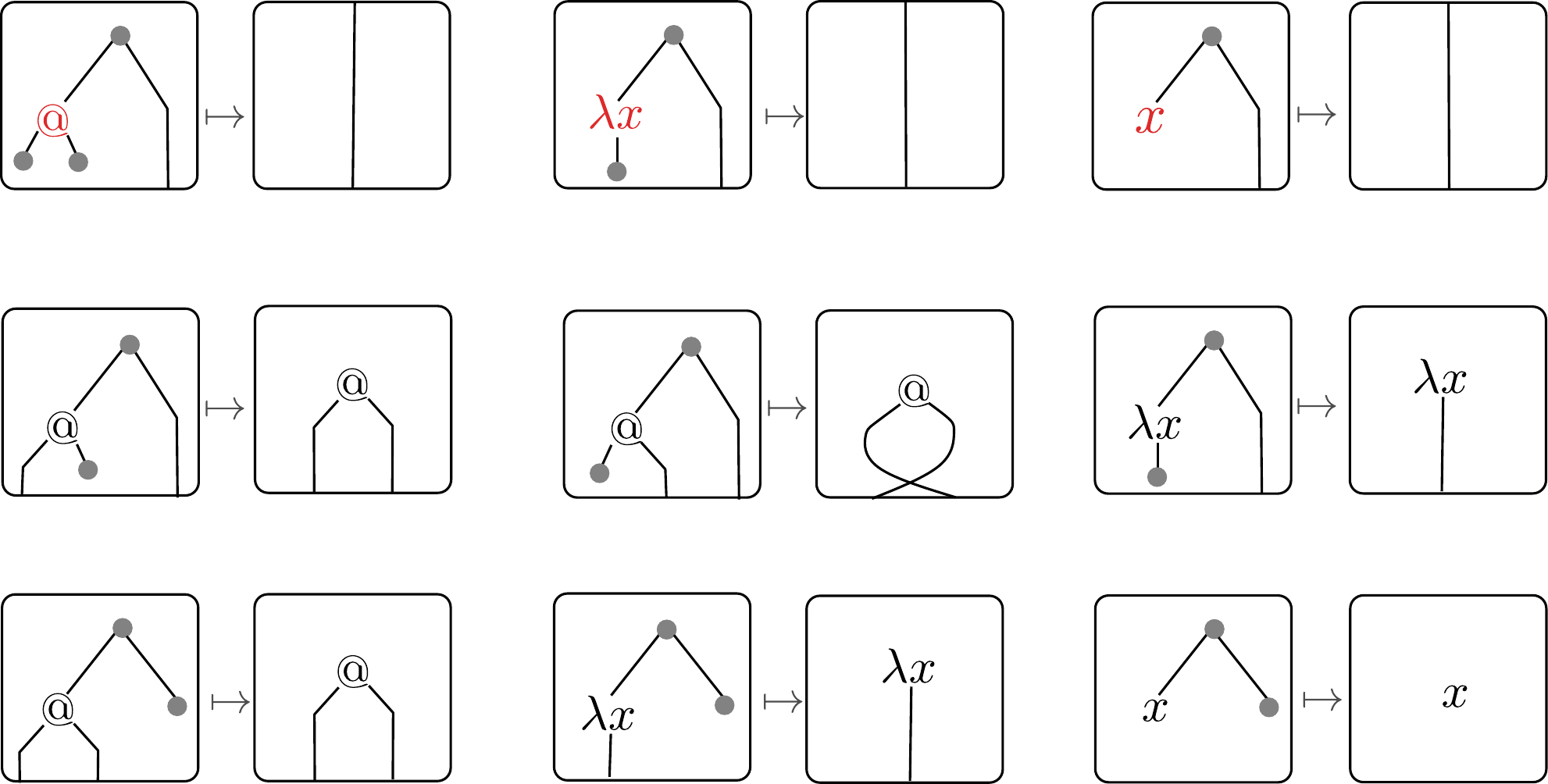}
\end{center}
\caption{Definition of the function $g$.} \label{fig:definition-g}
\end{figure*}
The red elements are those belonging to the first copy of $\ranked{\lamrank X}$.
\smallskip

Now back to our term, we replace the  $\ranked{\tmonad 1}$ factors by the empty term, and to the other factors we apply the function $g$. After that, we apply the function 
\begin{align*}
\ranked{\reduce 1\reduce 1 \Sigma \to \reduce 1\Sigma}
\end{align*}
which untwists two consecutive applications of $\reduce 1$. Doing so, we get a term of type  $$\ranked{\reduce 1\tmonad(X^\lambda+X^\lambda+0+2)}$$ which is  the normal form of $t$. Our running example becomes then
\begin{center}
\includegraphics[scale=.4]{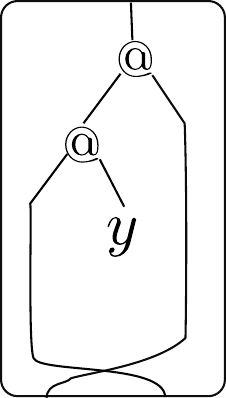}
\end{center}

\item Note that we obtained the desired term, but not with the desired type. To obtain a term in 
$\ranked{\reduce 1\tmonad{X^\lambda}}$, we get rid of the labels $0+2$ by transforming them respectively into variables and application nodes. The choice of which variables to choose is not important, since the only terms that will actually have $0+2$ in their results are \lambdaterms which are not thin or do not satisfy the conditions of Thm.~\ref{thm:normalise}.
\end{enumerate}
\end{proof}

\subsubsection{Factorising $\lambda$-terms into blocks of thin $\lambda$-terms}\label{subsub:facto}

\begin{proposition}\label{prop:FactoIntoThin} For every finite set of typed variables $X$, for every finite set of types $\Tt$ and for every $x\in X$, there is a factorisation $$\ranked{f:\tmonad \ranked{X^\lambda} \to \tmonad\tmonad \ranked{X^\lambda}+\bot}$$ 
which satisfies, for every \lambdaterm $t$ satisfying the conditions of Thm.~\ref{thm:normalise}, that
\begin{itemize}
\item[(1)] every full redex of $t$ is entirely contained in one of the factors of $f(t)$;
\item[(2)] the factors of $f(t)$ are thin.
\end{itemize}
and is undefined otherwise.
\end{proposition}

\begin{proof}
We define the function $\ranked{f}$ as the composition of the following three functions
\begin{align*}
\ranked{\tmonad \ranked{X^\lambda} \xrightarrow{\ g\ } \tmonad (\ranked{X^\lambda}+1) \xrightarrow{\ \mathsf{block}^\uparrow\ } \tmonad (\tmonad\ranked{X^\lambda}+\tmonad 1)
\xrightarrow{\ \mathsf{erase}\ } \tmonad \tmonad\ranked{X^\lambda}}
\end{align*}
The function $\ranked{g}$ will indicate, using the unary symbol $1$, the places wheres two distinct blocks of $\ranked{f}$ will be separated. We will describe it more precisely a bit later. The function $\mathsf{block}^\uparrow$ will create these blocks and finally, we erase all the factors $\ranked{\tmonad 1}$.

The function $\mathsf{block}^\uparrow$ is a prime function and  $\mathsf{erase}$ can be easily derivable. 
Let us show how to derive the function $\ranked{g}$, so that the $1$-nodes it introduces creates blocks satisfying the conditions (1) and (2) of Proposition~\ref{prop:FactoIntoThin} (when the input is a linear $\lambda$-term).  

We define $\ranked{g}$ as the composition of the characteristic function of three first-order unary queries: $\mathsf{@redex}, \mathsf{Right}$ and $\mathsf{Left}$, followed by a homomorphims $\ranked{h}$. We define them in the following:
\begin{itemize}
\item The property $\mathsf{App}$ checks whether a node is the application node of a redex. It can be easily expressed by a first-order formula.
\item The query $\mathsf{Right}$ (resp. $\mathsf{Left}$) checks if the node is an application node, which lies, together with his right (resp. left) child, between the binder of a redex  and the node it binds. Those properties can also be easily expressed by a first-order formula.
\end{itemize}

When we apply the characteristic functions of these queries to a term in $\ranked{\tmonad X^\lambda}$, each node will be decorated by three informations: whether is satisfies or not $\mathsf{App}$,  whether is satisfies or not $\mathsf{Right}$ and whether is satisfies or not $\mathsf{Left}$. Note that for linear $\lambda$-terms, some combinations of these properties cannot hold in the same node. For instance, a node cannot satisfy $\mathsf{Right}$ and $\mathsf{Left}$ simultaneously, as this would contradict linearity. 

Now we define the homomorphism $\ranked{h}$, which maps the $\lambda$-terms with these three informations to terms of $\ranked{\tmonad (\ranked{X^\lambda}+1)}$. We define the action of $\ranked{h}$ on each node, depending on its label and the three informations it contains:
\begin{itemize}
\item If the label of the node is $y$  for some variable $y\in X$, or if the label is $@$ and satisfies $\mathsf{App}$, then $\ranked{h}$ returns the same node (seen as a term), forgetting the extra three informations.
\item If the node is an application node satisfying 
\begin{align*}
\neg \mathsf{App} \wedge \neg \mathsf{Right} \wedge\neg \mathsf{Left} 
\end{align*} 
then $\ranked{h}$ adds $1$ to the two children of the node.
\item If the node is an application node satisfying 
\begin{align*}
\neg \mathsf{App} \wedge \mathsf{Right} \qquad\text{(resp. } \neg \mathsf{App} \wedge \mathsf{Left} \text{)}
\end{align*} 
then $\ranked{h}$ adds $1$ to the left (resp. right) child of the node. 
\end{itemize}
Let $t$ be a linear $\lambda$-term. We show that the factors induced by $g$ satisfy the two conditions of Proposition~\ref{prop:FactoIntoThin}. First of all, by analyzing the action of $\ranked{h}$ on each node, note that every application node will receive $1$ as one of its children, except when it is satisfies $\mathsf{App}$. Thus the only branching nodes in a factor are redexes, hence the factors are thin. Now suppose by contradiction that there is some full redex  of $t$ which is not entirely contained in a factor. This means that in $\ranked{g}(t)$ there is a $1$ between the application node of some redex  and its variable.
By construction of $\ranked{h}$, $1$ is the child of an application node (call it $n$). Suppose w.l.o.g. that it is the right child of $n$. The node $n$ cannot satisfy $\mathsf{App}$ because it got $1$ as a child by $\ranked{h}$. Is satisfies $\mathsf{Right}$ by the contradiction hypothesis. Thus its satisfies $\neg \mathsf{App} \wedge  \mathsf{Right}$, therefore it receives also $1$ as its left child by $h$. This means that $n$ received $1$ for its both children, and the only way to get that is to satisfy $\neg \mathsf{App} \wedge \neg \mathsf{Right} \wedge\neg \mathsf{Left}$, which gives a contradiction. 
\end{proof}
%
%
%
%
%

%% file: appendix-factfor.tex
\section{Decomposing the unfolding function}
\label{ap:matrix-power}
\newcommand{\treeunfold}{\mathrm{unfold}}

As discussed in the main body of the paper, the unfolding function may be regarded as unsatisfactory. In this section, we will decompose it into a collection of small functions containing no form of iteration.   

We present these new prime functions in Section~\ref{sec:functions-decomposing-unfolding}, and state the main result of this section which is that term unfolding  can be derived from these new prime functions (and the other prime functions of Section~\ref{sec:derivable-functions}). To prove this result, our strategy is to show that term unfolding can be derived for a restricted class of terms that we call \emph{homogeneous}, and then to show that every term can be factorised into  homogeneous terms. 

The notion of homogeneous terms, and the result about  decomposing arbitrary terms into homogeneous ones, are presented in Section~\ref{sec:factfor}. Next, in Section~\ref{sec:homo-unfold}, we show how term unfolding can be done for homogeneous inputs. Finally, in Section~\ref{sec:monotone-unfold-proof} we prove the main result of the section by combining  the results of Sections~\ref{sec:factfor} and~\ref{sec:homo-unfold}.

\subsection{New prime functions replacing the unfolding}\label{sec:functions-decomposing-unfolding}
In order to decompose the unfolding function, we enrich datatypes with the constructor of shallow terms introduced in Section~\ref{sec:shallow-terms}. 

We present the prime functions which will replace the unfolding in Figures~\ref{fig:prime-for-shallow-terms}--\ref{fig:weak-unfolding}. Prime functions of Figure~\ref{fig:prime-for-shallow-terms} describe the behaviour of the shallow term datatype. Figure~\ref{fig:additional-prime-for-fold} contains some additional laws for the fold datatype and Figure~\ref{fig:additional-distrib-prime} contains some new ditributivity laws. Prime functions of Figure~\ref{fig:weak-unfolding} are weak versions of the unfolding function, containing no form of iteration. 

Note that some of these functions were already presented in Appendix~\ref{sec:definition-of-shallow-unfolding} to define formally the unfolding function: distributivity of shallow terms over fold, distributivity of shallow terms over product (Figure~\ref{fig:additional-distrib-prime}), and the matching function (Figure~\ref{fig:weak-unfolding}). In appendix~\ref{sec:definition-of-shallow-unfolding}, those functions were introduced in a very formal (hence verbose) way. In this appendix, we made the opposite choice  of giving only informal definitions trough some hopefully clear and unambiguous pictures. 
\input{functions}

The main result of this section is that the unfolding can be replaced by the more atomic functions of Figures~\ref{fig:prime-for-shallow-terms}--\ref{fig:weak-unfolding}, in  presence of the prime functions presented in Section~\ref{sec:derivable-functions}, as stated in the following theorem
 
\begin{theorem}\label{thm:decompose-unfolding}
The unfolding function can be derived using the functions of Figures~\ref{fig:prime-for-shallow-terms}--\ref{fig:weak-unfolding} and the prime functions of Section~\ref{sec:derivable-functions}.
\end{theorem}

In the rest of Appendix~\ref{ap:matrix-power}, derivable means derivable from the prime functions of Figures~\ref{fig:prime-for-shallow-terms}--\ref{fig:weak-unfolding} and the prime functions of Section~\ref{sec:derivable-functions} except from unfolding.

\input{factfor}

\input{homo-unfold}

\subsection{Proof of Theorem~\ref{thm:decompose-unfolding}}
\label{sec:monotone-unfold-proof}
In this section, we complete the proof of Theorem~\ref{thm:decompose-unfolding}.  We say that a nested factorisation in $\tmonadn n \mati k \rSigma$ is \emph{monotone} if all of the labels from $\mati k \rSigma$ that appear in it are monotone. 
Consider the homomorphism which maps a branch to its corresponding twist, and which gives the completely undefined function in case the twist is not monotone.  The homomorphism uses an aperiodic monoid, as discussed in Example~\ref{ex:partial-monoton-functions}. 
 Apply the Factorisation Forest Theorem with respect to this homomorphism, yielding a derivable function
\begin{align*}
\ranked{ f : \tmonad \mati k \rSigma \to \tmonadn n \mati k \rSigma}
\end{align*}
which produces only nested factorisations that are  hereditarily homogeneous. (Also, because monotone functions are closed under composition, it follows that if  an input to $\ranked f$ is monotone, then the same is true for the output.) Therefore,  Theorem~\ref{thm:decompose-unfolding} follows by composing the function $\ranked f$ with the function $\ranked {g_n}$ from the following lemma. 

\begin{lemma}\label{lem:ind-homo-twist}
    For every finite ranked set $\rSigma$ and  $n \in \set{1,2,\ldots}$ there is a derivable function
    \begin{align*}
    \ranked{ g_n : \tmonadn n \mati k \rSigma \to \mati k \tmonad \rSigma}
    \end{align*}
    which makes the following diagram commute for inputs that are  monotone and  hereditarily homogeneous: 
    \begin{align*}
        \ranked{
            \xymatrix{
                \tmonadn n \mati k\rSigma \ar[d]_{\flatn n} \ar[rd]^g\\
                \tmonad  \mati k \rSigma \ar[r]_{\unfold}& \mati k \tmonad \rSigma 
            }
        }
    \end{align*}
\end{lemma}
\begin{proof}
    Induction on $n$. To make the induction pass through, we also show that each function $\ranked{g_n}$ is consistent wit the  twist homomorphism in the following sense: for every input $t \in \tmonadn n \mati k \rSigma$, and every port $i \in \set{1,\ldots,\arity t}$, the same value is obtained by: (a) recursive flattening $t$ and then composing all of the twists that are found on the path from the root to port $i$; (b) applying $\ranked{g_n}$ and then computing the twist corresponding to port $i$. 
    
    For the induction base $n=1$, hereditarily homogeneous inputs are units, and there are finitely many of them and the function can be derived on a case by case basis. 
    
    Consider the induction step, where the lemma has already been proved for $n$ and we want to prove it for $n+1$. The function is the composition
    \begin{align*}
        \ranked{
            \xymatrix@C=1cm{
                \tmonadn {n+1} \mati k \Sigma \ar[r]^{\tmonad g_n} & \tmonad \mati k{(\tmonad \Sigma)} \ar[r]^{\text{Lemma~\ref{lem:homo-twist}}}&  \mati k  {(\tmonad \tmonad \rSigma)} \ar[r]^{\mati k \flatt} & \mati k \rSigma
            }
        }
    \end{align*}
    Consider a  hereditarily homogeneous input $t   \in \ranked{\tmonadn{n+1} \mati k \Sigma}$. 
    \begin{enumerate}
            \item Apply the function from the induction assumption to every label of $t$, i.e.~apply 
        \begin{align*}
        \ranked{
            \xymatrix{
                \tmonadn {n+1} \mati k \Sigma \ar[r]^{\tmonad g_n} & \tmonad \mati k{(\tmonad \Sigma)}
            }
        }
        \end{align*}
        \item Let $t_1$ be the output from the previous step. Because $\ranked {g_n}$ is consistent with twists, and $t$ is hereditarily homogeneous, it follows that $t_1$  is either a shallow term, or it is homogeneous with respect to the twist homomorphism.  If $t_1$ is a shallow term, then we apply the shallow unfolding operation from .. . Otherwise, we $t_1$ is homogeneous
        , because $t$ is hereditarily homogeneous and $\ranked{g_n}$ is consistent with twists. Therefore, we can apply the function
        from Lemma~\ref{lem:homo-twist}, with the alphabet being $\tmonad \rSigma$. 
        \item The result of the previous step is a term $t_2 \in \mati k {(\tmonad \tmonad \rSigma)}$. To this term, we apply $\mati k \flatt$, yielding the final result.
    \end{enumerate}
    A routine check shows that the function $\ranked{g_{n+1}}$ defined above satisfies the property in the statement of the lemma, and that it is furthermore consistent with the twist homomorphism.     
\end{proof}

%% file: functions.tex
\begin{figure}
\fbox{
\begin{minipage}{1\linewidth}
\begin{itemize}
\item \textbf{Unit.}
$$
\begin{array}{c}
 \ranked{\Sigma} \ \ \ranked{\leftrightarrow} \ \    \ranked{ \Sigma \cdot 1}\\[5pt]
        {\includegraphics[scale=.4]{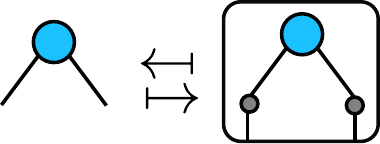}} 
\end{array}
$$
\item \textbf{Associativity.}
$$
\begin{array}{c}
 \ranked{(\Sigma \cdot \Gamma)\cdot \Delta } \ \ \ranked{\to} \ \    \ranked{ \Sigma \cdot (\Gamma \cdot \Delta)}\\[5pt]
        {\includegraphics[scale=.4]{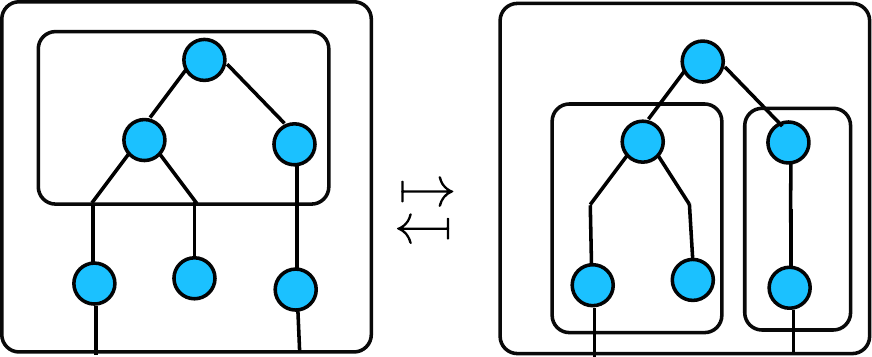}} 
\end{array}
$$
\item \textbf{Terms as shallow terms.}
$$
\begin{array}{c}
 \ranked {1 + \shallowterm \Sigma {\tmonad \Sigma}\ \ \leftrightarrow \ \  \tmonad \Sigma}\\[5pt]
        {
        \begin{tabular}{l}
            Every term is either just a port,\\ or has a root and child subterms.    
        \end{tabular}    
        }
\end{array}
$$
\item \textbf{Tensors as shallow terms.}
$$
\begin{array}{c}
\ranked {\Sigma^n \ \ \leftrightarrow \ \ \shallowterm n \Sigma}\\[5pt]
        {\includegraphics[scale=.4]{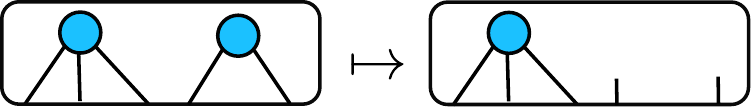}}
\end{array}
$$
\end{itemize}
\end{minipage}
}
\caption{Prime functions for shallow terms.}\label{fig:prime-for-shallow-terms}
\end{figure}

\begin{figure}
\fbox{
\begin{minipage}{1\linewidth}
\begin{itemize}
\item \textbf{Unit.}
$$
\begin{array}{c}
 \ranked{\Sigma \ \ \to \ \ \reduce k \Sigma^k}\\[5pt]
        {\includegraphics[scale=.4]{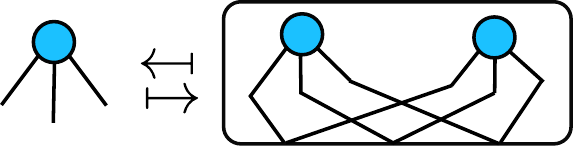}}
\end{array}
$$
\item \textbf{Increase fold.}
$$
\begin{array}{c}
 \ranked{\reduce k \Sigma \ \ \to \ \ \reduce {k+1}\Sigma}\\[5pt]
        {\includegraphics[scale=.4]{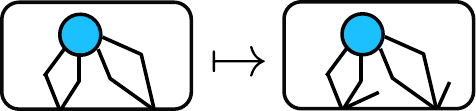}}	
\end{array}$$
\item \textbf{Decrease fold.}
$$
\begin{array}{c}
 \ranked{\reduce {k+1} \Sigma \ \ \to \ \ \reduce {k}\Sigma+\bot}\\[5pt]
        {\includegraphics[scale=.4]{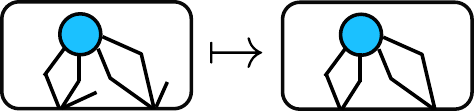}}	 
\end{array}$$
\item \textbf{Projection of products.}
$$
\begin{array}{c}
   \ranked{\Sigma\product \Sigma \ \ \to \ \ \reduce 1 \Sigma}\\[5pt]
        {\includegraphics[scale=.4]{pictures/tensor-projection-1}}	
\end{array}$$
\end{itemize}
\end{minipage}
}
\caption{Additional prime functions for folds.}\label{fig:additional-prime-for-fold}
\end{figure}

\begin{figure}
\fbox{
\begin{minipage}{1\linewidth}
\begin{itemize}
\item \textbf{Fold over coproduct.}
$$
\begin{array}{rll}
  \ranked{\reduce k (\Sigma_1 + \Sigma_2)}& \ranked{\to} & \ranked{\reduce k \Sigma_1 + \reduce k \Sigma_2}\\
         (a,i)/f & \mapsto & ((a/f),i)
\end{array}
$$
\item \textbf{Shallow terms over coproduct.}
$$
\begin{array}{rll}
  {\shallowterm {(\Sigma_1 + \Sigma_2)} \Gamma} & \ranked{\to} & \ranked{(\shallowterm {\Sigma_1} \Gamma) + (\shallowterm {\Sigma_2} \Gamma)}\\
        (a,i)\tensorpair{t_1,\dots,t_n} &\mapsto& (a\tensorpair{t_1,\dots,t_n},i)
\end{array}
$$
\item \textbf{Shallow terms over product.}
$$
\begin{array}{c}
  \ranked{{\shallowterm {(\Sigma_1 \product \Sigma_2)} \Gamma}
   \ \  \to \ \ {(\shallowterm {\Sigma_1} \Gamma) \product (\shallowterm {\Sigma_2} \Gamma)}}\\[5pt]
        {\includegraphics[scale=.4]{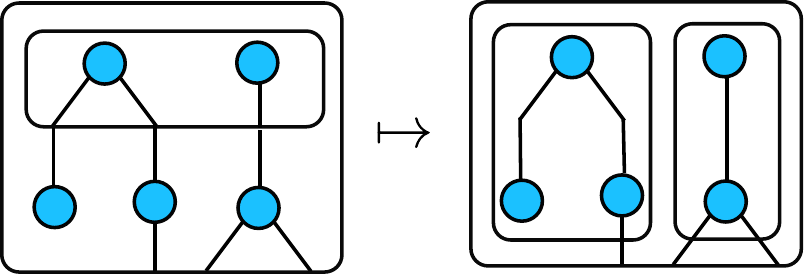}}
        \end{array}
$$
\item \textbf{Fold over product.}
$$
\begin{array}{c}
    \ranked{\reduce k (\Sigma_1 \product \Sigma_2) \ \ \to \ \ \reduce k ((\reduce k {\Sigma_1})\product (\reduce k \Sigma_2))}\\[5pt]
        {\includegraphics[scale=.4]{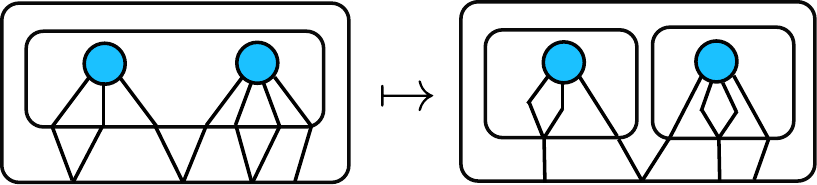}}   
\end{array}
$$
\item \textbf{Fold over product (bis).}
$$
\begin{array}{c}
\ranked{(\reduce k \Sigma_1) \product (\reduce k {\Sigma_2}) \ \ \to \ \reduce k (\Sigma_1 \product \Sigma_2)}\\[5pt]
        {\includegraphics[scale=.4]{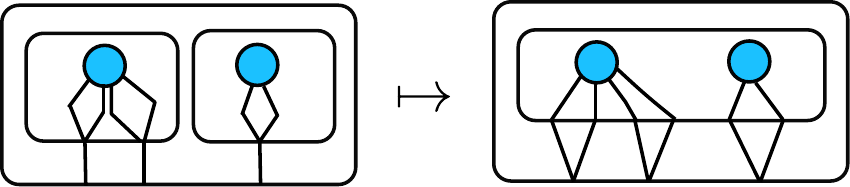}}  
\end{array}
$$
\item \textbf{Shallow terms over fold.}
$$
\begin{array}{c}
 \ranked{\shallowterm \Sigma {\reduce k \Gamma}\ \ \to \ \ \reduce k (\shallowterm \Sigma {\Gamma})}\\[5pt]
 {\includegraphics[scale=.4]{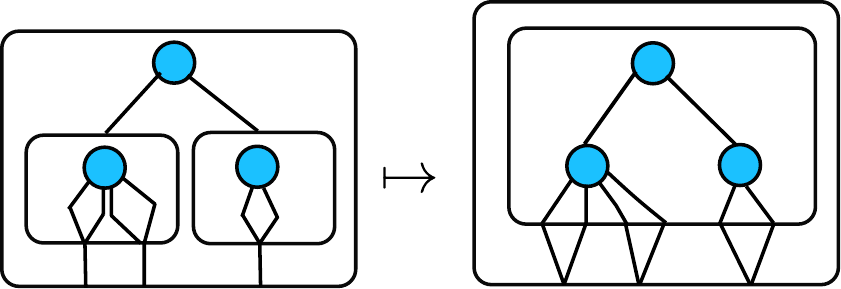}} 
\end{array}
$$
\item \textbf{Fold over shallow terms.} 
$$
\begin{array}{c}
 \ranked{\reduce k (\Sigma\cdot \Gamma)\ \ \to \ \ (\reduce k \Sigma) \cdot \mati k \Gamma}\\[5pt]
 {\includegraphics[scale=.4]{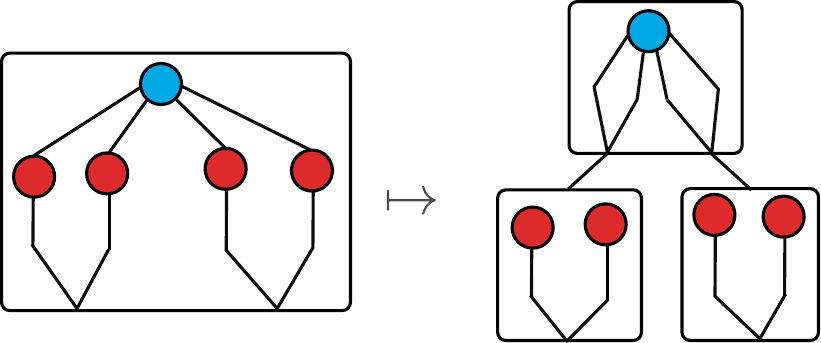}} 
\end{array}
$$
$\rGamma$ is a set of unary elements.
\end{itemize}
\end{minipage}
}
\caption{Additional distributivity prime functions.} \label{fig:additional-distrib-prime}
\end{figure}

\begin{figure}
\fbox{
\begin{minipage}{1\linewidth}
\begin{itemize}
\item \textbf{Untwist.}
$$
\begin{array}{c}
\ranked{\tmonad {\reduce 1\Sigma} \ \  \to \ \ \reduce 1 \tmonad \Sigma}\\[5pt]
         {\includegraphics[scale=.4]{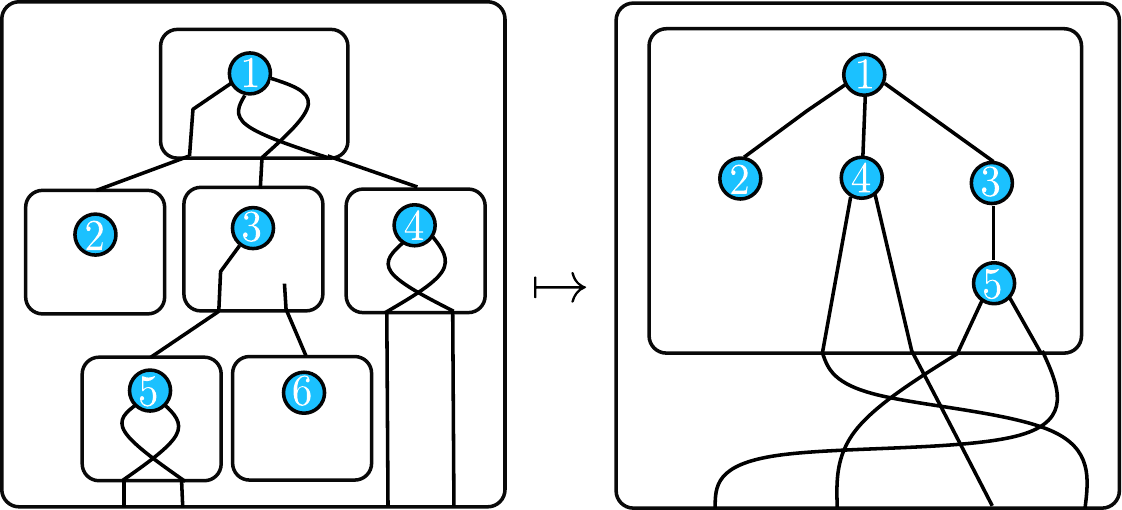}}
\end{array}
$$
\item \textbf{External fold.}
$$
\begin{array}{c}
\ranked{\tmonad {\reduce k\Sigma} \ \  \to \ \ \reduce k \tmonad \reduce k\Sigma}\\[5pt]
         {\includegraphics[scale=.43]{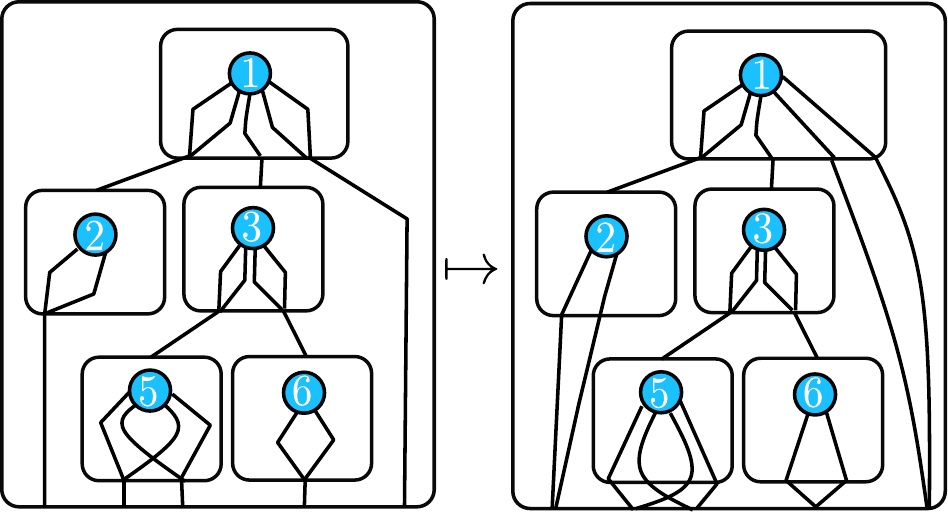}}
\end{array}
$$
\item \textbf{Matching.}
$$
\begin{array}{c}
\ranked{\shallowterm {\reduce k \Sigma}{\Gamma^k} \ \ \to \ \ \reduce 1(\shallowterm \Sigma \Gamma)}\\[5pt]
\includegraphics[scale=.3]{pictures/shallow-unfold}
\end{array}
$$
\end{itemize}
\end{minipage}
}
\caption{Weak forms of unfolding.}\label{fig:weak-unfolding}
\end{figure}

%% file: factfor.tex
\subsection{Factorisation forests}
\label{sec:factfor}
This section is devoted to stating and proving a tree version of the Factorisation Forest Theorem of Imre Simon.  Our result differs from the original Factorisation Forest Theorem in the following ways: (a) we consider trees instead of strings; (b) we use aperiodic finite monoids instead of arbitrary finite monoids; and (c) the factorisation in the conclusion of the theorem can be computed by a derivable function.  A tree generalisation of the Factorisation Forest Theorem was already proved by Colcombet~\cite[Theorem 1 and Section 3.3]{colcombetCombinatorialTheoremTrees2007}, but Colcombet's result is proved for monadic second-order logic, and therefore it does not satisfy condition (c).

\paragraph{Factorisation forests} The idea behind factorisation forests is to split a term into a nested factorisation, which is a term of terms of terms, and so on up to a certain depth.  
Define a \emph{nested factorisation} of depth $k \in \set{1,2,\ldots}$ over alphabet $\rSigma$ to be an element of $\tmonadn k \rSigma$ which is defined by
\begin{align*}
\tmonadn 0 \rSigma = \rSigma  \quad \text{and} \quad \tmonadn {k+1}\rSigma = \tmonad \tmonadn k \rSigma.
\end{align*}
Nested factorisations can be flattened to terms by using an  operation $\flatn k : \tmonadn k \rSigma \rto \tmonad \rSigma $ defined by 
\begin{align*}
     \flatn 1 = \text{\ranked{identity}} \quad \text{and} \quad  \flatn {k+1} \eqdef \flatt  \circ \tmonad \redpar { \flatn k}.
\end{align*}
An equivalent definition of $\flatn {k+1}$ would be $\flatn k \circ \tmonadn {k-1} \flatt$, the equivalence of these definitions corresponds to the fact that $\tmonad$ is a monad.

\paragraph{Branches and subbranches}
Define a \emph{branch} in a ranked set to be an element  of the ranked set together with a distinguished port. 
We draw branches like  this:
\mypic{82}
We write $\branches \rSigma$ for the (unranked) set of branches over a ranked set $\rSigma$. \label{page:branches}
For a term, we classify its edges as internal (linking a non-port node with a non-port child) and external (linking a non-port node with a child port). Each edge in a term $t \in \tmonad \rSigma$ corresponds to a branch over $\rSigma$, namely the branch which leads to the edge. Any branch obtained this way is called a \emph{subbranch} of $t$. Here is a picture of subbranches in the case of a term of terms:
\mypic{80} 

\newcommand{\hb}[2]{#2^{(#1)}}
Branches in  terms  form a monoid. Using the monoid structure of branches in terms, we can extend any function  $h : \branches \rSigma \to M$, with $M$ a monoid, to a  monoid homomorphism
\begin{align*}
\hb n h : \branches \tmonadn n \rSigma \to M
\end{align*}
which maps a branch of a term to the product -- in the monoid $M$ -- of all of its subbranches (after flattening).  A more formal definition is that $\hb 0 h$ is the same as $h$, while  $\hb {n+1} h$  is the unique monoid homomorphism  which makes the following diagram commute
\begin{align*}
\xymatrix{
    \branches \tmonadn n \rSigma \ar[dr]^{\hb n h} \ar[d]_{\branches \unit}\\
    \ar[r]_{\hb {n+1}  h}\branches \tmonadn {n+1} \rSigma & M
}
\end{align*}

The idea behind factorisation forests, as expressed in Definition~\ref{def:hom-for} below, is to factorise a term into a term of terms of terms (etc.) so that the depth of nesting is bounded, and at each level all branches behave regularly with respect to some monoid homomorphism. 

\begin{definition}[Homogeneous factorisations]\label{def:hom-for}
    Let $h : \branches \rSigma \to M$ be a function into a monoid $M$. 
    \begin{itemize}
\item     We say that a factorisation $t \in \tmonad \tmonad \rSigma$ is \emph{homogeneous with respect to $h$} if it either:
\begin{enumerate}
    \item \label{it:factfor-shallow} it is a shallow term (which means that all internal edges originate from the root); or 
    \item \label{it:factfor-allsame} all internal subbranches of $t$ have the same value under $\hb 1 h$; or
    \item \label{it:factfor-ab} if $a,b \in M$ appear as values -- under $\hb 1 h$ -- of internal branches in $t$, then $ab=a$. 
\end{enumerate}
\item We say that a nested factorisation  $t \in \tmonadn n \rSigma$ is \emph{hereditarily homogeneous with respect to $h$} if either $n=1$ and $t$ is the unit of a letter, or $n \ge 2$ and both:
\begin{enumerate}
    \item  it is homogeneous with respect to $\hb {n-1} h$; and 
    \item every node has a label in $\tmonadn {n-1} \rSigma$ that is hereditarily homogeneous with respect to $h$.
\end{enumerate}
    \end{itemize}
\end{definition}

Recall that a finite monoid is aperiodic if it has only trivial subgroups. An equivalent definition is that every element $m$ of the monoid satisfies 
\begin{align*}
  \exists n \in \set{1,2,\ldots}\   m^n = m^{n+1}.
\end{align*} 
A famous theorem of Sch\"utzenberger, McNaughton and Papert, see~\cite[Theorem VI.1.1]{straubingFiniteAutomataFormal1994} says that the languages  of words recognised by homomorphisms into finite aperiodic monoids are exactly  those that can be defined in first-order logic. This is the reason why  we consider aperiodic monoids.

\begin{example}\label{ex:partial-monoton-functions}
   Let $k \in \set{1,\ldots}$ and  consider the monoid of partial functions 
    \begin{align*}
    \set{1,\ldots,k} \to \set{1,\ldots,k}.
    \end{align*}
    This monoid is not aperiodic, because it contains the  group of all permutations of $\set{1,\ldots,k}$. Consider now the restriction of this monoid to partial functions which are monotone (this is a monoid, because such functions are closed under composition). This monoid is aperiodic, because if $f$ is a partial function, then for  every $i \in \set{1,2,\ldots,k}$ the sequence
    \begin{align*}
    f^1(k),f^2(k),f^3(k),\ldots
    \end{align*}
    reaches a fixpoint (or becomes undefined) in at most $k$ steps.
\end{example}
We are now ready to state our version of the Factorisation Forest Theorem. 
\begin{theorem}[Factorisation Forest Theorem]\label{thm:factfor}
    Let $\rSigma$ be a  ranked set and let $h : \branches  \rSigma \to M$ be a function into a finite aperiodic monoid $M$. There is some $n \in \set{1,2,\ldots}$ and a  function
    \begin{align*}
        \ranked {f : \tmonad \rSigma \to \tmonad^n \rSigma}  
    \end{align*}
such that $\flatn n \circ \ranked f$ is the identity on $\tmonad \rSigma$, and  all outputs of  $\ranked f$ are hereditarily homogeneous with respect to $h$. Furthermore, if $\rSigma$ is finite\footnote{This finiteness assumption could be relaxed by saying that $\rSigma$ is possibly infinite but the function $h$ is derivable, in the sense that a derivable function can decorate the ports of an element in $\rSigma$ by their values under $M$.} then $\ranked f$ is derivable. 
\end{theorem}

\newcommand{\hint}{\bar h}
\newcommand{\hintplus}{\bar h^+}
\newcommand{\branchesplus}{\mathsf B^+}

In the proof below, the constructions  are designed so that they can be formalised using derivable functions, however we leave the details of  the ``Furthermore'' part to the reader. 

Define a \emph{good set} to be any subset  $\ranked X \subseteq  \tmonad \rSigma$ which admits a function 
\begin{align*}
    \ranked {f : \tmonad \rSigma \to \tmonad^n \rSigma}   \qquad \text{for some $n \in \set{1,2,\ldots}$}
\end{align*}
such that $\flatn n \circ \ranked f$ is the identity on $\tmonad \rSigma$, and    $\ranked f$ restricted to $\ranked X$ produces only hereditarily homogeneous outputs. Our goal is to show that the entire set $\tmonad \rSigma$ is good. To prove this, we use a more refined result, stated below, which has a parameter  that can be used for  induction.  We say that a term $t \in \tmonad \rSigma$ uses $A \subseteq M$ for  internal subbranches if all  internal subbranch have image under $\hb 1 h$ that belongs to $A$. 

\begin{lemma}\label{lem:fact-ind}
    Let $\rSigma$ be a  ranked set and let $h : \branches \tmonad \rSigma \to M$ be a monoid homomorphism into a finite aperiodic monoid $M$. For every $A \subseteq M$, the  terms that use $A$ for inner subbranches  is good.
    \end{lemma}

\newcommand{\subgen}[1]{\langle #1 \rangle}
Theorem~\ref{thm:factfor} follows immediately from the  lemma, by taking $A$ to be the entire monoid. The rest of Section~\ref{sec:factfor} is therefore devoted to proving the lemma. The proof is by induction on  two parameters: (a) the size of $A$; and (b) 
the size of the semigroup $\subgen A \subseteq M$ that is  generated by $A$.
 These parameters are  ordered lexicographically, with the size of the semigroup being more important.

The induction base is when $A$ contains only one element $a$ of the monoid. If a term uses $\set a$ for  internal subbranches, then applying $\tmonad \unit$ leads to a factorisation that is  homogeneous according to item~\ref{it:factfor-allsame} of Definition~\ref{def:hom-for}, which is also hereditarily homogeneous because all nodes are labelled by units. This completes the proof of the induction base.

In the proof of the induction step, we consider two cases.
\begin{itemize}
    \item The first case is  when  every $a \in A$ satisfies 
    \begin{align*}
       \subgen{\set{ b a :  b \in \subgen A}} = \subgen A
  \end{align*}
  This means that every for every $a \in A$, the function $b \mapsto ba$ is a permutation of $\subgen A$. Since the monoid is aperiodic, this permutation must necessarily be the identity.  Therefore, we have $ab=a$ for every $a,b \in \subgen A$. This means that if all a term uses $A$ for  internal subbranches, then applying $\tmonad \unit$  gives a factorisation which is hereditarily homogeneous according to item~\ref{it:factfor-ab} of Definition~\ref{def:hom-for}. 
    \item If the previous item does not hold, then there is some $a \in A$ such that 
    \begin{align*}
         \subgen{\set{ b a :  b \in \subgen A}}
    \end{align*}
    is a proper subsemigroup of $\subgen A$.  Fix some such $a$.  Define a \emph{sensitive edge} in a term  $t \in \tmonad \rSigma$ to be any internal edge where the corresponding subbranch has value $a$ under  $\hb 1 h$. Call an internal edge \emph{post-sensitive} if it is not sensitive, but its parent edge is. Here is a picture:
    \begin{center}
\includegraphics[scale=.27, page=8]{pics}
\end{center}
    Define the \emph{split} of a term to be the factorisation which  cuts along post-sensitive  edges, as shown in the following picture:
    \begin{center}
\includegraphics[scale=.29, page=88]{pics}
\end{center}
    We only consider splits for terms which use $A$ for internal subbranches. Roughly speaking,  we will show that all factors in the split are good, and the split itself is good.  Combining these two observations, we will see that all terms 

    We begin by looking at the factors in the split (of a term where $A$ is used for internal subbranches). Here is a picture of such a factor:
    \mypic{89}
        If we follow a  branch in an factor of the split, from root to port, we first  have a sequence of non-sensitive  edges from the original term, followed by a sequence of sensitive edges. Group the non-sensitive edges together, and group the sensitive edges together, resulting in a shallow term from $\shallowterm{\tmonad \rSigma}{\tmonad \rSigma}$, which is illustrated in the following picture:
     \begin{center}
\includegraphics[scale=.3, page=90]{pics}
\end{center}
        In the resulting shallow term,  the root is labelled by a term without sensitive edges (i.e.~it is a term which uses $A - \set a$ for internal subbranches), while the children are labelled by terms where all edges are sensitive (i.e.~they are terms which use $\set a$ for internal subbranches). We can apply the induction in both cases, and combine the resulting nested factorisations using a shallow term, as in item~\ref{it:factfor-shallow} of Definition~\ref{def:hom-for}.

        Having established that the factors of the split are good, we turn to the split itself. By construction, every subbranch of the split is mapped by $\hb 1 h$ to the smaller semigroup 
        \begin{align*}
            \subgen{\set{ b a :  b \in \subgen A}},
       \end{align*}
       We can  view the  split as a term over alphabet $\rGamma = \tmonad \rSigma$.  Since all  internal subbranches of the split are in the smaller subsemigroup, we can apply the induction assumption of the lemma (with $\rGamma$ and $\hb 1 h$), showing that the split is good. More formally, the set 
       \begin{align*}
       \set{\text{split of $t$} : t \in \tmonad  \rGamma \text{ uses $A$ for internal subbranches}}
       \end{align*}
       is good. To show now that original set of terms $t$ that use $A$ for  internal branches is good, we first apply the split, then compute the nested factorisation for the split, and finally we compute the nested factorisations for the factors of the split (the letters from $\rGamma$.). 
\end{itemize}

%% file: homo-unfold.tex
\subsection{Term unfolding for homogeneous inputs}
\label{sec:homo-unfold}

The goal of this section is to show that term unfolding is derivable for homogeneous inputs. Actually, we will first show that unfolding is derivable for another particular case of inputs which we call \emph{constant-twists}. Then we will use this function as a macro to unfold the homogeneous inputs.  

\subsubsection{Unfolding constant-twist functions}

In the proof of this section, it will be sometimes convenient to manipulate \emph{partial shallow terms}, that is shallow terms where some children of the root maybe ports. We will define them more precisely, and show that unfolding matrix power of partial shallow terms is derivable.  

\paragraph*{Partial shallow unfold}
 If $\rGamma$ and $\rDelta$ are types, we define $\ranked{\Gamma\odot\Delta}$ to be $\ranked{\Gamma.(\Delta+\set{1})}$. We call its inhabitants the \emph{partial shallow} terms. A partial shallow term looks like this, where we omitted to draw the element $1$ 
\begin{center}
\includegraphics[scale=.4]{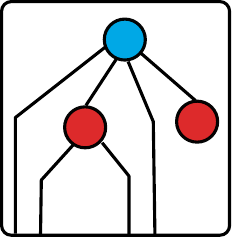}
\end{center}
We define the \emph{partial shallow unfolding} function as the extension  of the shallow unfold function of Figure~\ref{fig:weak-unfolding} to partial shallow terms. It is the function of type 
\begin{align*}
\ranked{{\reduce k \Sigma}\odot{\Gamma^k} \to \reduce k ({\Sigma}\odot{\Gamma})} 
\end{align*}
defined as in the following picture
 \begin{center}
\includegraphics[scale=.4]{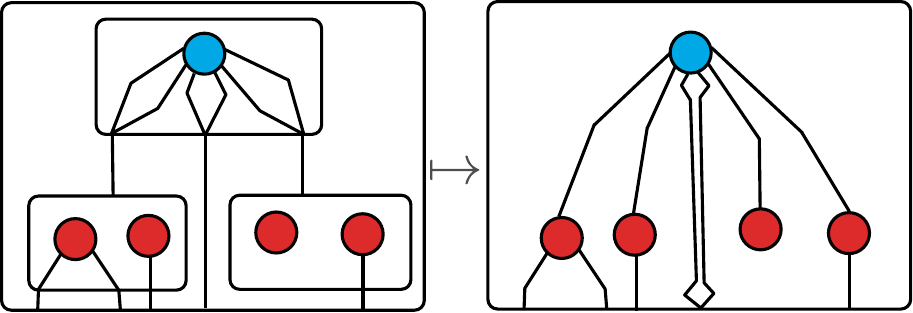}
 \end{center}
the partial shallow unfolding function can be derived as follows. Consider the functions $\ranked{f}$ and $\ranked{g}$ defined as follows
\begin{align*}
\ranked{f:\Gamma^k \xrightarrow{\text{Increse-fold}} \reduce k  \Gamma^k \xrightarrow{\reduce k (\iota_1)^k} \reduce k  (\Gamma+1)^k}  \\ 
\ranked{g:1 \xrightarrow{\text{k-Unit}} \reduce k 1^k \xrightarrow{\reduce k (\iota_2)^k} \reduce k  (\Gamma+1)^k }
\end{align*}
We start by lifting $f$ and $g$ as follows
\begin{align*}
\ranked{\reduce k \Sigma \odot \Gamma^k\overset{\text{{\tiny def}}}{=} \reduce k \Sigma \cdot (\Gamma^k+1) \to  \reduce k \Sigma \cdot \reduce k (\Gamma+1)^k}
\end{align*}
We compose the obtained function with the prime function which distributes   the shallow product over the fold:
\begin{align*}
\ranked{\reduce k \Sigma \cdot \reduce k (\Gamma+1)^k \to \reduce k (\reduce k \Sigma \cdot (\Gamma+1)^k)}
\end{align*}
Now we can apply the shallow unfold function, more precisely we lift it along the constructor $\reduce k$. Then we compose the result with the product of the graded monad:
\begin{align*}
\ranked{\reduce k (\reduce k \Sigma \cdot (\Gamma+1)^k) \xrightarrow{\reduce k \text{Shallow unfold}} \reduce k \reduce 1 ( \Sigma \cdot (\Gamma+1)) }
\end{align*}
Then we compose the result with the product of the graded monad, to obtain the desired function
\begin{align*}
\ranked{\reduce k \reduce 1 (\Sigma \cdot (\Gamma+1)) \xrightarrow{\flatt} \reduce k \Sigma \cdot (\Gamma+1) \overset{\text{{\tiny def}}}{=} \reduce k \Sigma \odot \Gamma}
\end{align*}

\paragraph*{Term unfolding for constant-twist inputs}

We say that a term $ t \in \tmonad \mati k \rSigma$ is a \emph{constant-twist term}  if each twist of an internal branches is a constant function. Note that the internal twists need not to be the same constant function. Here is an example of a constant-twist term

  \begin{center}
\includegraphics[scale=.35]{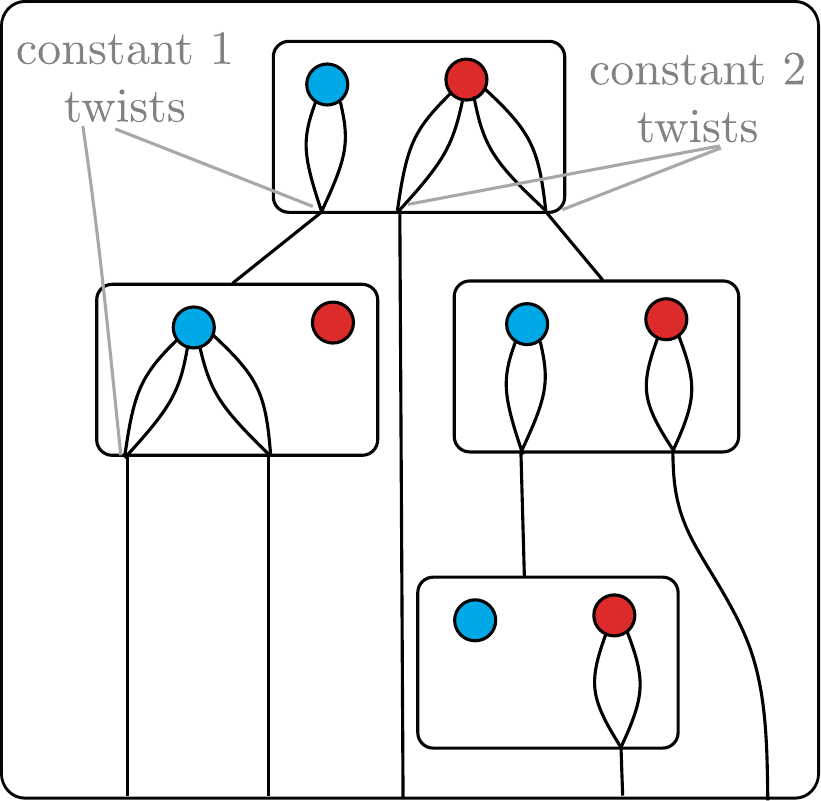}
 \end{center}
 This section is devoted to proving the following lemma. 

\begin{lemma}\label{lem:constant-twist}
    Let $k \in \set{1,2,\ldots}$. There is a derivable function 
    \begin{align*}
        \ranked{f : \tmonad \mati k \rSigma \rightharpoonup \mati k {(\tmonad \Sigma)} }
        \end{align*}      
which coincides with unfolding for all constant-twist inputs.
\end{lemma}

\begin{proof}
The function $f$ can be derived using the following steps. We use the example of the constant-twists term above as a running example.  
\begin{enumerate}
\item We start by applying the external unfolding function.  We get a term in $\ranked{\reduce k \tmonad \Sigma^{[k]}}$. Our example becomes like this
   \begin{center}
\includegraphics[scale=.35]{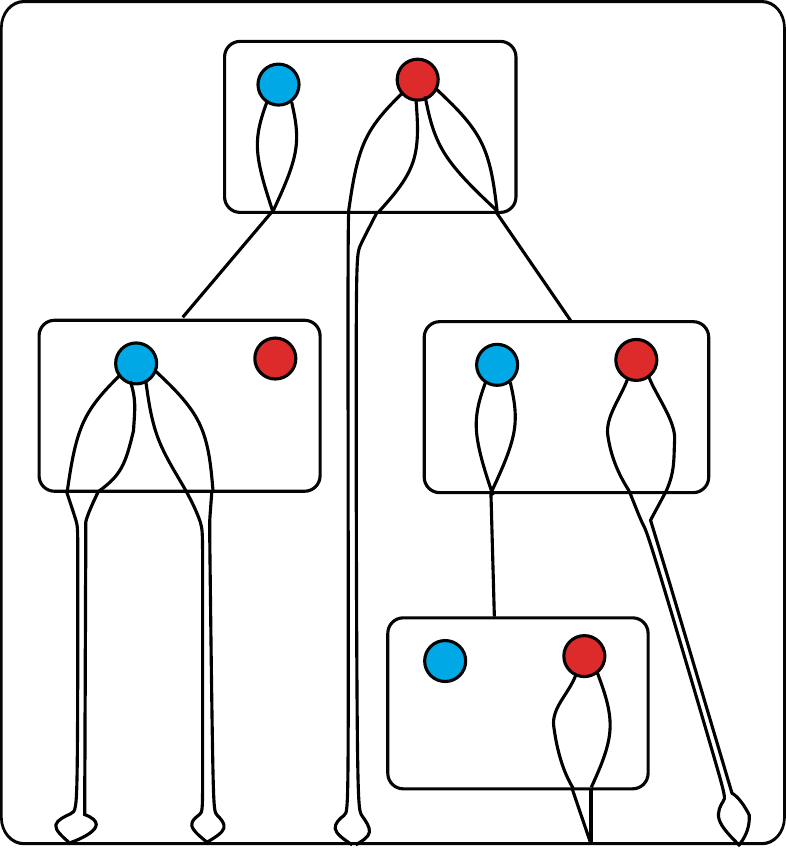}
 \end{center}
 \item Next, we will transform each matrix power node into a tensor product as follows 
 \begin{align*}
 \ranked{ \Sigma^{[k]} \rightarrow \reduce k (\reduce k \Sigma \product \dots \reduce k \Sigma) \to \reduce 1 (\reduce k \Sigma)^k}
 \end{align*}
 The idea here is that, since the image of each twist is a singleton,  the ports of the matrix power are independent. We can then transform safely each node into a tensor product. After that, we transform each tensor product $\ranked{(\reduce k \Sigma)^k}$ into a shallow term $\ranked{k.\reduce k \Sigma}$ which we see itself as a term of type $\ranked{\tmonad(k+\reduce k\Sigma)}$. After the application of the unfolding function $\mathrm{unfold}_1$ followed by a flattening, and the simplification of $\ranked{\reduce k \reduce 1 \tmonad (k+\reduce k \Sigma)}$ into $\ranked{\reduce k \tmonad (k+\reduce k \Sigma)}$ we get a term in $\ranked{\reduce k \tmonad(k+\reduce k \Sigma)}$. Our running example becomes as follows after this step
   \begin{center}
\includegraphics[scale=.35]{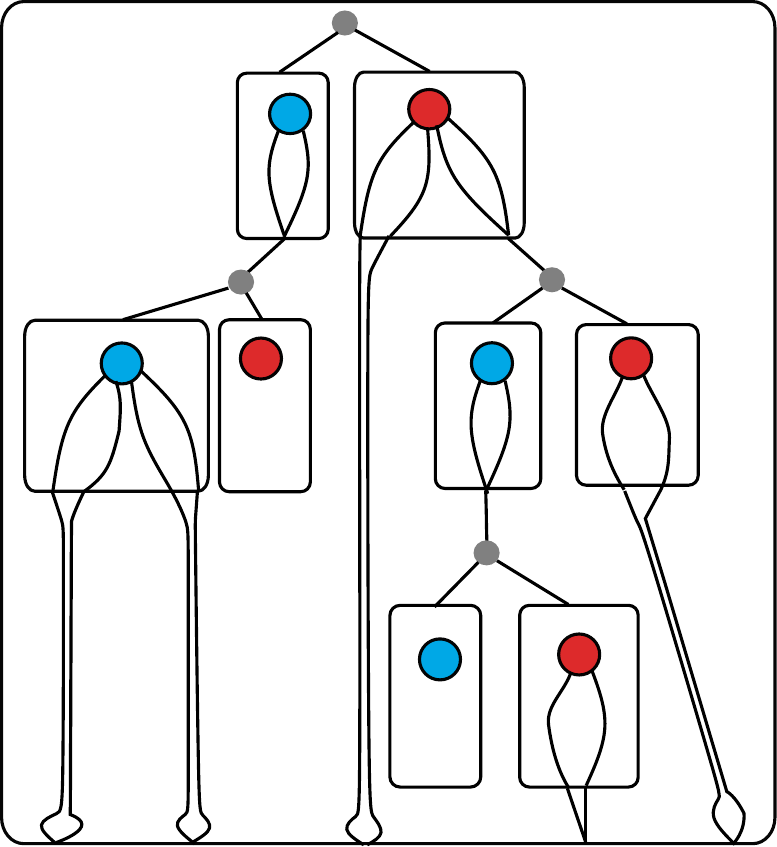}
 \end{center}
 \item Now we apply the factorization 
 \begin{align*}
 \ranked{\tmonad (k + \reduce k \Sigma)\to \tmonad \tmonad(k + \reduce k \Sigma)}
 \end{align*}
 which regroups each element $\ranked{\reduce k \Sigma}$ with its children of type $\ranked{k}$ in the same factor, and leaves the other nodes in isolated factors. At this point our term looks like this
   \begin{center}
\includegraphics[scale=.35]{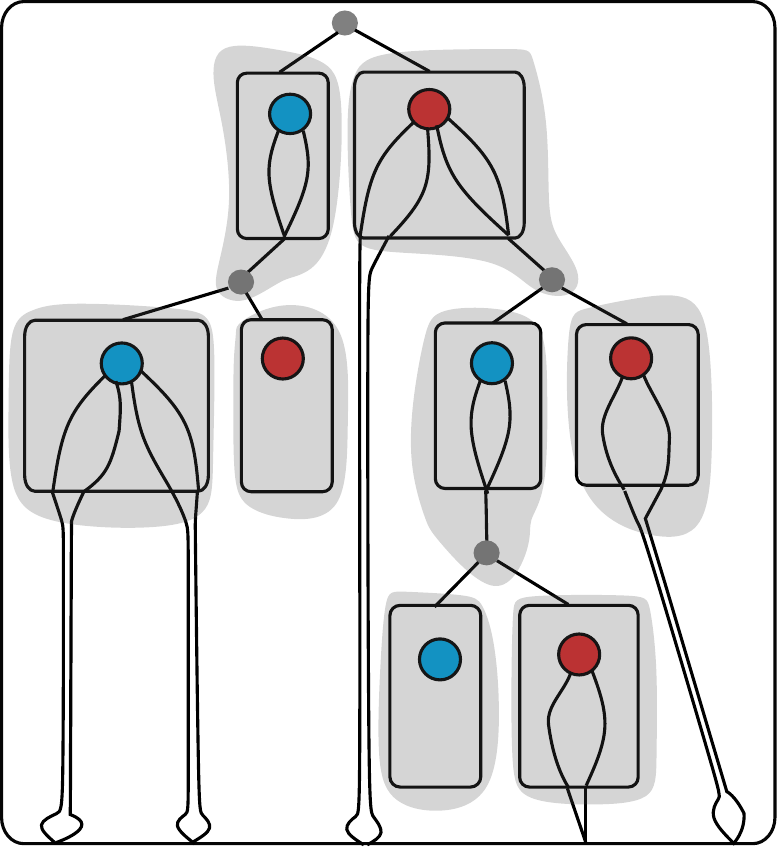}
 \end{center}
 Note that this factorization have the following shape: the root is labeled by $k$, and all the other nodes have labels in $\ranked{(\reduce k \Sigma) \odot k}$. We want to reflect this structure in  the type by applying the following function
 \begin{align*}
 \ranked{\tmonad \tmonad(k + \reduce k \Sigma) \to k. \tmonad ((\reduce k \Sigma)\odot k)}
 \end{align*}
 This function can be implemented easily using the decomposition function, the functions which maps every type to $\ranked{1}$ and mechanisms of raising errors. 
 \item In each node $\ranked{(\reduce k \Sigma)\odot k}$, we transform $\ranked{k}$ into $\ranked{1^k}$ (this function is basic since its domain is finite). After that, we apply the partial shallow unfold function, composed with the prime functions eliminating the $1$ and decreasing the fold:
\begin{align*}
\ranked{(\reduce k \Sigma)\odot k \to \reduce k (\Sigma \odot 1) \to \reduce k \Sigma \to \reduce 1 \Sigma}
\end{align*} 
  as illustrated by the following picture
   \begin{center}
\includegraphics[scale=.3]{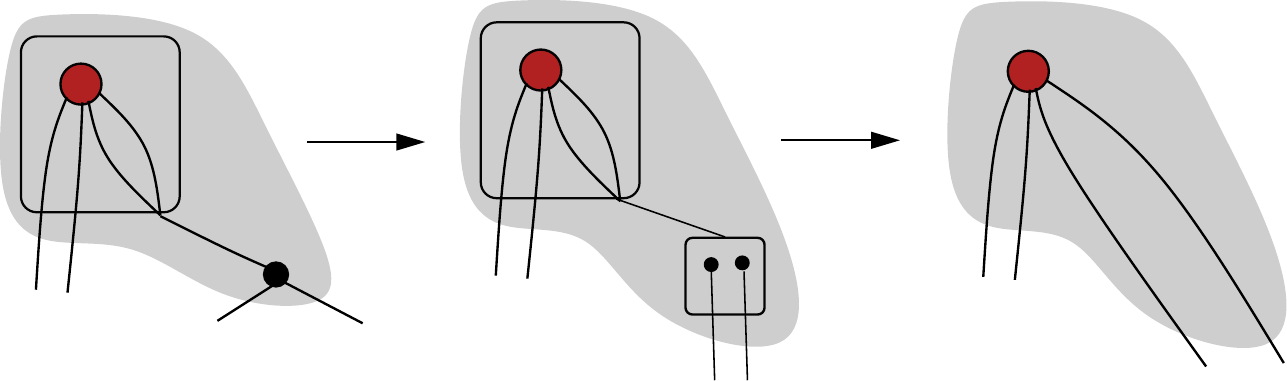}
 \end{center}
 \item At this point we have a term of type $\ranked{\reduce k (k\cdot \tmonad \reduce 1 \Sigma)}$. We apply the untwist function
 \begin{align*}
 \ranked{\tmonad \reduce 1 \Sigma \to \reduce 1 \tmonad \Sigma}
 \end{align*}
 then the function which transforms shallow terms into tensor product
 \begin{align*}
 \ranked{k \cdot \reduce 1 \tmonad \Sigma \to (\reduce 1 \tmonad \Sigma)^k}
 \end{align*}
 Now our term is of type $\ranked{\reduce k (\reduce 1 \tmonad \Sigma)^k}$. To conclude, we apply the prime function which permutes the tensor product with the fold, then we apply the product of the graded monad. 
 \end{enumerate}
\end{proof}

\subsubsection{Term unfolding for $\alpha$-homogeneous inputs}
\label{subsec:alpha-homo-unfold}

For a monotone function 
\begin{align*}
\alpha: \set{1,\ldots,k} \to \set{1,\ldots,k}
\end{align*}
we say that a term $ t \in \tmonad \mati k \rSigma$ is $\alpha$-homogeneous if all internal branches have twist $\alpha$. This section is devoted to proving the following lemma. 

\begin{lemma}\label{lem:homo-twist}
    Let $k \in \set{1,2,\ldots}$ and let $\alpha : \set{1,\ldots,k} \to \set{1,\ldots,k}$ be a monotone function. There is a derivable operation 
    \begin{align*}
        \ranked{f : \tmonad \mati k \rSigma \to \mati k {(\tmonad \Sigma)} }
        \end{align*}      
which coincides with term unfolding for all inputs which are $\alpha$-homogeneous.
\end{lemma}

\begin{proof}
We proceed by induction on $k$. When $k=1$, the unfolding coincides with the basic distributivity function 
\begin{align*}
\ranked{ \tmonad \reduce 1 \Sigma \to {\reduce 1 \tmonad \Sigma}}
\end{align*}
Let us treat the inductive case. For that, we introduce a tool that will be useful to analyze the function $\alpha$. For a function $$\alpha: \set{1,\ldots,k} \to \set{1,\ldots,k}$$ define  \emph{its graph} as the directed graph whose set of vertices is $\set{1,\ldots,k}$, and which contains an edge $i\rightarrow j$ if $\alpha(i)=j$. Note that the out-degree of the nodes is $1.$

\medskip
In the proof of the inductive case, we distinguish two cases. The first one is when the graph of $\alpha$ is not weakly connected. In this case, by monotonicity of $\alpha$, we can find $m\in \set{1,k-1}$ such that $\alpha(\set{1,m})\subseteq\set{1,m}$ and $\alpha(\set{m+1,k})\subseteq\set{m+1,k}$. The idea is then to create two copies of the original term: in the first one we keep only the first $m$ elements of the tensor product of each node, and in the second one we keep the last $k-m$ copies. Then we unfold these terms by applying the induction hypothesis, and  finally we gather them to obtain the  unfolding of the original term. 

Let us now implement the ideas we discussed above. We start by unfolding the external twists, using the basic external unfold function. This way, the domain of every external twist  cannot be shared by the two disconnected components of the domain of $\alpha$.
Then, we duplicate the input term using the basic function 
\begin{align*}
\ranked{\tmonad \mati k \Sigma\to \reduce 2 (\tmonad \mati k \Sigma\product \tmonad \mati k \Sigma)}
\end{align*}
To the first copy, we apply the function  
\begin{align*}
\ranked{f_1:\tmonad \mati k \Sigma \to \mati m {(\tmonad \Sigma)}}
\end{align*}
which keeps only the first $m$ elements of the tensor product, then applies the induction hypothesis to the obtained term.
To the second copy, we apply the function 
\begin{align*}
\ranked{f_2:\tmonad \mati k \Sigma \to \mati {k-m} {(\tmonad \Sigma)}}
\end{align*}
which keeps only the last $k-m$ elements of the tensor product, then applies the induction hypothesis to the obtained term.

 The function $\ranked{f_1}$ can be derived using the tensor projection function, the merge of folds, then reducing the fold and finally invoking the induction hypothesis. 

When we apply $\ranked{f_1}$ and $\ranked{f_2}$ to the two copies of the original term, we get a term of type 
\begin{align*}
\ranked{ \reduce 2 (\mati m {(\tmonad \Sigma)}\product  \mati {k-m} {(\tmonad\Sigma)})}
\end{align*}
At this point, we are almost done, we only need to transform the type in order to match the desired type. For that we increase the fold by applying the following prime functions
\begin{align*}
\ranked{\reduce m (\tmonad \Sigma)^m \to \reduce k (\tmonad \Sigma)^m \qquad \reduce {k-m} (\tmonad \Sigma)^{k-m} \to \reduce k (\tmonad \Sigma)^{k} }
\end{align*}
We swap the fold with the tensor product using the corresponding prime function, then we decrease the fold. This concludes the proof of the first case.
\medskip

Now consider the case where the graph of $\alpha$ is weakly connected. By monotonicity, we can show that either
\begin{align*}
\alpha^{-1}(1)=\emptyset\qquad\text{ or }\qquad\alpha^{-1}(k)=\emptyset
\end{align*}
By symmetry, we suppose wlog that $\alpha^{-1}(k)=\emptyset$. We suppose also that $\alpha(k)=k-1$, the general case can be treated in a similar way.
We consider as example the following function $\alpha$, whose graph, drawn below, is weakly connected
\begin{center}
\includegraphics[scale=.4]{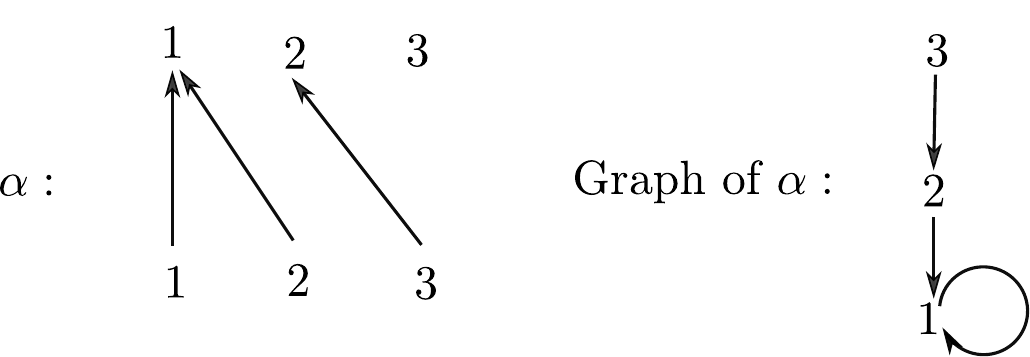}
\end{center}
Consider the function
\begin{align*}
\ranked{\reduce k \Sigma^k \to \reduce 2 (\reduce {k-1} \Sigma^{k-1}\product \Sigma)}
\end{align*}
which acts as in the following picture
\begin{center}
\includegraphics[scale=.4]{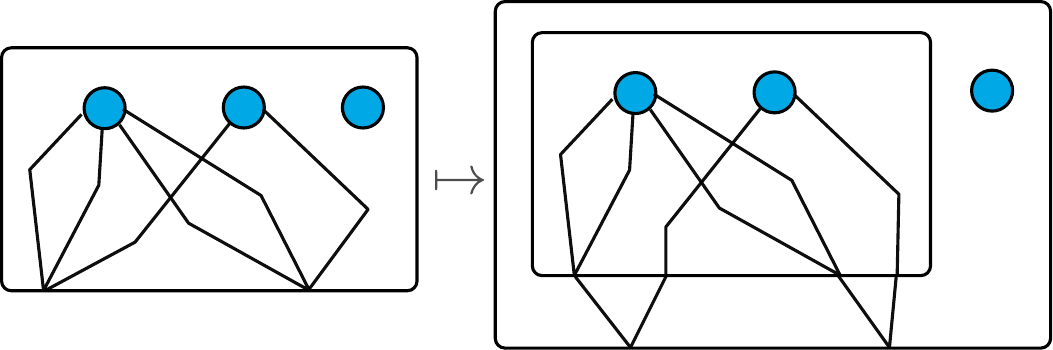}
\end{center}
If we inject both $\ranked{\reduce {k-1} \Sigma^{k-1}}$ and $\ranked{\Sigma}$ into $\rDelta\eqdef\ranked{\reduce {k-1} \Sigma^{k-1}+\Sigma}$, we get a term in the matrix power $\mati 2 \rDelta$. Note that the twists of such elements are the constant 1. We can then apply the unfolding function for constant-twists inputs
\begin{align*}
\ranked{\tmonad \mati 2 \Delta \to \mati 2 {\tmonad \Delta}}
\end{align*}
We can decompose the term $\tmonad \rDelta$ into  $\ranked{(\tmonad \reduce {k-1} \Sigma^{k-1})\odot \Sigma}$, by analyzing its structure. Now we can apply the induction hypothesis to unfold  $\ranked{\tmonad \reduce {k-1} \Sigma^{k-1}}$ into $\ranked{\reduce {k-1} (\tmonad \Sigma)^{k-1}}$. By applying the prime functions which permute the fold with the tensor product then increase the fold, we obtain the desired term.
\end{proof}

\subsubsection{Term unfolding for homogeneous inputs}
\label{subsec:something-homo-unfold}
we say that a term $ t \in \tmonad \mati k \rSigma$ is homogeneous if for every two internal branches $b_1, b_2$ having twists $\alpha_1, \alpha_2$ respectively and such that $b_2$ is a child of $b_1$, we have that
\begin{align*}
\alpha_1\alpha_2=\alpha_1
\end{align*}

The rest of this section is devoted to proving the following lemma. 

\begin{lemma}\label{lem:homo-2-twist}
    Let $k \in \set{1,2,\ldots}$. There is a derivable operation 
    \begin{align*}
        \ranked{f : \tmonad \mati k \rSigma \to \mati k {(\tmonad \Sigma)} }
        \end{align*}      
which coincides with term unfolding for all inputs which are homogeneous.
\end{lemma}

\begin{lemma}\label{lem:decomp-twist-2}
Let $\alpha:[1,k]\to [1,k]$ and $\beta:[1,k]\to [1,k]$ be two monotone functions such that
\begin{align*}
\alpha\beta=\alpha.
\end{align*}
 If the graph of $\alpha$ is not weakly connected, then so is the graph of $\beta$. Moreover, if $m\in \set{1,\dots, k}$ is such that 
\begin{align*}
\alpha[1,m] \subseteq [1,m]  \text{ and } 
\alpha[m+1,k]\subseteq [m+1,k]
\end{align*}
then we have also 
\begin{align*}
\beta[1,m]\subseteq [1,m] \text{ and }
\beta[m+1,k]\subseteq [m+1,k]
\end{align*}

 If the graphs of $\alpha$ and $\beta$ are both weakly connected, then  $\alpha$ and  $\beta$ are both constant functions.
\end{lemma}

\begin{proof}[Proof of Lemma~\ref{lem:homo-2-twist}]
We proceed by induction on $k$. The base case, ie when $k=1$ is realized by the untwist prime function. Let us treat the inductive case. First, we factorize our term in such a way that in each factor, either all the internal twits are weakly connected, or all of them are not weakly connected. To realize this factorization, it is enough to detect the first nodes (that is the closest to the root) where the twist becomes not weakly connected. Indeed, by Lemma\ref{lem:decomp-twist-2}, we know that the twits of the sub-tree rooted in such nodes are all not weakly connected. To detect these node, the following prime function is of particular interest
\begin{center}
\includegraphics[scale=.4]{pictures/last-prime-function}
\end{center}
If we analyze this factorization, it has the form $\ranked{\tmonad \mati k \Sigma \odot \tmonad \mati k \Sigma}$, where the root contains only connected internal twists and the leaves only weakly connected internal twists. By Lemma~\ref{lem:decomp-twist-2}, we know that the internal twists of the root  are constant, and that in each leave, there is an integer $m$ such that every internal twist $\alpha$ satisfies
\begin{align*}
\alpha[1,m] \subseteq [1,m]  \text{ and } 
\alpha[m+1,k]\subseteq [m+1,k].
\end{align*}
To unfold the root, we apply Proposition\ref{lem:constant-twist}. To unfold each leave, we proceed by induction, in the exact same way as the non-connected case in the proof of Proposition~\ref{lem:homo-twist}. Finally to untwist the whole term, we apply the prime shallow unfold function. 
\end{proof}

%% file: appendix-chain.tex
\section{Chain logic and general unfold}
\label{sec:appendix-chain}
In this section, we prove Theorem~\ref{thm:chain-transductions}, which says that adding general unfold to \mso yields exactly the chain logic tree-to-tree transductions. For the rest of this section, we use the word ``derivable'' to mean derivable in the extension of Definition~\ref{def:derivable-function} where general unfold is used instead of monotone unfold. 

To prove that every derivable function is a chain logic transduction, we use the same proof as in Appendix~\ref{sec:to-logic}. The only difference is that we need to deal with  general unfolding instead of monotone unfolding. For general unfolding, we use the same proof as in Section~\ref{sec:fo-transduction-for-unfolding}, with the only difference being in Lemma~\ref{lem:counter-free}. As opposed to the monotone case in Lemma~\ref{lem:counter-free}, we need to compose  not necessarily monotone partial functions. In the presence of non-monotone functions, the language corresponding to $L$ from Lemma~\ref{lem:counter-free} is no longer first-order definable, but it is still a regular language, and therefore it is definable in \mso. Chain logic can evaluate arbitrary \mso properties on paths in a tree, and therefore a formula of chain logic can be used to compute the twist function between two nodes in an input tree.

There rest of this appendix is devoted to the converse implication in Theorem~\ref{thm:chain-transductions}, which says that every chain logic tree-to-tree transduction is derivable, in the presence of general unfolding.

\paragraph*{Chain logic relabellings.}
Define  \emph{chain logic relabellings} in the same way as the first-order relabellings from Definition~\ref{def:forat}, except that chain logic is used instead of first-order logic.   As in Theorem~\ref{thm:mso-transductions} about \mso transductions, we push all of the power of chain logic into  tree relabellings.

\begin{lemma}\label{lem:chain-colcombet}
    Every   chain logic tree-to-tree transduction can be decomposed as: (a) a chain logic relabelling; followed by (b) a first-order tree-to-tree transduction. 
\end{lemma}
\begin{proof}[Proof sketch.]
    Same proof as in~\cite[Corollary 1]{colcombetCombinatorialTheoremTrees2007}, except that \mso is replaced  by chain logic. The key property is that the compositionality method, which is used in Lemmas 1 and 2 of~\cite{colcombetCombinatorialTheoremTrees2007},  also works for chain logic. 
\end{proof}

Thanks to the above lemma, and derivability of first-order tree-to-tree transductions from our main theorem, 
in order to finish the proof of Theorem~\ref{thm:chain-transductions}, it suffices to show that every chain logic relabelling is derivable.  
To prove derivability of chain logic relabellings, we  decompose them into simpler pieces. Unlike for first-order relabellings, where the decomposition was based on Schlingloff's theorem about temporal logic, in the case of chain logic we use an approach based on top-down tree automata\footnote{The results of this section could be translated into an apparently new result, which says that chain logic has the same expressive power as an  extension of Schlingloff's logic obtained by adding group modalities as defined by Baziramwabo, McKenzie and  Th{\'e}rien in~\cite[Section 4]{baziramwabo1999modular}.}. 

\paragraph*{Top-down tree automata.}
We begin by defining top-down tree automata. 
These are automata which process the input tree in a deterministic top-down (i.e.~root-to-leaves)  pass. Since we do not use nondeterministic top-down tree automata, we implicitly assume that the automata are deterministic.

\begin{definition}
    A \emph{top-down tree automaton}  is given by:
    \begin{enumerate}
        \item an \emph{input alphabet} $\rSigma$, which is a finite ranked set;
        \item a  finite unranked set of \emph{states} $Q$;
        \item a designated initial state in $Q$;
        \item \label{it:top-down-transition} for each input letter $a \in \rSigma$, a transition function
        \begin{align*}
        \delta_q : Q \to Q^{\text{arity of $a$}};
        \end{align*}
        \item an \emph{accepting set}, which is a subset of 
        \begin{align*}
        Q \times \text{(input letters of arity zero)}.
        \end{align*}
    \end{enumerate}
\end{definition}
For an input tree $t \in \trees \rSigma$, the \emph{run} of the automaton is defined to be the labelling of the nodes by states, which is  defined as follows by induction on the distance from the root. The state in the root is the initial state. Suppose that we have already defined the state $q$ in a node $x$ of the input tree. Apply the transition function, corresponding to the label of node $x$, to the state $q$, yielding a tuple of states $q_1,\ldots,q_n$. These are the states of the run in the children of node $x$. An input tree is accepted if for every leaf, the accepting set contains the pair (state in the leaf, label of the leaf).

\begin{definition}[Tree relabellings associated to a top-down tree automaton]
    \label{def:tree-relabellings-for-a-top-down-tree-automaton}
    We associate two tree-to-tree functions to a top-down tree automaton $\Aa$ with input alphabet $\rSigma$. Each  of these is a special cases of a chain logic relabelling. 
\begin{itemize}
    \item The \emph{ancestor relabelling}, is denoted by 
     \begin{align*}
        \Aa^\uparrow: \trees \rSigma \to \trees \ranked{(\black Q \times \rSigma)},
        \end{align*}
        where  $Q \ranked{\times \Sigma}$ is be the ranked set which consists of one copy of the alphabet $\rSigma$ for each state. The ancestor relabelling simply extends the input tree with the run of the automaton. Note that the accepting set  of the automaton does not play a role in the definition of the ancestors relabelling.
    \item The \emph{descendant relabelling}, denoted by
    \begin{align*}
        \Aa^\downarrow : \trees \rSigma \to \trees\ranked{(\rSigma + \rSigma)},
        \end{align*}
    is the characteristic function, in the sense of Section~\ref{sec:fo-translation}, of the query which selects nodes whose subtree is accepted by $\Aa$. In other words, for each node $x$ in the input tree, its label is replaced by the corresponding label in the first copy of $\rSigma$ if the subtree of $x$ is accepted by $\Aa$, and otherwise it is replaced by the corresponding label in the second copy of $\rSigma$. 
\end{itemize}
\end{definition}

The reason for  notation in the above definition is that, in the ancestor relabelling, the label of a node depends on its ancestors, while in the descendant  relabelling, the label of a node depends on its descendants. It is worth pointing out that many different runs of the automaton are used in the descendant relabelling, because for each node the automaton is started again with the initial state in that node. 

We begin with the following lemma, which states a connection between chain logic and (nestings of) top-down tree automata that was described in~\cite{bojanczykDecidablePropertiesTree2004}.
\begin{lemma}\label{lem:chain-phd}
    Every chain logic relabelling is a composition of functions which are either:
    \begin{itemize}
        \item[(a)] a letter-to-letter homomorphism; or 
        \item[(b)] the descendant relabelling of  a  top-down tree automaton.
    \end{itemize}
\end{lemma}
\begin{proof}
    Adjusting for a slightly different terminology, this lemma is the same as ~\cite[Theorem 2.5.9]{bojanczykDecidablePropertiesTree2004}. To help with the terminology, we note that the wordsum automata (WS) from~\cite{bojanczykDecidablePropertiesTree2004}  are the same as top-down tree automata here, while the cascade product of wordsum automata  is the same as composing descendant relabellings.
\end{proof}
Since letter-to-letter homomorphisms are derivable, in order to finish the proof of Theorem~\ref{thm:chain-transductions}, it  remains to prove that every function of kind (b) in the above lemma is  derivable. We prove this by doing a further decomposition, which reduces  the descendant relabelling  to the ancestor relabelling.

\begin{lemma}\label{lem:reduce-descendant-to-ancestor}
    For every top-down tree automaton, its descendant relabelling is  a composition of functions which are either:
    \begin{itemize}
        \item[(c)] a first-order relabelling; or 
        \item[(d)] the ancestor relabelling of a  top-down tree automaton.
    \end{itemize}
\end{lemma}
\begin{proof} In this proof, we use the forward Ramseyan splits  of Colcombet~\cite{colcombetCombinatorialTheoremTrees2007}.

    Fix a top-down tree automaton $\Aa$. 
      For the proof of this lemma, as well as for subsequent results, it will be convenient to use a different perspective on top-down tree automata, which uses automata on words. Recall the set of branches $\branches \rSigma$ that was defined in page~\pageref{page:branches}: a branch is a letter together with a distinguished port. Define the \emph{branch automaton} of $\Aa$ to be the deterministic word automaton, where the input alphabet is $\branches \rSigma$, the states are the same, the initial state is the same as in $\Aa$, and the transition function is defined by 
    \begin{align*}
    (\overbrace q^{Q}, \overbrace{(a,i)}^{\branches \rSigma}) \qquad \mapsto \qquad \text{$i$-th state in the tuple $\delta_a(q)$.}
    \end{align*}
    The branch automaton does not have accepting states.  Roughly speaking, the run of a top-down tree automaton corresponds to running the branch automaton on every root-to-leaf path in the tree. This correspondence is spelled out in more detail below.

    Consider two nodes in a tree, called the \emph{source} and \emph{target}, such that the source is an ancestor of the target. The source can be equal to the target. The \emph{path} between these two nodes is defined to be the set of edges in the tree which connects them. We can view the path as a word over the alphabet $\branches \rSigma$, as illustrated in the following picture:
    \mypic{114}
    The correspondence between the top-down tree automaton $\Aa$ and its  branch automaton can now be phrased as follows: for a node $x$, the state of the top-down tree automaton in node $x$ is the same as the state of the branch automaton after reading the (word corresponding to the) path from the root to node $x$.  

    Equipped with the above terminology, we complete the proof of the lemma. For a path in an input tree, define its  \emph{state transformation} to be the function of type $Q \to Q$ which describes the state transformation of the branch automaton over the (word corresponding to the) path.  By Colcombet's results on forward Ramseyan splits~\cite[Lemma 3]{colcombetCombinatorialTheoremTrees2007},   there is a top-down tree automaton $\Bb$ with input alphabet $\rSigma$  and a family of first-order formulas
\begin{align*}
\set{\varphi_f(x,y)}_{f : Q \to Q}
\end{align*}
with the following property. For every input tree $t \in \trees \rSigma$ and nodes $x \le y$ in that tree, the state transformation for the path from $x$ to $y$ is equal to $f$ if and only if 
\begin{align*}
\Bb^{\uparrow}(t) \models \varphi_f(x,y).
\end{align*}
The idea is that the  top-down tree automaton $\Bb$ computes the forward Ramseyan split associated to state transformations in the branch automaton of $\Aa$. It follows that there is a formula $\varphi(x)$ of first-order logic such that for every $t \in \trees \rSigma$, 
\begin{align*}
\Bb^{\uparrow}(t) \models \varphi(x)
\end{align*}
holds if and only if  the subtree of  node $x$ is accepted by the automaton $\Aa$. The formula says that for all leaves  $y \le x$, the corresponding state transformation of the branch automaton leads to an accepting state. Therefore, the descendant relabelling of $\Aa$ can be computed by first applying the ancestor relabelling of $\Bb$, and then a first-order relabelling.
\end{proof}

We now show that the ancestor relabellings produced by the previous lemma can be further decomposed, so that the underlying automata are reversible. 
Call a top-down tree automaton \emph{reversible} if the corresponding branch automaton, as defined in the proof of Lemma~\ref{lem:reduce-descendant-to-ancestor}, is reversible, which means that for every input letter the corresponding transition function is a permutation of the states. 
\begin{lemma}\label{lem:reduce-to-reversible}
    For every  top-down automaton, its ancestor function is a composition of functions which are either:
    \begin{itemize}
        \item[(c)] a first-order relabelling; or 
        \item[(e)] the ancestor relabelling of a reversible top-down automaton.
    \end{itemize}  
\end{lemma}
\begin{proof}
    A corollary of the original Krohn-Rhodes theorem. 
    
    Define a \emph{Mealy machine} to be a string-to-string transducer, which is obtained from a deterministic word automaton by adding an output function, which maps every transition to a letter of an output alphabet.  The original Krohn-Rhodes theorem says that every Mealy machine is a composition of Mealy machines where the underlying automaton is either aperiodic or reversible.  
    
    Take a top-down tree automaton. We can view its associated branch automaton  as a Mealy machine  which decorates each position in the input word by the state after reading the input word up to and including that position. To this Mealy machine apply the Krohn-Rhodes theorem. The relabellings for the aperiodic Mealy machines can be computed by the functions of kind (c), while the relabellings for the reversible ones correspond to kind (e).
\end{proof}

Putting together Lemmas~\ref{lem:chain-phd}, \ref{lem:reduce-descendant-to-ancestor} and~\ref{lem:reduce-to-reversible}, we see that every chain logic relabelling is a composition of functions which have kinds (c) or (e) as in the statement of Lemma~\ref{lem:reduce-to-reversible}. Since first-order relabellings are derivable, it remains to derive the functions of kind (e). 

\begin{lemma}
    For every reversible top-down tree automaton, its ancestor relabelling is derivable (in the presence of general unfolding).
\end{lemma}
\begin{proof}
    Let the states of the automaton be $Q = \set{q_1,\ldots,q_k}$. We assume that $q_1$ is the initial state.
    Consider an input letter $a$, and its associated transition function as in item~\ref{it:top-down-transition}. Here is a picture of such a transition function, where the letter $a$ is binary and the number of states is $k=3$.
    \mypic{116}
    In terms of the above picture, the reversibility of the automaton can be described as follows:
    \mypic{117}
    We can represent the above transition function as an element of the $k$-th matrix power of $Q$ copies of the states, denoted by 
    \begin{align*}
    \hat a \in \ranked{\mati k {(\black Q \times \rSigma)}},
    \end{align*}
    which is illustrated in the following picture:
    \mypic{115}
    More formally, $\hat a$ is defined so that for every port $i$ of the letter $a$, the $i$-th twist function (see Section~\ref{sec:unfolding}) is equal to the state transformation of the branch automaton when reading the letter $(a,i) \in \branches \rSigma$. The twist functions need not be monotone, since the branch automaton need not be monotone. 
    The reversibility of the automaton is crucial here; for a non-reversible automaton we might need to use a sub-port of the matrix power several times. 

    The transformation $a \mapsto \hat a$ is defined so that unfolding the matrix power captures exactly run computation in the top-down tree automaton, as described in the following commuting  diagram
    \begin{align*}
    \xymatrix@C=3cm{
        \trees \rSigma \ar[r]^{\trees(a \mapsto \hat a)} \ar[dr]_{\Aa^\uparrow}& 
        \trees \ranked{(\mati k {(\black Q \times \rSigma)})} \ar[d]^{
            \substack{
                \text{unfold and}\\
                \text{take coordinate $1$}
            }}\\
        & \trees{\ranked{(\black Q \times \rSigma)}}
    }
    \end{align*}
This completes the proof of the lemma. Note how general unfolding is used, since the twists involved need not be monotone.
\end{proof}